\documentclass[nofootinbib]{revtex4}
\usepackage{graphicx}
\usepackage{dcolumn}
\usepackage{amsmath,amssymb,epsfig}
\usepackage{paralist}
\usepackage{comment}
\usepackage{graphicx}
\usepackage{wrapfig}
\usepackage{multirow}
\usepackage{color,soul}
\usepackage{suffix}
\usepackage{mathtools}

\allowdisplaybreaks

\renewcommand{\vec}[1]{\boldsymbol{\mathrm{#1}}}

\let\Re\relax
\DeclareMathOperator{\Re}{Re}

\begin{document}

\title{Image formation for extended sources with the solar gravitational lens}

\author{Slava G. Turyshev$^{1}$, Viktor T. Toth$^2$}

\affiliation{\vskip 3pt
$^1$Jet Propulsion Laboratory, California Institute of Technology,\\
4800 Oak Grove Drive, Pasadena, CA 91109-0899, USA}

\affiliation{\vskip 3pt
$^2$Ottawa, Ontario K1N 9H5, Canada}

\date{\today}

\begin{abstract}

We study the image formation process with the solar gravitational lens (SGL) in the case of an extended, resolved source. An imaging telescope, modeled as a convex lens, is positioned within the image cylinder formed by the light received from the source. In the strong interference region of the SGL, this light is greatly amplified, forming the Einstein ring around the Sun, representing a distorted image of the extended source. We study the intensity distribution within the Einstein ring observed in the focal plane of the convex lens.  For any particular telescope position in the image plane, we model light received from the resolved source as a combination of two signals: light received from the directly imaged region of the source and light from the rest of the source. We also consider the case when the telescope points away from the extended source or, equivalently, it observes light from sources in sky positions that are some distance away from the extended source, but still in its proximity.  At even larger distances from the optical axis, in the weak interference or geometric optics regions, our approach recovers known models related to microlensing, but now obtained via the wave-optical treatment. We then derive the power of the signal and related photon fluxes within the annulus that contains the Einstein ring of the extended source, as seen by the imaging telescope. We discuss the properties of the deconvolution process, especially its effects on noise in the recovered image. We compare anticipated signals from realistic exoplanetary targets against estimates of noise from the solar corona and estimate integration times needed for the recovery of high-quality images of faint sources. The results demonstrate that the SGL offers a unique, realistic capability to obtain resolved images of exoplanets in our galactic neighborhood.

\end{abstract}


\maketitle

\section{Introduction}
\label{sec:aintro}

As a consequence of the gravitational diffraction of light \cite{Turyshev:2017,Turyshev-Toth:2017}, electromagnetic (EM) waves traveling from distant sources in the close proximity of the Sun are focused by the solar gravitational field at heliocentric distances beyond  $\overline z\simeq b^2/(2r_g)\gtrsim 547.6 \,(b/R_\odot)^2$ astronomical units (AU), where $b$ is a light ray's impact parameter, $r_g=2GM_\odot/c^2$ is the Schwarzschild radius of the Sun and $R_\odot$ is its radius. This diffraction process is characterized by truly remarkable properties: At optical or near infrared wavelengths, it offers light amplification of up to a factor of $4\pi^2 r_g/\lambda\simeq 2.1\times 10^{11}\, (1\,\mu{\rm m}/\lambda)$, and angular resolution of up to $\simeq0.38\,{\lambda}/{b}=0.10\,({\lambda}/{1\,\mu{\rm m}})(R_\odot/b)$ nanoarcseconds (nas) \cite{Turyshev:2017,Turyshev-Toth:2017,Turyshev-Toth:2019-extend}.

\begin{figure}
\includegraphics[scale=0.25]{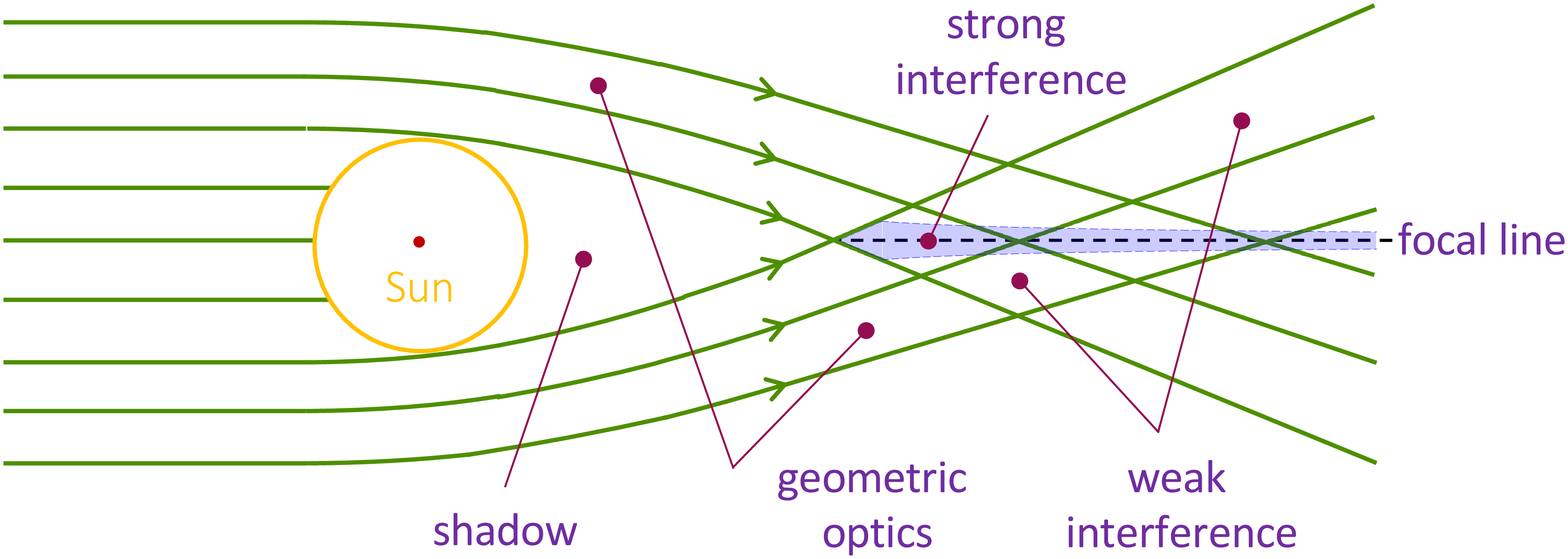}
\caption{\label{fig:regions}The different optical regions of the SGL
(adapted from \cite{Turyshev-Toth:2019-extend}).
}
\end{figure}

The resulting solar gravitational lens (SGL) allows for extraordinary observational capabilities, including, for instance, direct high-resolution imaging and spectroscopy of Earth-like exoplanets  \cite{Turyshev-etal:2018}. We can benefit from this unique natural `instrument' with the help of a meter-class telescope, equipped with a solar coronagraph (which is needed to block the solar light), and positioned in the strong interference region of the SGL (see Fig.~\ref{fig:regions}) with respect to the intended imaging target. Until recently such deep space missions were hard to contemplate, but with recent reports on the Voyager 1 spacecraft reaching distances beyond 140 AU while still transmitting valuable data after more than 42 years of continuous operation, and with advances in spacecraft miniaturization and progress in propulsion technologies, efforts to explore the space outside our solar system have intensified \cite{KISS:2015,Turyshev-etal:2018}.

Recognizing its value for astronomy and astrophysics, recently we investigated the optical properties of the SGL and developed its wave-optical treatment \cite{Turyshev-Toth:2017,Turyshev-Toth:2019,Turyshev-Toth:2019-extend}. With this knowledge, we studied photometric imaging with the SGL \cite{Turyshev-Toth:2019-blur}, estimating the total power that is incident on the aperture of an imaging telescope, thus measuring the amplitude of the incident signal. As part of the investigation,  we studied the fact that imaging of extended sources with the SGL is affected by blurring, due to the SGL's inherent spherical aberration. With these results at hand, we  investigated the process of image formation of point sources using an optical telescope placed in the SGL focal region \cite{Turyshev-Toth:2019-image}. We derived analytical expressions that can be used to model extended sources using numerical tools.

In the present paper, we investigate the image formation process by an optical telescope in the SGL focal region, viewing an extended, resolved source positioned at a large, but finite distance from the Sun. This investigation of the imaging process requires knowledge not only of the amplitude of the signal, but also its phase.  Our objective is to derive analytical expressions that may be used to evaluate signals from realistic targets, which is important for a variety of potential astronomical applications of the SGL. To assess realistic observing scenarios in the context of a potential deep space mission, we also study the process of deconvolving blurred SGL images under realistic conditions in the presence of various sources of noise. We provide the theoretical foundation to address these important questions. Our ultimate goal is to offer analytical tools to compute photon fluxes from realistic sources, to estimate detection SNRs, required integration times for a given observing scenario, to evaluate the quality of reconstructed images and, by doing so, to move the concept of imaging with the SGL from a domain of theoretical physics to the mainstream of astronomy and astrophysics.

Our paper is organized as follows:
Section~\ref{sec:im-form} introduces the SGL and the solution for the EM field in the image plane in the strong interference region behind the Sun.
Section \ref{sec:image-sens} discusses the modeling of the intensity distribution observed  in the focal plane behind the convex lens. We present the total signal received from the extended source as consisting of two parts: the signal from the directly imaged region of the source and the blur received from the rest of the source. Although our basic results are generic, to allow for the analytic evaluation of realistic observing scenarios, we model the source as a uniformly illuminated disk. This approach allows us to develop analytical expressions to estimate the total photon flux received by the telescope.
In Section \ref{sec:weak-int} we study image formation in the geometric optics and weak interference regions, thus extending our results to all the optical regions behind the Sun and demonstrating the compatibility of our results with known microlensing models.
In Section~\ref{sec:power} we derive the power deposited in the focal plane of the imaging telescope from the directly imaged region of the target object, the rest of the target and also light contamination from off-target sources. We estimate the photon flux received at the detector from a realistic distant target for various cases of the image-telescope geometries. We estimate the resulting SNRs in the presence of light from the solar corona, which is the dominant source of noise.
In Section \ref{sec:convolve} we develop an approach to evalaute the ``deconvolution penalty'', the amount by which measurement noise is amplified by the deconvolution process that is used to recover a high-quality image from observations blurred by the SGL. We evaluate the integration times needed to obtain direct, high-quality resolved images of exoplanets, and demonstrate the superiority of the SGL compared to exoplanet imaging scenarios unaided by the SGL.
In Section \ref{sec:disc} we discuss results and explore avenues for the next phase of our investigation of imaging and spectroscopy of exoplanets with the SGL.
Finally, Appendix \ref{sec:model} contains a brief analysis of the solar corona using the same methodology applied in the rest of the paper, offering a suitable basis for comparison. In Appendix~\ref{sec:PSF-average} we derive a form of the point-spread function of the SGL that is averaged  over the aperture of an optical telescope and discuss the properties of this averaged formulation.

\section{Image formation process with the SGL}
\label{sec:im-form}

\subsection{The EM field in the strong interference region}
\label{sec:EM-field}

\begin{figure}[h]
\includegraphics[scale=0.7]{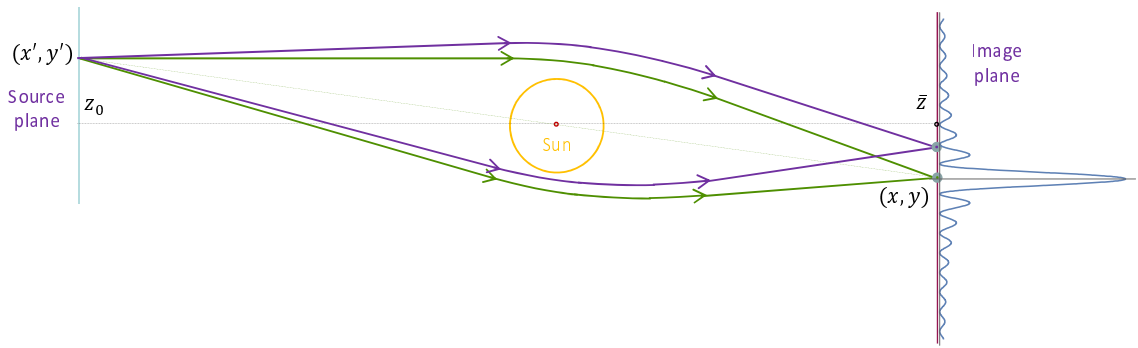}
\caption{\label{fig:imaging-geom}The geometry of imaging a point source with the SGL. A point source with coordinates $(x',y')$ is positioned in the source plane, at the distance $z_0$ from the Sun. The SGL image plane is at the heliocentric distance ${\overline z}$. Rays with different optical paths produce a diffraction pattern in the SGL image plane that is observed by an imaging telescope.}
\end{figure}

In \cite{Turyshev-Toth:2019-extend}, we considered light from an extended source at a finite distance, $z_0$ from the Sun. We parameterize the problem using a heliocentric spherical coordinate system $(r,\theta,\phi)$ that is aligned with a preferred axis: a line connecting a preselected (e.g., central) point in the source to the center of the Sun, as shown in Fig.~\ref{fig:imaging-geom}. We also use of a cylindrical coordinate system $(\rho,z,\phi)$, with the $z$-axis corresponding to the preferred axis. Furthermore, we characterize points in the image plane and the source plane (both perpendicular to the $z$-axis) using 2-dimensional vector coordinates $\vec{x}$ and $\vec{x}'$, respectively.

We consider light, modeled as a monochromatic high-frequency EM wave (i.e., neglecting terms $\propto(kr)^{-1}$ where $k=2\pi/\lambda$ is the wavenumber) coming from a source at the distance of $r_0=(z_0^2+|{\vec x}'|^2)^\frac{1}{2}\simeq z_0\gg r_g$ from the Sun (see Fig.~\ref{fig:imaging-geom}) and received on the opposite side of it at the heliocentric distance of $r=(\overline z^2+|{\vec x}|^2)^\frac{1}{2}\simeq \overline z\gg r_g$, we derived the components of the EM field near the optical axis in the strong interference region of the SGL (see Fig.~\ref{fig:regions}). Up to terms of ${\cal O}(\rho^2/z^2, \sqrt{2r_g\overline z}/z_0)$, the components of such an EM field take the form \cite{Turyshev-Toth:2019-extend,Turyshev-Toth:2019-blur,Turyshev-Toth:2019-image}
{}
\begin{eqnarray}
    \left( \begin{aligned}
{E}_\rho& \\
{H}_\rho& \\
  \end{aligned} \right) =    \left( \begin{aligned}
{H}_\phi& \\
-{E}_\phi& \\
  \end{aligned} \right)&=&
  \frac{E_0}{z_0}  \sqrt{2\pi kr_g}e^{i\sigma_0}
  J_0\Big(\frac{2\pi}{\lambda}
\sqrt{\frac{2r_g}{\overline z}}
|{\vec x}+\frac{\overline z}{{ z}_0}{\vec x'}|\Big)
    e^{i\big(k(r+r_0+r_g\ln 2k(r+r_0))-\omega t\big)}
 \left( \begin{aligned}
 \cos\phi& \\
 \sin\phi& \\
  \end{aligned} \right),
  \label{eq:DB-sol-rho}
\end{eqnarray}
where the $z$-components of the EM wave behave as $({E}_z, {H}_z)\sim {\cal O}({\rho}/{z}, \sqrt{2r_g\overline z}/z_0)$. The quantity $\overline z=z(1+z/z_0+{\cal O}(z^2/z_0^2))$ denotes heliocentric distances along the line connecting the point source and the center of the Sun (see Fig.~\ref{fig:imaging-geom}). Note that these expressions are valid for forward scattering when $\theta+ \sqrt{2r_g\overline z}/z_0\approx 0$, or when $0\leq \rho\leq r_g$.

We can describe the imaging of an extended source. For that, we use the solution for the EM field (\ref{eq:DB-sol-rho}) and study the Poynting vector, ${\vec S}=(c/4\pi)\big<\overline{[\Re{\vec E}\times\Re{\vec H}]}\big>$, that describes the energy flux in the image plane \cite{Wolf-Gabor:1959,Richards-Wolf:1959,Born-Wolf:1999}. Normalizing this flux to the time-averaged value that would be observed if the gravitational field of the Sun were absent, $|\overline{\vec S}_0|=(c/8\pi)E_0^2/z_0^2$, we define the amplification factor of the SGL, ${ \mu}_{\tt SGL}=|{\vec S}|/|\overline{\vec S}_0|$:
{}
\begin{eqnarray}
{ \mu}_{\tt SGL}({\vec x},{\vec x}')&=&
\mu_0J^2_0\Big(\frac{2\pi}{\lambda}
\sqrt{\frac{2r_g}{\overline z}}
|{\vec x}+\frac{\overline z}{{ z}_0}{\vec x'}|\Big),
\qquad {\rm with} \qquad
\mu_0=\frac{4\pi^2}{1-e^{-4\pi^2 r_g/\lambda}}\frac{r_g}{\lambda}\simeq1.17\times 10^{11}\,
\Big(\frac{1\,\mu{\rm m}}{\lambda}\Big).
\label{eq:S_z*6z-mu2}
\end{eqnarray}

The angular resolution of the SGL is determined by the first zero of the Bessel function $J_0(x)$ in (\ref{eq:S_z*6z-mu2}), which occurs at $x=2.4048$ and yields
{}
\begin{eqnarray}
R_{\tt SGL}=\big|\frac{{\vec x}}{\overline z}+\frac{{\vec x'}}{{ z}_0}\big|=0.38 \frac{\lambda}{\sqrt{2r_g\overline z}}=0.10\Big(\frac{\lambda}{1\,\mu{\rm m}}\Big)\Big(\frac{650\,{\rm AU}}{\overline z}\Big)^\frac{1}{2}~{\rm nas}.
\label{eq:S_=}
\end{eqnarray}
Note that by setting ${\vec x}'=0$ in (\ref{eq:S_=}), we recover the SGL's resolution for point sources \cite{Turyshev-Toth:2017}. Let us compare the SGL to a conventional optical telescope with aperture $d$ and focal length of $f$. Its light amplification is known to be   \cite{Born-Wolf:1999,Goodman:2017} (see also the relevant derivations in Appendix~\ref{sec:model}, for instance, (\ref{eq:pow-cor})):
{ }
\begin{eqnarray}
\mu_{\tt tel}({\vec x},{\vec x}')&=&
i_0 \Big( \frac{2
J_1\big(u\frac{1}{2}d\big)}{u\frac{1}{2}d}\Big)^2, \qquad {\rm with}\qquad  i_0= \Big(\frac{kd^2}{8f}\Big)^2 \qquad {\rm and} \qquad u =\frac{\pi d}{\lambda}\big|\frac{{\vec x}}{\overline z}+\frac{{\vec x'}}{{ z}_0}\big|.
  \label{eq:amp=*}
\end{eqnarray}
As it is well known, it is the first zero of the Bessel function $J_1(x)$ at $x=1.220\pi$ in (\ref{eq:amp=*}) that determines the telescope's resolution:
{}
\begin{eqnarray}
R_{\tt tel}=\big|\frac{{\vec x}}{\overline z}+\frac{{\vec x'}}{{ z}_0}\big|=1.22\, \frac{\lambda}{d}=0.21\Big(\frac{\lambda}{1\,\mu{\rm m}}\Big)\Big(\frac{1\,{\rm m}}{d}\Big)~{\rm as},
\label{eq:S_=0}
\end{eqnarray}
which is more than $2\times10^9$ times less than that of the SGL. Again, by setting ${\vec x}'=0$ in (\ref{eq:S_=0}), we recover the familiar expression for the angular resolution of an optical telescope for point sources \cite{Born-Wolf:1999,Goodman:2017}.

However, the impressive amplification and angular resolution of the SGL (\ref{eq:S_=})  come at a price, which is the spherical aberration inherent in the SGL's optical properties \cite{Turyshev-Toth:2019-blur}. To discuss the impact of this aberration on the prospective imaging with the SGL, it is convenient to introduce its point-spread function (PSF),  given by ${\rm PSF}={ \mu}_{\tt SGL}({\vec x},{\vec x}')/\mu_0=J^2_0\big(({2\pi}/{\lambda}) \sqrt{{2r_g}/{\overline z}} |{\vec x}+({\overline z}/{{ z}_0}){\vec x'}|\big)$.  This expression (\ref{eq:S_z*6z-mu2}) is the PSF of the SGL, scaled by the amplification factor on the optical axis,  $\mu_0$. (Note that (\ref{eq:amp=*}) does the same, by scaling the PSF of an optical telescope, $\propto (2J_1(x)/x)^2$, using the intensity at the center, $i_0$.)

The PSF concept is used in Fourier optics to describe the properties of an imaging system characterized by its diffraction pattern \cite{Born-Wolf:1999,Goodman:2017}. In fact, the imaging system's resolution can be limited either by aberration or by diffraction causing blurring of the image. These two phenomena have different origins and are unrelated. The PSF describes the interplay between diffraction and aberration: the smaller the aperture of a lens the more likely the PSF is dominated by diffraction. As was discussed in \cite{Turyshev-Toth:2019-extend}, the  PSF of the SGL is rather broad, behaving as $\propto 1/\rho$, as the distance from the optical axis, $\rho=|{\vec x}+({\overline z}/{{ z}_0}){ \vec x'}|$, increases. The PSF of an optical telescope (\ref{eq:amp=*})  falls off much faster, behaving as $\propto 1/\rho^3$. It is this behavior of the monopole SGL that is responsible for the considerable blurring of any image that forms in the SGL's image plane. However, given that the PSF of the SGL is known, its inverse can be used to reconstruct the original image \cite{Turyshev-etal:2018}. Below we will consider the impact of the SGL blur on the image quality.

Examining (\ref{eq:S_z*6z-mu2}) and recognizing that (\ref{eq:S_=}) is extremely small, we see that a monopole gravitational lens acts as a convex lens by focusing light, according to
{}
\begin{equation}
{\vec x}=-\frac{\overline z}{z_0}{\vec x}' \qquad \rightarrow \qquad x=-\frac{\overline z}{z_0}x', \qquad y=-\frac{\overline z}{z_0}y'.
\label{eq:mapping}
\end{equation}
These expressions imply that the SGL focuses light in the opposite quadrant in the image plane while also reducing the size of the image compared to the source by a factor of ${\overline z}/{z_0}\sim1.0\times 10^{-4}\,({\overline z}/650 ~{\rm AU}) (30~{\rm pc}/z_0)$. For an exoplanet with radius $R_\oplus$, positioned at a distance of $z_0$ from the Sun, the image of this target at a heliocentric distance of ${\overline z}$, will be compressed to a cylinder with radius
{}
\begin{equation}
r_\oplus=\frac{\overline z}{z_0}R_\oplus=669.98\,\Big(\frac{\overline z}{650 ~{\rm AU}}\Big) \Big(\frac{30~{\rm pc}}{z_0}\Big)~{\rm m}.
\label{eq:rE}
\end{equation}
A telescope with aperture $d\ll r_\oplus$ would have to scan this image by traversing and sampling the image plane at multiple locations to recover the image.

\begin{figure}
\includegraphics[scale=0.9]{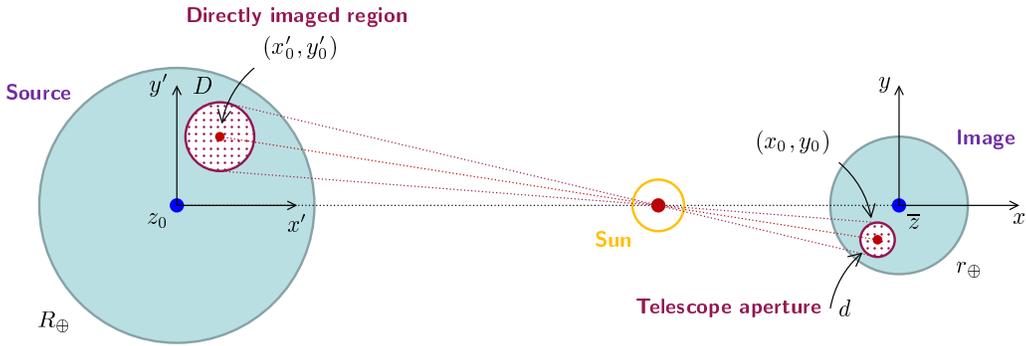}
\caption{\label{fig:dir-image}Imaging of extended resolved sources with the SGL. The SGL is a convex lens, producing inverted images of a source.}
\end{figure}

Consider the process of imaging an extended, resolved source. In the most widely considered practical scenario, the kilometer-scale image plane is sampled by a telescope with a meter-scale aperture. Such a telescope has the resolution required to employ a coronagraph, but it is otherwise used as a photometric detector,
measuring the brightness of the Einstein ring that forms around the Sun from light originating from the exoplanet. First, we recognize that the telescope's aperture is much smaller than the image size, $d\ll 2r_\oplus$. This leads us to separate the received signal into two parts: the signal received from the directly imaged region that corresponds to the telescope location, and the blur due light received from the rest of the source. Based on the SGL's mapping (\ref{eq:mapping}) for a given point $(x_0,y_0)$ in the image plane (Fig.~\ref{fig:dir-image}), the directly imaged region will be in the vicinity of the point $(x'_0,y_0')=-(z_0/\overline z)(x_0,y_0)$ in the source plane.  Furthermore, given the telescope aperture $d$, the directly imaged region in the source plane has the diameter
{}
\begin{equation}
D=\frac{z_0}{\overline z}d =9.52\,\Big(\frac{d}{1 ~{\rm m}}\Big) \Big(\frac{650 ~{\rm AU}}{\overline z}\Big) \Big(\frac{z_0}{30~{\rm pc}}\Big)~{\rm km},
\label{eq:Dd}
\end{equation}
centered at $(x'_0,y_0')$. The signal that is received from the areas outside of $D$ on the source is causing the blur \cite{Turyshev-Toth:2019-blur}. Using (\ref{eq:rE}) and (\ref{eq:Dd}), we see that a telescope with the aperture $d$ could resolve an exoplanet whose radius is $R_{\tt exo}$ with $N_d$ linear resolution elements (see Fig.~\ref{fig:dir-image}) given by
{}
\begin{equation}
N_d=\frac{2R_\oplus}{D}\Big(\frac{R_{\tt exo}}{R_\oplus}\Big)=\frac{2r_\oplus}{d}\Big(\frac{R_{\tt exo}}{R_\oplus}\Big)=1339.95\,\Big(\frac{1 ~{\rm m}}{d}\Big) \Big(\frac{\overline z}{650 ~{\rm AU}}\Big) \Big(\frac{30~{\rm pc}}{z_0}\Big)\Big(\frac{R_{\tt exo}}{R_\oplus}\Big).
\label{eq:res-el}
\end{equation}

\subsection{Image formation by an optical telescope in the SGL image plane}
\label{sec:image-form-Fourier}

\begin{figure}
\includegraphics[scale=0.65]{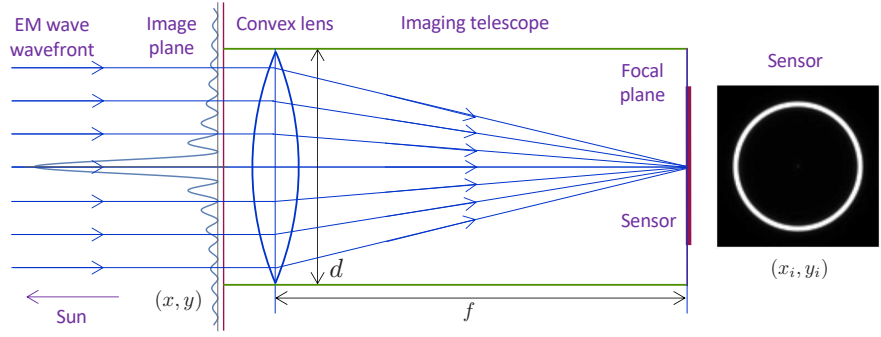}
\caption{\label{fig:imaging-sensor}Imaging a point source with the SGL with a telescope. The telescope is positioned on the optical axis that connects the source and the Sun and it ``sees'' the Einstein ring. The telescope is represented by a convex lens with a diameter $d$ and a focal length $f$. Positions in the SGL image plane, $(x,y)$, and the optical telescope's focal plane, $(x_i,y_i)$, are also shown. }
\end{figure}

To produce images of faint, distant objects with the SGL, we represent an imaging telescope by a  convex lens with aperture $d$ and focal distance $f$; see Fig.~\ref{fig:imaging-sensor}. We position the telescope at a point with coordinates ${\vec x}_0$ in the image plane in the strong interference region of the lens (Fig.~\ref{fig:regions}) \cite{Nambu:2013,Kanai-Nambu:2013,Nambu:2013b,Born-Wolf:1999,Turyshev-Toth:2019-extend}.  To stay within the image, ${\vec x}_0$ is within the range:  $|{\vec x}_0|+d/2\leq r_\oplus$. The amplitude of the EM wave just in front of the telescope aperture, from (\ref{eq:DB-sol-rho}), is given as
{}
\begin{eqnarray}
{\cal A}({\vec x},{\vec x}_0, {\vec x}')&=&
\sqrt{\mu_0} J_0\Big(k
\sqrt{\frac{2r_g}{\overline z}} |{\vec x}+{\vec x}_0+\frac{\overline z}{{ z}_0}{\vec x}'|\Big).
  \label{eq:amp-w}
\end{eqnarray}

The presence of a convex lens is equivalent to a Fourier transform of the wave (\ref{eq:amp-w}).  The focal plane of the optical telescope is located at the focal distance $f$ of the lens, centered on ${\vec x}_0$. Using the Fresnel--Kirchhoff diffraction formula, the amplitude of the image field in the optical telescope's focal plane at a location ${\vec x}_i=(x_i,y_i)$ is given by  \cite{Wolf-Gabor:1959,Richards-Wolf:1959,Born-Wolf:1999}:
{}
\begin{eqnarray}
{\cal A}({\vec x}_i,{\vec x}_0, {\vec x}')=\frac{i}{\lambda}\iint \displaylimits_{|{\vec x}|^2\leq (d/2)^2} \hskip -7pt  {\cal A}({\vec x},{\vec x}_0, {\vec x}')e^{-i\frac{k}{2f}|{\vec x}|^2}\frac{e^{iks}}{s}d^2{\vec x}.
  \label{eq:amp-w-f0}
\end{eqnarray}
The function $e^{-i\frac{k}{2f}|{\vec x}|^2}=e^{-i\frac{k}{2f}(x^2+y^2)}$ represents the action of the convex lens that transforms incident plane waves to spherical waves, focusing at the focal point. Assuming that the focal length is sufficiently greater than the radius of the lens, we may approximate the optical path $s$ as $s=\sqrt{(x-x_i)^2+(y-y_i)^2+f^2}\sim f+\big((x-x_i)^2+(y-y_i)^2\big)/2f$. This allows us to present (\ref{eq:amp-w-f0}) as
{}
\begin{eqnarray}
{\cal A}({\vec x}_i,{\vec x}_0,{\vec x}')&=&
-  \sqrt{\mu_0} \frac{e^{ikf(1+{{\vec x}_i^2}/{2f^2})}}{i\lambda f}\iint\displaylimits_{|{\vec x}|^2\leq (\frac{1}{2}d)^2} d^2{\vec x}
  J_0\Big(k
\sqrt{\frac{2r_g}{\overline z}} |{\vec x}+{\vec x}_0+\frac{\overline z}{{ z}_0}{\vec x}'|\Big) e^{-i\frac{k}{f}({\vec x}\cdot{\vec x}_i)}.
  \label{eq:amp-w-f}
\end{eqnarray}

To account for the propagation distance between the source and the image plane,  we recognize that the field strength, $E_0/z_0$, of the plane wave in (\ref{eq:DB-sol-rho}) is a function of the coordinates on the source plane, namely $E_0({\vec x}')/{\bar r}$, where $\bar r$ is distance between a point on the source plane with coordinates of $({\vec x}',-z_0)$ and a point on the image plane with coordinates of $({\vec x}+{\vec x}_0,\overline z)$, namely $\bar r=(({\vec x}+{\vec x}_0-{\vec x}')^2+({\overline z}+z_0)^2)$. Given the fact that $z_0\gg \{|{\vec x}'|, {\overline z},|{\vec x}+{\vec x}_0|\}$, we may approximate $r\simeq z_0+{\cal O}({\overline z}^2/z_0^2)$, yielding the transformation of the field strength as $E_0/z_0\rightarrow E_0({\vec x}')/z_0$. Note that we do not approximate the phase of the EM wave (\ref{eq:DB-sol-rho}), only its amplitude. This is because the phase  is the quantity of our primary interest for the SGL, thus, we need to know it with the most available precision.

Next, with the amplitude ${\cal A}({\vec x}_i,{\vec x}_0,{\vec x}')$ given by (\ref{eq:amp-w-f}), the EM field (\ref{eq:DB-sol-rho}) in the focal plane of the telescope (indicated by subscript ${\vec x}_i$) produced by a point source positioned in the  source plane at coordinates ${\vec x}'$ (Figs.~\ref{fig:imaging-geom}, \ref{fig:dir-image}) is given as
{}
\begin{eqnarray}
    \left( \begin{aligned}
{E}_\rho& \\
{H}_\rho& \\
  \end{aligned} \right)_{\hskip -3pt {\vec x}_i} =    \left( \begin{aligned}
{H}_\phi& \\
-{E}_\phi& \\
  \end{aligned} \right)_{\hskip -3pt \vec x_i} &=&\frac{{E}_0({\vec x}')}{z_0}
  {\cal A}({\vec x}_i,{\vec x}_0,{\vec x}')
    e^{i\big(k(r+r_0+r_g\ln 2k(r+r_0))-\omega t\big)}
 \left( \begin{aligned}
 \cos\phi& \\
 \sin\phi& \\
  \end{aligned} \right).
  \label{eq:DB-sol-rho2}
\end{eqnarray}

With this expression, we may compute the Poynting vector of the EM field that originates at a point source at coordinates $\vec x'$ in the source plane, is captured by a telescope with aperture $d$ in the image plane centered on coordinates ${\vec x}_0$, and is finally received in the telescope's image plane at ${\vec x}_i$. Given the form (\ref{eq:DB-sol-rho2}) of the EM field, the Poynting vector will have only one nonzero component, $S_z$. With overline and brackets denoting time-averaging and ensemble averaging (over the source's surface), correspondingly, and defining $\Omega(t)=k(r+r_0+r_g\ln 2k(r+r_0))-\omega t$, we compute $S_z$ as
 {}
\begin{eqnarray}
S_z({\vec x}_i,{\vec x}_0,{\vec x}')=\frac{c}{4\pi}\big<\overline{[\Re{\vec E}\times\Re{\vec H}]}_z\big>=\frac{c}{4\pi}\frac{E_0^2}{z_0^2}
\big<\overline{\big(\Re\big[{\cal A}({\vec x}_i,{\vec x}_0,{\vec x}')e^{i\Omega(t)}\big]\big)^2}\big>.
  \label{eq:Pv}
\end{eqnarray}

Dividing this expression by the time-averaged Pointing vector of a spherical EM wave propagating in the absence of gravity that would be received at the same location but before entering the telescope \cite{Born-Wolf:1999}, $|\overline{{\vec S}}_0|=({c}/{8\pi}) E_0^2/z_0^2,$ we obtain the  amplification factor, $\mu({\vec x}_i,{\vec x}_0,{\vec x}')=S_z({\vec x}_i,{\vec x}_0,{\vec x}')/|\overline{{\vec S}}_0|$  of the optical system consisting of the SGL and an imaging telescope, i.e., the convolution of the PSF of the SGL with that of an optical telescope:
 {}
\begin{eqnarray}
\mu({\vec x}_i,{\vec x}_0,{\vec x}')=2
\big<\overline{\big(\Re\big[{\cal A}({\vec x}_i,{\vec x}_0,{\vec x}')e^{i\Omega(t)}\big]\big)^2}\big>.
  \label{eq:psf}
\end{eqnarray}

To compute the intensity distribution corresponding to the light received from the entire extended source and received in the focal plane of the imaging telescope, we need to  form a product of the source's surface brightness per unit area, $B_{\tt s}({\vec x}')\propto {E}^2_0({\vec x}')$ with dimensions of ${\rm W \,m}^{-2}{\rm sr}^{-1}$, and the PSF from (\ref{eq:psf}), and integrate the result over the entire surface of the source.  Therefore, the intensity distribution on the detector at the focal plane of the optical telescope that is positioned on the image plane in the strong interference region of the SGL, may be presented as
  {}
\begin{eqnarray}
I({\vec x}_i,{\vec x}_0) =\frac{1}{z^2_0} \iint d^2{\vec x}'  B_{\tt s}({\vec x}') \mu({\vec x}_i,{\vec x}_0,{\vec x}'),
  \label{eq:power}
\end{eqnarray}
which accounts for the fact that the EM field originating at the extended source is not spatially coherent.

As a result, to compute the power received by a detector in the focal plane of an imaging telescope positioned it the SGL image plane, we need to first compute the Fourier transform of the complex amplitude of the EM field (\ref{eq:amp-w-f}) and then follow the process that is outlined above and is captured by (\ref{eq:psf}) and (\ref{eq:power}). This approach allows one to employ the powerful tools of  Fourier optics (e.g.,  \cite{Goodman:2017})  to develop practical applications of the SGL.

\section{Modeling the signal in the focal plane of an optical telescope}
\label{sec:image-sens}

In the previous section we obtained expressions that characterize the intensity distribution of light originating at a distant, extended source and received by an imaging telescope in the image plane.  We now consider the intensity distribution in the focal plane of an optical telescope. We recognize that an actual astrophysical telescope is a complex instrument and has physical limitations related to its design and manufacturing specifications. In our present analysis, we use an idealized model in the form of an optically perfect convex thin lens. This is sufficient to study the principles of image formation in the telescope image plane.

\subsection{Complex amplitude in the focal plane}
\label{sec:extended-image}

Expression (\ref{eq:amp-w-f}) is rather complex and cannot be evaluated analytically in the general case. Such expressions are usually evaluated numerically instead, often in the spatial frequency domain after a Fourier-transform \cite{Goodman:2017}. However, some useful analytical approximations do exist, which we explore here.

To simplify the discussion, it is convenient to express the position ${\vec x}_0$ of the telescope in the SGL image plane via the coordinates ${\vec x}_0'$ of the corresponding central position of the directly imaged region in the source plane (see Fig.~\ref{fig:dir-image}).  Using the mapping (\ref{eq:mapping}), this can be done as
{}
\begin{equation}
{\vec x}_0=-\frac{\overline z}{z_0}{\vec x}_0'.
\label{eq:mapping*}
\end{equation}
As a result, (\ref{eq:amp-w-f}) takes the following equivalent form:
{}
\begin{eqnarray}
{\cal A}({\vec x}_i,{\vec x}_0',{\vec x}')&=&-\sqrt{\mu_0}
\frac{e^{ikf(1+{{\vec x}_i^2}/{2f^2})}}{i\lambda f}
 \iint\displaylimits_{|{\vec x}|^2\leq (\frac{1}{2}d)^2}\hskip -8pt
  d^2{\vec x}
  J_0\Big(k
\sqrt{\frac{2r_g}{\overline z}} |{\vec x}+\frac{\overline z}{{z}_0}({\vec x}'-{\vec x}_0')|\Big) e^{-i\frac{k}{f}({\vec x}\cdot{\vec x}_i)}.
  \label{eq:amp-w-fd3*}
\end{eqnarray}

Because the spatial frequency $\alpha$ is high, the Bessel function $J_0(\alpha\rho)$ in (\ref{eq:amp-w-fd3*}) oscillates rapidly as the distance from the optical axis $\rho$ increases, but the overall behavior of this function diminishes rather slowly, $\propto 1/\sqrt{\rho}$. Such a behavior of $J_0$ in the complex amplitude of the EM wave (\ref{eq:amp-w-fd3*}) is the source of a significant imaging blur \cite{Turyshev-Toth:2019-extend,Turyshev-Toth:2019-blur}. In other words, a telescope with aperture $d\ll r_\oplus$ in the focal region of the SGL receives light not only from the directly imaged region with diameter of $D=(z_0/{\overline z}) d \leq R_\oplus$ on the surface of a resolved source, but also from the rest of that surface that lies outside the region with the diameter $D$.

Following \cite{Turyshev-Toth:2019-blur}, we recognize that for any given location of the telescope in the image plane, the total EM field at the telescope's focal plane from an exoplanet, ${\cal A}_{\tt source}$, is the sum of two contributions: the EM field received from the directly imaged region, ${\cal A}_{\tt dir}$, and the blur from the rest of the source, ${\cal A}_{\tt blur}$. We therefore need to evaluate the integral in (\ref{eq:amp-w-fd3*}) in these two regions:
{ }
\begin{eqnarray}
{\cal A}_{\tt source}({\vec x}_i,{\vec x}_0',{\vec x}')&=&{\cal A}_{\tt dir}({\vec x}_i,{\vec x}_0',{\vec x}')+{\cal A}_{\tt blur}({\vec x}_i,{\vec x}_0',{\vec x}').
  \label{eq:amp-total}
\end{eqnarray}
In this expression, the directly imaged region is given by expression  (\ref{eq:amp-w-fd3*}) for all the points on the source, ${\vec x}'$, that lie within  the range $|{\vec x}'-{\vec x}'_0|\leq \frac{1}{2}D$.
In addition, blur from the rest of the source is also given by expression (\ref{eq:amp-w-fd3*}), but for $|{\vec x}'-{\vec x}'_0|\geq \frac{1}{2}D$, $|{\vec x}'|<\rho_\oplus$, where $\rho_\oplus$ is the radius of the source, as measured from the origin of the coordinate system.

Although the expressions for ${\cal A}_{\tt dir}({\vec x}_i,{\vec x}_0',{\vec x}')$ and ${\cal A}_{\tt blur}({\vec x}_i,{\vec x}_0',{\vec x}')$ have identical analytical form, the amplitudes of the EM waves in these expressions correspond to different regions with different intensities, ${E}^{\tt dir}_0$ and ${E}^{\tt blur}_0$. Radiation received from these two regions is spatially incoherent, $\big<{E}^{\tt dir}_0{E}^{\tt blur}_0\big>=0$, where $\big<...\big>$ denotes spatial averaging.

To compute ${\cal A}_{\tt dir}$ and ${\cal A}_{\tt blur}$, we need to evaluate the  double integral over $d^2 {\vec x}$ for two different regions. To do this, we introduce two-dimensional coordinates to describe points in the source plane, $\vec x'$; the position of the telescope in the image plane,  $\vec x_0$; points in the image plane within the telescope's aperture,  $\vec x$; and points in the optical telescope's focal plane ${\vec x}_i$. These are given as  follows:
\begin{eqnarray}
\{{\vec x}'\}&\equiv& (x',y')=\rho'\big(\cos\phi',\sin\phi'\big)=\rho'{\vec n}',
\label{eq:x'}\\
\{{\vec x}_0\}&\equiv& (x_0,y_0)=\rho_0\big(\cos\phi_0,\sin\phi_0\big)=\rho_0{\vec n}_0, \label{eq:x0}\\
\{{\vec x}\}&\equiv& (x,y)=\rho\big(\cos\phi,\sin\phi\big)=\rho\,{\vec n}, \label{eq:x}\\
 \{{\vec x}_i\}&\equiv& (x_i,y_i)=\rho_i\big(\cos\phi_i,\sin\phi_i\big)=\rho_i{\vec n}_i.
  \label{eq:p}
\end{eqnarray}
We introduce the following notations for the two relevant  spatial frequencies and a useful ratio for convenience:
{}
\begin{eqnarray}
\alpha=k \sqrt{\frac{2r_g}{\overline z}}, \qquad \eta_i=k\frac{\rho_i}{f}, \qquad \beta=\frac{\overline z}{{z}_0}.
  \label{eq:alpha-mu}
\end{eqnarray}
The quantities  $\alpha$ and $\eta_i$ are the spatial frequencies involved in the image formation process with the SGL using a convex lens at the image plane. The frequency $\alpha$ is fixed and is determined by the chosen observation wavelength and the heliocentric distance. The frequency $\eta_i$ is variable: in addition to the observing wavelength and the focal length of the optical telescope, the subscript $i$ serves as a reminder that it depends also on the position $\vec{x}_i$ in the optical telescope's focal plane. The quantity $\beta$ is a scale factor that accounts for the finite distance to the source and heliocentric distance to the image plane.

With the notations (\ref{eq:alpha-mu}), the integral in (\ref{eq:amp-w-fd3*}) present in the expressions for both complex amplitudes takes the form
{}
\begin{eqnarray}
\iint\displaylimits_{|{\vec x}|^2\leq (\frac{1}{2}d)^2}
  J_0\Big(\alpha \big|{\vec x}+\beta({\vec x}'-{\vec x}'_0)\big|\Big) e^{-i\eta_i({\vec x}\cdot{\vec n}_i)}.
  \label{eq:amp-w-d*}
\end{eqnarray}
By evaluating this integral for different regions in the source plane, we can compute the amplitudes ${\cal A}_{\tt dir}({\vec x}_i,{\vec x}_0',{\vec x}')$ and ${\cal A}_{\tt blur}({\vec x}_i,{\vec x}_0',{\vec x}')$ that are needed to evaluate the signal received form the entire source.

\subsection{Complex amplitude of the EM field received from the directly imaged region}
\label{sec:power-dim}

We first consider the directly imaged region (see Fig.~\ref{fig:regions}.) Assuming that $\beta|{\vec x'}-{\vec x}'_0|\ll |{\vec x}|$ everywhere in this region, we may evaluate (\ref{eq:amp-w-d*}) by keeping only the leading term in the series expansion with respect to the small parameter $\beta|{\vec x}'-{\vec x}'_0|/|{\vec x}|$, which implies that the EM field here may be approximated by light coming from the central point, ${\vec x}'={\vec x}'_0$, in that unresolved spot with diameter of $D$ in the source plane. With this assumption and notations (\ref{eq:x}), the integral (\ref{eq:amp-w-d*}) may be easily evaluated:
{}
\begin{eqnarray}
\int_0^{\frac{1}{2}d}\hskip -8pt \rho d\rho
 \int_0^{2\pi}  \hskip -8pt d\phi\,
  J_0(\alpha \rho) e^{-i\eta_i \rho\cos(\phi-\phi_i)}=
  \pi\Big(\frac{d}{2}\Big)^2 \frac{2}{(\alpha^2-\eta_i^2){\textstyle\frac{1}{2}}d}   \Big(\alpha J_0(\eta_i {\textstyle\frac{1}{2}}d) J_1(\alpha {\textstyle\frac{1}{2}}d)-\eta_i J_0(\alpha {\textstyle\frac{1}{2}}d) J_1(\eta_i {\textstyle\frac{1}{2}}d)\Big),
  \label{eq:amp-B-res}
\end{eqnarray}
where $\alpha$ and $\eta_i$ are given by (\ref{eq:alpha-mu}).
This result allows use to present the complex amplitude of the EM field received from the directly imaged region, ${\cal A}_{\tt dir}({\vec x}_i,{\vec x}_0,{\vec x}')$, which can be derived from (\ref{eq:amp-w-fd3*}) in the following form:
{ }
\begin{eqnarray}
{\cal A}_{\tt dir}({\vec x}_i,{\vec x}_0',{\vec x}')&=&i
\sqrt{\mu_0}e^{ikf(1+{{\vec x}_i^2}/{2f^2})}
 \Big(\frac{kd^2}{8 f}\Big) \frac{2}{(\alpha^2-\eta_i^2){\textstyle\frac{1}{2}}d}   \Big(\alpha J_0(\eta_i {\textstyle\frac{1}{2}}d) J_1(\alpha {\textstyle\frac{1}{2}}d)-\eta_i J_0(\alpha {\textstyle\frac{1}{2}}d) J_1(\eta_i {\textstyle\frac{1}{2}}d)\Big).
  \label{eq:amp-dir3}
\end{eqnarray}

We now can compute the Poynting vector of a plane wave that travels through the gravitational field of the Sun and is received in the focal plane of a convex lens placed in the focal region of the SGL.  For this, we substitute the result (\ref{eq:amp-dir3}) into  (\ref{eq:Pv}). After temporal averaging, we obtain the following expression for the Poynting vector for an EM wave that depends only on the radial position $\rho_i$ in the focal plane of the lens (where from (\ref{eq:alpha-mu}) we have $\eta_i=\eta_i(\rho_i)$):
 {}
\begin{eqnarray}
S_{\tt dir}(\rho_i)=\frac{c}{8\pi}
E_0^2
 \Big(\frac{kd^2}{8f}\Big)^2 \mu_0\Big(\frac{2}{(\alpha^2-\eta_i^2){\textstyle\frac{1}{2}}d}   \Big(\alpha J_0(\eta_i {\textstyle\frac{1}{2}}d) J_1(\alpha {\textstyle\frac{1}{2}}d)-\eta_i J_0(\alpha {\textstyle\frac{1}{2}}d) J_1(\eta_i {\textstyle\frac{1}{2}}d)\Big)\Big)^2.
  \label{eq:Pv1}
\end{eqnarray}
Substituting this result in (\ref{eq:psf}), we derive the PSF of an imaging system that relies on the SGL and a convex lens, scaled by the Fresnel number and the gain of the SGL on the optical axis:
 {}
\begin{eqnarray}
\mu_{\tt dir}(\rho_i)=\mu_0 \Big(\frac{kd^2}{8f}\Big)^2\Big(\frac{2}{(\alpha^2-\eta_i^2){\textstyle\frac{1}{2}}d}   \Big(\alpha J_0(\eta_i {\textstyle\frac{1}{2}}d) J_1(\alpha {\textstyle\frac{1}{2}}d)-\eta_i J_0(\alpha {\textstyle\frac{1}{2}}d) J_1(\eta_i {\textstyle\frac{1}{2}}d)\Big)\Big)^2,
  \label{eq:psf1}
\end{eqnarray}
with $\alpha$ and $\eta_i$ given by (\ref{eq:alpha-mu}). This imaging PSF is a result of a convolution of two point-spread functions: the PSF of the SGL (\ref{eq:S_z*6z-mu2}) and that of the convex lens, behaving as $\propto (2J_1(x)/x)^2$. This expression shows that the PSF for an unresolved source does not depend on the source's position in the source plane; nor does it depend on the telescope's position in the image plane. It is determined entirely by the parameters of the imaging telescope \cite{Turyshev-Toth:2019-image}.

\begin{figure}
\includegraphics[width=\linewidth]{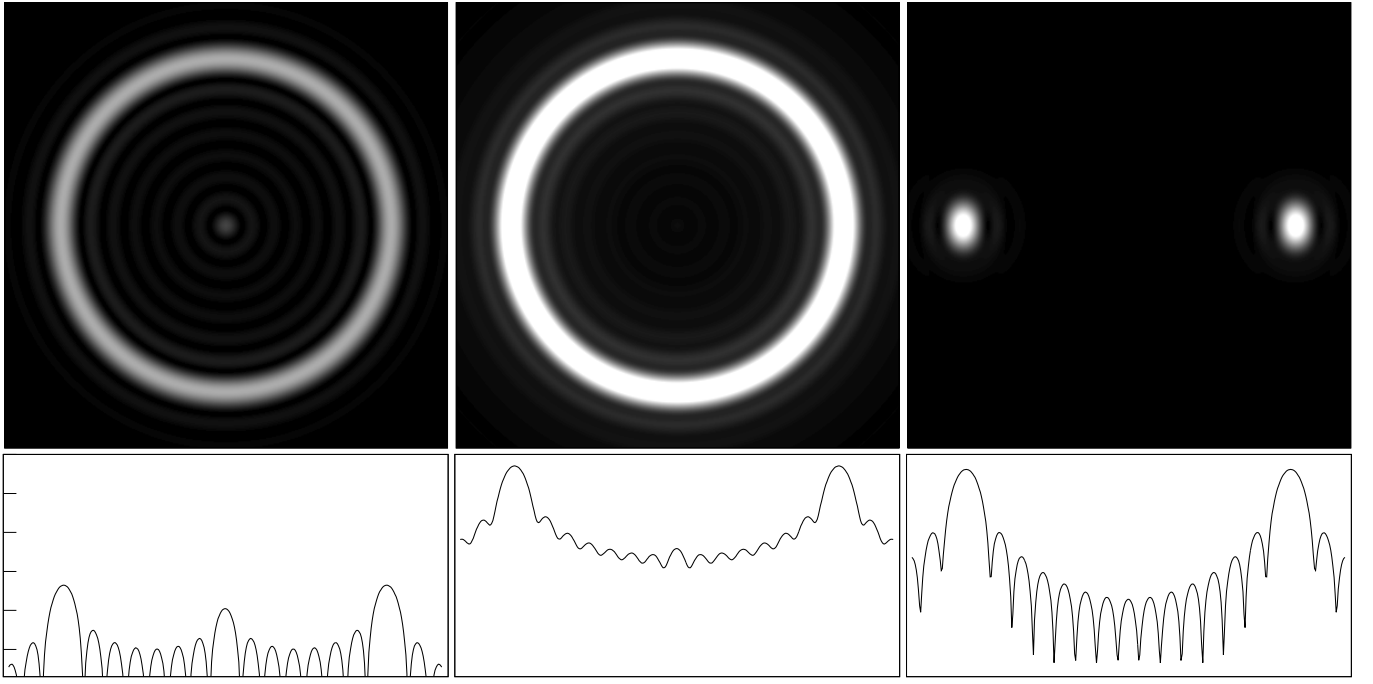}
\caption{\label{fig:images}Top row: Density plots simulating images that appear in the focal plane of the optical telescope. Left: the directly imaged region, in accordance with Eq.~(\ref{eq:pow-dirD}). The brightness of this image is exaggerated to ensure that the Einstein ring and diffraction artifacts remain visible. Center: Light from the rest of the source, in accordance with Eq.~(\ref{eq:P-blur*2*}). This is the dominant light contribution, yielding a much brighter Einstein ring with less prominent diffraction artifacts. Right: image contamination due to a nearby source of light, in accordance with Eq.~(\ref{eq:P-blur*off4*}), showing light from another uniformly illuminated disk of the same size, offset horizontally by ten radii. Bottom row: corresponding dimensionless intensities depicted on a decimal logarithmic scale. The contribution from the directly imaged region is ${\cal O}(10^3)$ less than the contribution from the rest of the source. Contribution from a nearby object is of similar intensity, but confined to narrow sections of the Einstein ring.
}
\end{figure}

Substituting result (\ref{eq:psf1}) into (\ref{eq:power}), we derive the intensity distribution for light received from the directly imaged region, which is  determined by the following expression:
  {}
\begin{eqnarray}
I_{\tt dir}(\rho_i) =\frac{1}{z^2_0}
\int_0^{2\pi} \hskip -8pt d\phi' \int^{\textstyle\frac{D}{2}}_0\hskip -8pt \rho' d\rho'
 B_{\tt s}({\vec x}') \mu_{\tt dir}(\rho_i).
  \label{eq:pow-dir}
\end{eqnarray}
Assuming that the surface brightness within the directly imaged region  is uniform, $ B_{\tt s}({\vec x}')=B_{\tt s}$, the integrals in (\ref{eq:pow-dir}) are easily computed. As a result, we obtain the following intensity distribution for the light received from this region:
  {}
\begin{eqnarray}
I_{\tt dir}(\rho_i) =\pi B_{\tt s} \Big(\frac{kd^2}{8f}\Big)^2\frac{\mu_0d^2}{4{\overline z}^2}
\Big(\frac{2}{(\alpha^2-\eta_i^2){\textstyle\frac{1}{2}}d}   \Big(\alpha J_0(\eta_i {\textstyle\frac{1}{2}}d) J_1(\alpha {\textstyle\frac{1}{2}}d)-\eta_i J_0(\alpha {\textstyle\frac{1}{2}}d) J_1(\eta_i {\textstyle\frac{1}{2}}d)\Big)\Big)^2,
  \label{eq:pow-dirD}
\end{eqnarray}
where we accounted for (\ref{eq:Dd}). We note that (\ref{eq:pow-dirD}) agrees with a similar expression given by Eq. (15) in \cite{Turyshev-Toth:2019-image} (which was obtained for imaging a point source), by extending it to the case of an extended source at a large, but finite distance. Fig.~\ref{fig:images} (left) shows the characteristic behavior\footnote{See also \url{https://youtu.be/wdFEM9KiMZU} for a video simulation.} captured by (\ref{eq:pow-dirD}).

To study the behavior of (\ref{eq:pow-dirD}) at the Einstein ring, we take the limit of $\eta_i\rightarrow \alpha$,  that results in
{}
\begin{eqnarray}
I_{\tt dir}(\rho_i^{\tt ER}) =\pi B_{\tt s} \Big(\frac{kd^2}{8f}\Big)^2\frac{\mu_0d^2}{4{\overline z}^2}
\Big(J^2_0(\alpha {\textstyle\frac{1}{2}}d)+J^2_1(\alpha {\textstyle\frac{1}{2}}d)\Big)^2.
  \label{eq:pow-dirD+}
\end{eqnarray}

To take the next step, we use well-known approximations for the Bessel functions for large arguments \cite{Abramovitz-Stegun:1965}, given as
\begin{eqnarray}
J_0(x)\simeq \sqrt{\frac{2}{\pi x}}\cos(x-{\textstyle\frac{\pi}{4}})+{\cal O}\big(x^{-1}\big)
\qquad {\rm and} \qquad
J_1(x)\simeq \sqrt{\frac{2}{\pi x}}\sin(x-{\textstyle\frac{\pi}{4}})+{\cal O}\big(x^{-1}\big).
\label{eq:BF}
\end{eqnarray}
These approximations lead to the following approximation for (\ref{eq:pow-dirD}), which describes  the intensity distribution  on the Einstein ring resulting from light originating in the  directly imaged region:
{}
\begin{eqnarray}
I_{\tt dir}(\rho_i^{\tt ER}) =B_{\tt s} \Big(\frac{kd^2}{8f}\Big)^2 \frac{4\mu_0}{\pi\alpha^2{\overline z}^2}=B_{\tt s} \Big(\frac{kd^2}{8f}\Big)^2\frac{4}{ k{\overline z}},
  \label{eq:pow-dirD2}
\end{eqnarray}
where we used the definitions for $\mu_0$ and $\alpha$ given by (\ref{eq:S_z*6z-mu2}) and (\ref{eq:alpha-mu}), correspondingly.

We note that the intensity distribution for the EM image field received from the directly imaged region does not explicitly depend on the Schwarzschild radius of the gravitational lens as it is implicitly encoded in the position of the Einstein ring in the focal plane. In addition, there is no dependence on the distance to the source or position of the telescope in the image plane. However, as expected, the distribution strongly depends on the telescope aperture and is slowly decreasing with increase of the heliocentric distance.

\subsection{Amplitude of the EM field received from outside the directly imaged region}
\label{sec:power-blur}

We now consider light originating from the areas within the source that are outside the directly imaged region (Fig.~\ref{fig:regions}), but still deposited in the focal plane of the optical lens because of the PSF (\ref{eq:S_z*6z-mu2}). This process is represented by the complex amplitude ${\cal A}_{\tt blur}({\vec x}_i,{\vec x}_0',{\vec x}')$ in (\ref{eq:amp-total}). To compute ${\cal A}_{\tt blur}$, we again use (\ref{eq:amp-w-fd3*}), but this time, we assume that the directly imaged region is very small compared to the rest of the planet, so that outside the directly imaged region the following inequality holds $|{\vec x}|\ll\beta|{\vec x}'-{\vec x}'_0|$. For most of this region,  in (\ref{eq:amp-w-d*}), the Bessel function, $J_0$, may be approximated by taking its asymptotic behavior for large arguments (\ref{eq:BF}), yielding
 {}
\begin{eqnarray}
  J_0\Big(\alpha|{\vec x}+\beta({\vec x}'-{\vec x}'_0)|\Big)
=\frac{1}{\sqrt{2\pi \alpha |{\vec x}+\beta({\vec x}'-{\vec x}'_0)|}}\Big(e^{i\big(\alpha |{\vec x}+\beta({\vec x}'-{\vec x}'_0)|-\frac{\pi}{4}\big)}+e^{-i\big(\alpha |{\vec x}+\beta({\vec x}'-{\vec x}'_0)|-\frac{\pi}{4}\big)}\Big).
  \label{eq:bf0}
\end{eqnarray}

To evaluate (\ref{eq:bf0}), we rely on  (\ref{eq:x'})--(\ref{eq:p}), but slightly redefining them  by introducing
{}
\begin{eqnarray}
\{({\vec x}'-{\vec x}'_0)\}={\vec x}''=\rho''{\vec n}''= \rho''(\cos\phi'',\sin\phi'').
  \label{eq:coord2}
\end{eqnarray}
Next, given the fact that $|{\vec x}|\leq \beta|{\vec x}'-{\vec x}'_0|$,  we expand $ |{\vec x}+\beta({\vec x}'-{\vec x}'_0)|$ to first order in ${\vec x}$:
 {}
\begin{eqnarray}
 |{\vec x}+\beta({\vec x}'-{\vec x}'_0)|=\beta |{\vec x}'-{\vec x}'_0|+({\vec x}\cdot {\vec n}'') +{\cal O}(\rho^2)=\beta \rho''+\rho\cos(\phi-\phi'') +{\cal O}(\rho^2).
  \label{eq:mod}
\end{eqnarray}

With these definitions, the double integral (\ref{eq:amp-w-d*}) takes the form
{}
\begin{eqnarray}
\iint\displaylimits_{|{\vec x}|^2\leq (\frac{1}{2}d)^2}\hskip -8pt
  d^2{\vec x}
  J_0\Big(\alpha|{\vec x}+\beta({\vec x}'-{\vec x}'_0)|\Big) e^{-i\frac{k}{f}({\vec x}\cdot{\vec x}_i)}&=&\frac{1}{\sqrt{2\pi \alpha \beta \rho'}}
  \int_0^{2\pi} \hskip -8pt d\phi \int_0^{d/2} \hskip -8pt \rho d\rho\,
\Big(1-\frac{\rho\cos(\phi-\phi'')}{\beta \rho''}\Big)\times\nonumber\\
&&\hskip -110pt
\times\,
\Big(e^{i\big(\alpha \beta \rho''-\frac{\pi}{4}+\alpha \rho\cos(\phi-\phi'')\big)}+e^{-i\big(\alpha \beta \rho''-\frac{\pi}{4}+\alpha \rho\cos(\phi-\phi'')\big)}\Big)e^{-i\rho \eta_i\cos(\phi-\phi_i)}+{\cal O}(\rho^2).~~~
  \label{eq:am-3*2*}
\end{eqnarray}
The phases of these two integrals may be given as
{}
\begin{eqnarray}
\varphi_\pm({\vec x})&=&
\pm(\alpha \beta \rho''-{\textstyle\frac{\pi}{4}})+u_\pm\,\rho\cos\big(\phi-\epsilon_\pm\big)+{\cal O}(\rho^2),
  \label{eq:ph4}
\end{eqnarray}
where $u_\pm$ has the form
{}
\begin{eqnarray}
u_\pm=\sqrt{\alpha^2\mp2\alpha\eta_i\cos\big(\phi''-\phi_i\big)+\eta_i^2},
  \label{eq:upm}
\end{eqnarray}
and the angles $\epsilon_\pm$ are given by the following relationships:
{}
\begin{eqnarray}
\cos\epsilon_\pm=\frac{\pm\alpha  \cos\phi''-\eta_i\cos\phi_i}{u_\pm}, \qquad
\sin\epsilon_\pm=\frac{\pm\alpha  \sin\phi''-\eta_i\sin\phi_i}{u_\pm}.
  \label{eq:eps}
\end{eqnarray}

With this, the two integrals present in (\ref{eq:am-3*2*}) may be evaluated as
{}
\begin{eqnarray}
I^\pm({\vec x}_i, {\vec x}'')&=& \pi\Big(\frac{d}{2}\Big)^2\frac{1}{\sqrt{2\pi \alpha \beta \rho''}} e^{\pm i\big(\alpha \beta \rho''-{\textstyle\frac{\pi}{4}}\big)}\Big\{\Big( \frac{2
J_1\big(u_\pm\frac{1}{2}d\big)}{u_\pm\frac{1}{2}d}\Big)-i\frac{d\cos(\phi''-\epsilon_\pm)}{2\beta \rho'}\Big(\frac{2J_2(u_\pm\rho)}{u_\pm\frac{1}{2}d}\Big)\Big\}.
  \label{eq:I120}
\end{eqnarray}

Substituting  expressions (\ref{eq:I120}) in (\ref{eq:am-3*2*}) and then using the result in (\ref{eq:amp-w-fd3*}), we derive the amplitude ${\cal A}_{\tt blur}({\vec x}_i,{\vec x}'_0,{\vec x}')$:
{ }
\begin{eqnarray}
{\cal A}_{\tt blur}({\vec x}_i,{\vec x}'')&=&
ie^{ikf(1+{{\vec x}_i^2}/{2f^2})}
 \Big(\frac{kd^2}{8f}\Big) \frac{\sqrt{\mu_0}}{\sqrt{2\pi \alpha \beta \rho''}} \times\nonumber\\
 &&\hskip 0pt\times\,
 \bigg\{e^{i\big(\alpha \beta \rho''-{\textstyle\frac{\pi}{4}}\big)}
 \Big[\Big( \frac{2
J_1\big(u_+\frac{1}{2}d\big)}{u_+\frac{1}{2}d}\Big)-i\frac{d\cos(\phi''-\epsilon_+)}{2\beta \rho''}\Big(\frac{2J_2(u_+\rho)}{u_+\frac{1}{2}d}\Big)\Big]+
\nonumber\\
 &&\hskip 20pt\,
 +e^{-i\big(\alpha \beta \rho''-{\textstyle\frac{\pi}{4}}\big)}
 \Big[\Big( \frac{2
J_1\big(u_-\frac{1}{2}d\big)}{u_-\frac{1}{2}d}\Big)-i\frac{d\cos(\phi''-\epsilon_-)}{2\beta \rho''}\Big(\frac{2J_2(u_-\rho)}{u_-\frac{1}{2}d}\Big)\Big]
\bigg\}.
  \label{eq:amp-blur3}
\end{eqnarray}

We may now compute the Poynting vector of a plane wave originating from outside the directly imaged region, traveling through the gravitational field in the vicinity of the Sun, arriving in the focal plane of an imaging telescope.  For this, similarly to the derivation of (\ref{eq:Pv1}), we substitute (\ref{eq:amp-blur3}) into  (\ref{eq:Pv}). After temporal averaging, we obtain the following expression (similar to that obtained in \cite{Turyshev-Toth:2019-image} for point sources):
 {}
\begin{eqnarray}
S_{\tt blur}({\vec x}_i,{\vec x}'')&=&\frac{c}{8\pi}
{E}_{\tt dir}^2({\vec x}')
 \Big(\frac{kd^2}{8f}\Big)^2\frac{\mu_0}{2\pi \alpha \beta \rho''}
 \times\nonumber\\
 &&\hskip -40pt \times\,
 \bigg\{\Big( \frac{2
J_1\big(u_+\frac{1}{2}d\big)}{u_+\frac{1}{2}d}\Big)^2+\Big( \frac{2
J_1\big(u_-\frac{1}{2}d\big)}{u_-\frac{1}{2}d}\Big)^2+
2\sin(2\alpha\beta\rho'')\Big( \frac{2
J_1\big(u_+\frac{1}{2}d\big)}{u_+\frac{1}{2}d}\Big)\Big( \frac{2
J_1\big(u_-\frac{1}{2}d\big)}{u_-\frac{1}{2}d}\Big)-\nonumber\\
 &-&
 \frac{d\cos(2\alpha\beta\rho'')}{\beta\rho''}\Big\{
 \frac{\alpha-\eta_i\cos(\phi''-\phi_i)}{u_+}\Big( \frac{2
J_1\big(u_-\frac{1}{2}d\big)}{u_-\frac{1}{2}d}\Big)\Big( \frac{2
J_2\big(u_+\frac{1}{2}d\big)}{u_+\frac{1}{2}d}\Big)+
\nonumber\\
&&\hskip 80pt +\,
\frac{\alpha+\eta_i\cos(\phi''-\phi_i)}{u_-}\Big( \frac{2
J_1\big(u_-\frac{1}{2}d\big)}{u_-\frac{1}{2}d}\Big)\Big( \frac{2
J_2\big(u_-\frac{1}{2}d\big)}{u_-\frac{1}{2}d}\Big)\Big\}+{\cal O}\Big(\frac{d^2}{\beta^2\rho'^2}\Big)
\bigg\}.
  \label{eq:Sblur}
\end{eqnarray}

As outside the directly imaged region the ratio $d/(\beta\rho'')$ is very small, we may neglect this term in the expression above. Substituting the result in (\ref{eq:psf}), we compute the PSF for the SGL's blur for a resolved source:
 {}
\begin{eqnarray}
\mu_{\tt blur}({\vec x}_i,{\vec x}'')=\frac{\mu_0}{2\pi \alpha \beta \rho''} \Big(\frac{kd^2}{8f}\Big)^2
 \bigg\{\Big( \frac{2
J_1\big(u_+\frac{1}{2}d\big)}{u_+\frac{1}{2}d}\Big)^2+\Big( \frac{2
J_1\big(u_-\frac{1}{2}d\big)}{u_-\frac{1}{2}d}\Big)^2+
2\sin(2\alpha\beta\rho'')\Big( \frac{2
J_1\big(u_+\frac{1}{2}d\big)}{u_+\frac{1}{2}d}\Big)\Big( \frac{2
J_1\big(u_-\frac{1}{2}d\big)}{u_-\frac{1}{2}d}\Big)
\bigg\}.~~
  \label{eq:psf-bl1}
\end{eqnarray}

Using this result (\ref{eq:psf-bl1}) in (\ref{eq:power}), we derive the expression that may be used to determine the intensity distribution for the signal received from the area outside the directly imaged region:
  {}
\begin{eqnarray}
I_{\tt blur}({\vec x}_i,{\vec x}_0) =\frac{1}{z^2_0}
\iint d^2{\vec x}''  B_{\tt s}({\vec x}'')  \mu_{\tt blur}({\vec x}_i,{\vec x}'').
  \label{eq:pow-blur}
\end{eqnarray}
This integral must be evaluated for two different regions corresponding to the telescope pointing within the image and outside of it, as was done in \cite{Turyshev-Toth:2019-blur}, where we considered the photometric signal (or the  power of the signal just before the telescope's aperture.)

\subsubsection{Intensity distribution for light from outside of the directly imaged region}
\label{sec:blur-in}

Expression (\ref{eq:pow-blur}) allows us to compute the power received from the resolved source from the area lying outside the directly imaged region.  To do that, we introduce a new coordinate system in the source plane, ${\vec x}''$, with the origin at the center of the directly imaged region: ${\vec x}'-{\vec x}'_0={\vec x}''$. As vector ${\vec x}'_0$ is constant, $dx'dy'=dx''dy''$. Next, in the new coordinate system, we use polar coordinates $(x'',y'')\rightarrow (r'',\phi'')$. In these coordinates, the circular edge of the source, $R_\oplus$, is no longer a circle but a curve, $\rho_\oplus(\phi'')$, the radial distance of which is given by the following relation:
{}
\begin{eqnarray}
\rho_\oplus(\phi'')
&=&\sqrt{R_\oplus^2-{\rho'_0}^2\sin^2\phi''}-\rho'_0\cos\phi''.
\label{eq:rho+}
\end{eqnarray}

For an actual astrophysical source, $B_s({\vec x}')$ is, of course, an arbitrary function of the coordinates ${\vec x}'$ and thus the integral can only be evaluated numerically. However, we can obtain an analytic result in the simple case of a disk of uniform brightness, characterized by ${B}_s({\vec x}') ={B}_s$. In this case, we integrate (\ref{eq:pow-blur}):
  {}
\begin{eqnarray}
I_{\tt blur}({\vec x}_i,{\vec x}_0) &=&\frac{1}{z^2_0}
\int_0^{2\pi} \hskip -6pt d\phi'' \int_{\textstyle\frac{D}{2}}^{\rho_\oplus}\hskip -4pt \rho'' d\rho''
 B_{\tt s}({\vec x}') \mu_{\tt blur}({\vec x}_i,{\vec x}_0,{\vec x}')=\frac{B_{\tt s}}{z^2_0} \Big(\frac{kd^2}{8f}\Big)^2\frac{\mu_0}{2\pi \alpha \beta} \times\nonumber\\
&&\hskip-40pt
\times\,
\int_0^{2\pi} \hskip -8pt d\phi'' \int_{\textstyle\frac{D}{2}}^{\rho_\oplus}\hskip -4pt d\rho''
 \bigg\{\Big( \frac{2
J_1\big(u_+\frac{1}{2}d\big)}{u_+\frac{1}{2}d}\Big)^2+\Big( \frac{2
J_1\big(u_-\frac{1}{2}d\big)}{u_-\frac{1}{2}d}\Big)^2+
2\sin(2\alpha\beta\rho'')\Big( \frac{2
J_1\big(u_+\frac{1}{2}d\big)}{u_+\frac{1}{2}d}\Big)\Big( \frac{2
J_1\big(u_-\frac{1}{2}d\big)}{u_-\frac{1}{2}d}\Big)
\bigg\}.
  \label{eq:P-blur*=0}
\end{eqnarray}

The integral over $d\rho''$ in (\ref{eq:P-blur*=0}) can be easy evaluated, resulting in
 {}
\begin{eqnarray}
I_{\tt blur}({\vec x}_i,{\vec x}_0) &=&
\frac{B_{\tt s}}{{\overline z}^2} \Big(\frac{kd^2}{8f}\Big)^2\frac{\mu_0 d}{2\alpha} \times
\nonumber\\
&&\hskip-50pt\times\,
\bigg\{\frac{1}{2\pi}
\int_0^{2\pi} \hskip -8pt d\phi'' \bigg(\frac{2r_\oplus}{d}
\Big(\sqrt{1-\big(\frac{\rho_0}{r_\oplus}\big)^2\sin^2\phi''}-\frac{\rho_0}{r_\oplus}\cos\phi''\Big)-1\bigg)
 \bigg(\Big( \frac{2
J_1\big(u_+\frac{1}{2}d\big)}{u_+\frac{1}{2}d}\Big)^2+\Big( \frac{2
J_1\big(u_-\frac{1}{2}d\big)}{u_-\frac{1}{2}d}\Big)^2\bigg)-\nonumber\\
&&\hskip-60pt
-\,\frac{2}{\alpha d}\,\frac{1}{2\pi}
\int_0^{2\pi} \hskip -8pt d\phi'' \Big(\cos\Big[2\alpha r_\oplus
\Big(\sqrt{1-\big(\frac{\rho_0}{r_\oplus}\big)^2\sin^2\phi''}-\frac{\rho_0}{r_\oplus}\cos\phi''\Big)\Big]-\cos[\alpha d]\Big)\Big( \frac{2
J_1\big(u_+\frac{1}{2}d\big)}{u_+\frac{1}{2}d}\Big)\Big( \frac{2
J_1\big(u_-\frac{1}{2}d\big)}{u_-\frac{1}{2}d}\Big)
\bigg\}.~~~
  \label{eq:P-blur*}
\end{eqnarray}

We observe that the ratios involving the Bessel functions in the expression (\ref{eq:P-blur*}) above are at most $2J_1(x)/x=1$, at $x=0$. Given the fact that the spatial frequency $\alpha$ is quite high, for most values of the argument these ratios become negligible. In addition, the last term in this expression is at most $\propto 2/\alpha d$, which is negligibly small even compared to the next smallest term (i.e., that does not contain $r_\oplus$) in the first integral in this expression. Therefore, the last term in this expression can be omitted, and expression (\ref{eq:P-blur*}) takes the form
  {}
\begin{eqnarray}
I_{\tt blur}({\vec x}_i,{\vec x}_0) &=&
\frac{B_{\tt s}}{{\overline z}^2} \Big(\frac{kd^2}{8f}\Big)^2\frac{\mu_0 d}{2\alpha}
\frac{1}{2\pi}
\int_0^{2\pi} \hskip -8pt d\phi'' \bigg(\frac{2r_\oplus}{d}
\sqrt{1-\big(\frac{\rho_0}{r_\oplus}\big)^2\sin^2\phi''}-1\bigg)
 \bigg(\Big( \frac{2
J_1\big(u_+\frac{1}{2}d\big)}{u_+\frac{1}{2}d}\Big)^2+\Big( \frac{2
J_1\big(u_-\frac{1}{2}d\big)}{u_-\frac{1}{2}d}\Big)^2\bigg),~~~~~
  \label{eq:P-blur*2*}
\end{eqnarray}
where we obtained the final form of the equation by dropping the $(\rho_0/r_\oplus)\cos\phi''$ term in the first integral in (\ref{eq:P-blur*}), as this term, multiplied by the squared Bessel-function terms that have the same periodicity by virtue of the dependence of $u_\pm$ on $\phi''$, vanishes identically when integrated over a full $2\pi$ period.

Expression (\ref{eq:P-blur*2*}) describes the blur contribution to the intensity distribution in the focal plane, corresponding to the image of an object of uniform brightness. Fig.~\ref{fig:images} (center) shows the characteristic behavior presented in this expression. This result is in a good agreement with a similar one given by Eq.~(33) of \cite{Turyshev-Toth:2019-image}, but extends the latter on the case of an extended, resolved source positioned at a large, but finite distance from the SGL.  Considering the terms remaining in (\ref{eq:P-blur*2*}), we note that the spatial frequency  $u_\pm$, as a function of $\phi''$ and $\phi_i$,  is given by expression  (\ref{eq:upm})  as $u_\pm=(\alpha^2\mp2\alpha\eta_i\cos\big(\phi_i-\phi''\big)+\eta_i^2)^\frac{1}{2}$. To study the behavior of  $P_{\tt blur}({\vec x}_i,{\vec x}_0)$ at the Einstein ring, we take the limit $\eta_i\rightarrow \alpha$ to present the ratios of the Bessel functions as
 {}
\begin{eqnarray}
\Big( \frac{2
J_1\big(u_+\frac{1}{2}d\big)}{u_+\frac{1}{2}d}\Big)^2+\Big( \frac{2
J_1\big(u_-\frac{1}{2}d\big)}{u_-\frac{1}{2}d}\Big)^2 ~~~\rightarrow~~~
\Big( \frac{2
J_1\big(\alpha d\sin{\textstyle\frac{1}{2}}(\phi_i-\phi'')\big)}{\alpha d\sin{\textstyle\frac{1}{2}}(\phi_i-\phi'')}\Big)^2+\Big( \frac{2
J_1\big(\alpha d\cos{\textstyle\frac{1}{2}}(\phi_i-\phi'')\big)}{\alpha d\cos{\textstyle\frac{1}{2}}(\phi_i-\phi'')}\Big)^2.
  \label{eq:Bess-rat}
\end{eqnarray}
Given that $\alpha d \gg 1$, these expressions suggest that for any value of $\phi''$ they will uniquely select such a value for $\phi_i$ that would make $u_\pm=0$ and thus, the arguments of the Bessel functions vanish. When this happens, the ratios of the Bessel functions reach their maximal value of 1, resulting in two peaks positioned at the azimuthal angles $\phi_i=\phi''$ and $\phi_i=\phi''+\pi$ (similar observation was made in \cite{Turyshev-Toth:2019-image}).

This observation greatly simplifies (\ref{eq:P-blur*2*}) (and (\ref{eq:P-blur*})), resulting in the following compact form for the intensity distribution  for light received from the Einstein ring in the focal plane of the telescope:
 {}
\begin{eqnarray}
I_{\tt blur}(\rho_i^{\tt ER},\rho_0) &=&
\frac{B_{\tt s}}{{\overline z}^2} \Big(\frac{kd^2}{8f}\Big)^2\frac{\mu_0 d}{\alpha}
\Big(\frac{2r_\oplus}{d}\epsilon(\rho_0)-1\Big)=\pi
B_{\tt s} \Big(\frac{kd^2}{8f}\Big)^2\frac{d}{{\overline z}}
\sqrt{\frac{2r_g}{{\overline z}}}\Big(\frac{2r_\oplus}{d}\epsilon(\rho_0)-1\Big),
  \label{eq:P-blur*2}
\end{eqnarray}
where the blur factor $\epsilon(\rho_0)$  is given by the following expression \cite{Turyshev-Toth:2019-blur} (see also Fig.~\ref{fig:epsilon-beta}):
{}
\begin{eqnarray}
\epsilon(\rho_0)
&=&\frac{1}{2\pi}\int_0^{2\pi} \hskip -3pt d\phi''\sqrt{1-\Big(\frac{\rho_0}{r_\oplus}\Big)^2\sin^2\phi''}=\frac{2}{\pi}{\tt E}\Big[\Big(\frac{\rho_0}{r_\oplus}\Big)^2\Big],
\label{eq:eps_r0}
\end{eqnarray}
where ${\tt E}[x]$ is the elliptic integral \cite{Abramovitz-Stegun:1965}.

As a result, we see that the intensity distribution describing the signal received in the focal plane of the telescope is given as a sum of the intensities of the signal received from the directly imaged region (\ref{eq:pow-dirD2}) and that received from the rest of the source (\ref{eq:P-blur*2}) (similarly to the result derived in \cite{Turyshev-Toth:2019-blur} for photometric signals), which, in terms of the intensity distribution, takes the form
{}
\begin{eqnarray}
I_{\tt source}(\rho_i^{\tt ER}, \rho_0)&=&I_{\tt dir}(\rho_i^{\tt ER},0)+I_{\tt blur}(\rho_i^{\tt ER},\rho_0)=\nonumber\\
&=&\pi
B_{\tt s}\Big(\frac{kd^2}{8f}\Big)^2\Big\{\frac{2R_\oplus}{z_0}\sqrt{\frac{2r_g}{\overline z}}\epsilon(\rho_0)+\frac{4}{\pi k{\overline z}}-\frac{d}{{\overline z}}
\sqrt{\frac{2r_g}{{\overline z}}}\Big\}\simeq \pi
B_{\tt s}\Big(\frac{kd^2}{8f}\Big)^2\frac{2R_\oplus}{z_0}\sqrt{\frac{2r_g}{\overline z}}\epsilon(\rho_0),
\label{eq:dir+blur}
\end{eqnarray}
where we neglected the two terms in the middle expression, as their magnitudes are negligible in comparison to the leading term.

\subsubsection{Blur at an off-image telescope position}
\label{sec:extend-photo-vic}

As discussed in \cite{Turyshev-Toth:2019-blur}, in the case of the SGL, blur from an extended source is present even outside the direct image of the source. Therefore, even a telescope positioned at $\rho_0\geq r_\oplus$ will receive light from the source. In this case, the blur for the off-image position, $I_{\tt off}({\vec x}_0)$, is obtained by integrating (\ref{eq:pow-blur}) over the surface of the source as it is seen from an off-image coordinate system.

The same conditions to derive (\ref{eq:P-blur*}) are valid, so the power received by the telescope takes the same form. The only difference comes from the fact that we are outside the image, thus, the integration limits change. First, we note that the circular edge of the source, $R_\oplus$, is given by a curve, $\rho_\oplus(\phi'')$, the radial distance of which in this polar coordinate system is given as
\begin{eqnarray}
\rho_\oplus(\phi'') &=&\pm \sqrt{R_\oplus^2-{\rho'_0}^2\sin^2\phi''}+\rho'_0\cos\phi'',
\label{eq:rho++}
\end{eqnarray}
with the angle $\phi''$ in this case is defined so that $\phi''=0$ when pointing at the center of the source. The angle $\phi''$ varies only within the range $\phi''\in [\phi_-,\phi_+]$, with $\phi_\pm=\pm \arcsin ({R_\oplus}/{\rho'_0})$. Given the sign in front of the square root in (\ref{eq:rho++}), for any angle $\phi''$ there will be two solutions for $\rho_\oplus(\phi'')$, given as $(\rho^-_\oplus,\rho^+_\oplus)$.

\begin{figure}
\rotatebox{90}{\hskip 40pt {\footnotesize $\epsilon(\rho_0)+\beta(\rho_0)$}}
\includegraphics[width=0.40\linewidth]{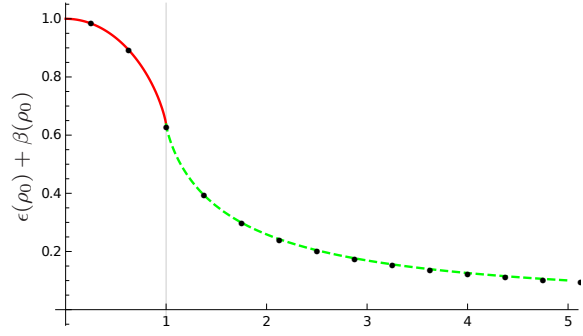}
\caption{\label{fig:epsilon-beta}Combined behavior of $\epsilon(\rho_0)$ (\ref{eq:eps_r0}), for $0\leq \rho_0/r_\oplus\leq 1$ (solid red curve) and $\beta(\rho_0)$ (\ref{eq:beta_r0}), for $\rho_0/r_\oplus\geq 1$ (dashed green curve). The horizontal axis is in units of $\rho_0/r_\oplus$. The dots represent values obtained from a numerical  simulation of (\ref{eq:power}) with (\ref{eq:amp-w-d*}).
}
\end{figure}

Assuming that the brightness of the source in this region is uniform, $B(x',y')=B_{\tt s}$, we use (\ref{eq:rho++}) and evaluate (\ref{eq:pow-blur}) for this set of conditions:
  {}
\begin{eqnarray}
I_{\tt off}({\vec x}_i,{\vec x}_0) &=&\frac{1}{z^2_0}
\int_{\phi_-}^{\phi_+} \hskip -3pt d\phi''
\int_{\rho^-_\oplus}^{\rho^+_\oplus}\hskip -3pt \rho'' d\rho''
 B_{\tt s}({\vec x}'') \mu_{\tt blur}({\vec x}_i,{\vec x}'')=\frac{B_{\tt s}}{z^2_0} \Big(\frac{kd^2}{8f}\Big)^2\frac{\mu_0}{2\pi \alpha \beta} \times\nonumber\\
&&\hskip-40pt
\times\,
\int_{\phi_-}^{\phi_+} \hskip -3pt d\phi''
\int_{\rho^-_\oplus}^{\rho^+_\oplus}\hskip -3pt d\rho''
 \bigg\{\Big( \frac{2
J_1\big(u_+\frac{1}{2}d\big)}{u_+\frac{1}{2}d}\Big)^2+\Big( \frac{2
J_1\big(u_-\frac{1}{2}d\big)}{u_-\frac{1}{2}d}\Big)^2+
2\sin(2\alpha\beta\rho'')\Big( \frac{2
J_1\big(u_+\frac{1}{2}d\big)}{u_+\frac{1}{2}d}\Big)\Big( \frac{2
J_1\big(u_-\frac{1}{2}d\big)}{u_-\frac{1}{2}d}\Big)
\bigg\}.~~~
  \label{eq:P-blur*off}
\end{eqnarray}
The integral over $d\rho''$ can be easy evaluated, resulting in
 {}
\begin{eqnarray}
I_{\tt off}({\vec x}_i,{\vec x}_0) &=&
\frac{B_{\tt s}}{z_0^2} \Big(\frac{kd^2}{8f}\Big)^2\frac{2\mu_0}{\alpha\beta} \times
\nonumber\\
&&\hskip-70pt\times\,
\bigg\{\frac{R_\oplus}{2\pi}
\int_{\phi_-}^{\phi_+} \hskip -8pt d\phi'' \Big(\sqrt{1-\big(\frac{{\rho}_0}{r_\oplus}\big)^2\sin^2\phi''}\Big)
 \bigg(\Big( \frac{2
J_1\big(u_+\frac{1}{2}d\big)}{u_+\frac{1}{2}d}\Big)^2+\Big( \frac{2
J_1\big(u_-\frac{1}{2}d\big)}{u_-\frac{1}{2}d}\Big)^2\bigg)-\nonumber\\
&&\hskip-55pt
-\,\frac{1}{\alpha \beta}\,\frac{1}{2\pi}
\int_{\phi_-}^{\phi_+} \hskip -8pt d\phi''\sin\big(2\alpha \rho_0\cos\phi''\big)
 \sin\Big[2\alpha r_\oplus
\Big(\sqrt{1-\big(\frac{\rho_0}{r_\oplus}\big)^2\sin^2\phi''}\Big)\Big]\Big( \frac{2
J_1\big(u_+\frac{1}{2}d\big)}{u_+\frac{1}{2}d}\Big)\Big( \frac{2
J_1\big(u_-\frac{1}{2}d\big)}{u_-\frac{1}{2}d}\Big)
\bigg\}.
  \label{eq:P-blur*off4}
\end{eqnarray}

Similarly to the approach that we used in evaluating the magnitude of the terms in (\ref{eq:P-blur*}) we we may drop the second term in this expression transforming (\ref{eq:P-blur*off4}) into
 {}
\begin{eqnarray}
I_{\tt off}({\vec x}_i,{\vec x}_0) &=&
\frac{B_{\tt s}}{z_0^2} \Big(\frac{kd^2}{8f}\Big)^2\frac{2\mu_0R_\oplus}{\alpha\beta}
\frac{1}{2\pi}
\int_{\phi_-}^{\phi_+} \hskip -8pt d\phi'' \Big(\sqrt{1-\Big(\frac{\rho_0}{r_\oplus}\Big)^2\sin^2\phi''}\Big)
 \bigg(\Big( \frac{2
J_1\big(u_+\frac{1}{2}d\big)}{u_+\frac{1}{2}d}\Big)^2+\Big( \frac{2
J_1\big(u_-\frac{1}{2}d\big)}{u_-\frac{1}{2}d}\Big)^2\bigg).~~~
  \label{eq:P-blur*off4*}
\end{eqnarray}
Fig.~\ref{fig:images} (right) shows the behavior captured in this expression that is characterized by two peaks of light deposited at the Einstein ring.  Such a behavior is expected for sources of light external to the target, including its parent star. Specifically, the light from the parent star is not a significant source of light contamination, as its signal will be deposited in just two compact spots on the image plane (as seen in Fig.~\ref{fig:images} (right)), which can be easily blocked.

Next, using similar arguments that led to result (\ref{eq:Bess-rat}) (but taking only one of the ratios), we present  (\ref{eq:P-blur*off4*}), as
 {}
\begin{eqnarray}
I_{\tt off}(\rho_i^{\tt ER},\rho_0) &=&\pi B_{\tt s}
 \Big(\frac{kd^2}{8f}\Big)^2\frac{2R_\oplus}{z_0} \frac{\mu_0}{\pi\alpha \overline z} \beta(\rho_0)=\pi
B_{\tt s} \Big(\frac{kd^2}{8f}\Big)^2\frac{2R_\oplus}{z_0} \sqrt{\frac{2r_g} {\overline z}}\beta(\rho_0),
  \label{eq:P-blur*off4=}
\end{eqnarray}
with the factor $\beta (\rho_0)$ given by the following expression:
{}
\begin{eqnarray}
\beta (\rho_0)
&=&\frac{1}{\pi}\int_{\phi_-}^{\phi_+} \hskip -3pt d\phi''\sqrt{1-\Big(\frac{\rho_0}{r_\oplus}\Big)^2\sin^2\phi''}=\frac{2}{\pi}{\tt E}\Big[\arcsin \frac{r_\oplus}{\rho_0},\Big(\frac{\rho_0}{r_\oplus}\Big)^2\Big],
\label{eq:beta_r0}
\end{eqnarray}
where ${\tt E}[a,x]$ is the incomplete elliptic integral \cite{Abramovitz-Stegun:1965}. This result is also similar to that obtained for the case of  photometric imaging with the SGL discussed in  \cite{Turyshev-Toth:2019-blur}. The combined behavior of this factor and $\epsilon(\rho_0)$ (given by Eq.~(\ref{eq:eps_r0})) is shown in Fig.~\ref{fig:epsilon-beta}.

Expressions (\ref{eq:dir+blur}) and (\ref{eq:P-blur*off4=}) are our main results that may be used to evaluate the signals to be expected for imaging with the SGL.  The describe the intensity distribution in the focal plane  of an imaging telescope that is positioned in the image plane in the strong interference region of the SGL. As such, these results are helpful for the ongoing instrument and mission design studies \cite{Turyshev-etal:2018}.

\section{Image formation in the geometric optics and weak interference regions}
\label{sec:weak-int}

As the optical telescope is moved farther away from the optical axis,  it enters the weak interference region and eventually the region of geometric optics.  It is important to study the image formation process in these regions, as modeling the magnitude of the signals detected here is useful to develop realistic SNR estimates that account for background noise. These models can also to be used in the development of autonomous navigation algorithms, required to navigate a space-based telescope towards the SGL's optical axis with respect to an imaging target such as an exoplanet \cite{Turyshev-etal:2018}.

\subsection{EM field in the geometric optics and weak interference regions}
\label{sec:EM-field-gowi}

The solution for the EM field in the geometric optics and weak interference regions consists of a combination of the gravity-modified incident wave and also the scattered wave that results from the diffraction of the incident wave on the solar gravity field \cite{Turyshev-Toth:2017,Turyshev-Toth:2019-extend}. Following the approach presented in  \cite{Turyshev-Toth:2017,Turyshev-Toth:2019-extend,Turyshev-Toth:2019-image}, we use the method of stationary phase to develop a solution for the incident and scattered EM fields that in the spherical coordinate system $(r,\theta,\phi)$, to the order of ${\cal O}\big(r_g^2, \theta^2, \sqrt{2r_g\tilde r}/z_0\big)$, take the form
{}
\begin{eqnarray}
    \left( \begin{aligned}
{D}_\theta& \\
{B}_\theta& \\
  \end{aligned} \right)_{\tt \hskip -2pt in/sc} =    \left( \begin{aligned}
{B}_\phi& \\
-{D}_\phi& \\
  \end{aligned} \right)_{\tt \hskip -2pt in/sc}&=&
  \frac{E_0}{z_0}
  {\cal A}_{\tt in/sc} (\tilde r,\theta) e^{i\big(k(r+r_0+r_g\ln 4k^2rr_0)-\omega t\big)}
 \left( \begin{aligned}
 \cos \phi& \\
 \sin \phi& \\
  \end{aligned} \right),
  \label{eq:DB-sol-in}
\end{eqnarray}
with the complex amplitudes ${\cal A}_{\tt in}$ and ${\cal A}_{\tt sc}$ (shorthanded as $A_{\tt in/sc}$  with the upper and lower signs are for the ``{\tt in}'' and ``{\tt sc}'' waves, correspondingly)  given as
{}
\begin{eqnarray}
{\cal A}_{\tt in/sc}(\tilde r,\theta)&=&
a_{\tt in/sc}(\tilde r,\theta)
\exp\Big[{-ik\Big\{{\textstyle\frac{1}{4}} \theta \big(\tilde r \theta\pm \sqrt{\tilde r^2 \theta^2+8r_g \tilde r}\big)\big)-r_g+2r_g\ln {\textstyle\frac{1}{2}} k \big(\tilde r \theta\pm \sqrt{\tilde r^2 \theta^2+8r_g \tilde r}\big)\Big\}}\Big],
  \label{eq:DB-sol-inA}
\end{eqnarray}
where the real-valued amplitude factors $a_{\tt in}$  and $a_{\tt sc}$ have the form
{}
\begin{eqnarray}
a^2_{\tt in/sc}(\tilde r,\theta)&=&
\frac{\big({\textstyle\frac{1}{2}} (\sqrt{1+{8r_g}/{\tilde r\theta^2}}\pm1)\big)^2}{\sqrt{1+{8r_g}/{\tilde r\theta^2}}},
\label{eq:a_insc*}
\end{eqnarray}
with the radial components of both EM waves behave as $({E}_r, {H}_r)_{\tt \hskip 0pt in/sc} \sim {\cal O}({\rho}/{r},\sqrt{2r_g\tilde r}/r_0)$. Also, the effective distance $\tilde r$ is given as $\tilde r=z_0{\overline z}/(z_0+{\overline z})$ (see details in \cite{Turyshev-Toth:2019-extend}). Note that for large angles $\theta\gg \sqrt{2r_g/\tilde r}$, expression (\ref{eq:a_insc*}) results in the known forms of the amplitude factors $a^2_{\tt in}(\tilde r,\theta)=1+{\cal O}(r_g\theta^2,r_g^2)$ and $a^2_{\tt sc}(\tilde r,\theta) =(2r_g/\tilde r\theta^2)^2 +{\cal O}(r_g\theta^2,r_g^2)$, see \cite{Turyshev-Toth:2019-extend}. However, expression (\ref{eq:a_insc*}) allows studying the case when  $\theta\simeq \sqrt{2r_g/\tilde r}$.

Since we are concerned with the EM field in the image plane, it is convenient to transform solution (\ref{eq:DB-sol-in}) to cylindrical coordinates $(\rho,\phi,z)$, as was done in  \cite{Turyshev-Toth:2017,Turyshev-Toth:2019-extend}. As result, the components of this EM field, to ${\cal O}(r_g^2, \theta^2)$, take the form
{}
\begin{eqnarray}
    \left( \begin{aligned}
{E}_\rho& \\
{H}_\rho& \\
  \end{aligned} \right)_{\tt \hskip -2pt in/sc} =    \left( \begin{aligned}
{H}_\phi& \\
-{E}_\phi& \\
  \end{aligned} \right)_{\tt \hskip -2pt in/sc}&=&
  \frac{E_0}{z_0}
  {\cal A}_{\tt in/sc}\big(\tilde r,\theta \big)e^{i\big(k(r+r_0+r_g\ln k^2rr_0)-\omega t\big)}
 \left( \begin{aligned}
 \cos \phi& \\
 \sin \phi& \\
  \end{aligned} \right),
  \label{eq:DB-sol-in-cc}
\end{eqnarray}
where the $z$-components of the EM waves behave as $({E}_z, {H}_z)_{\tt \hskip 0pt in/sc} \sim {\cal O}({\rho}/{z},b/z_0)$.

Expressing the combination $\tilde r\theta$ via the angle $\delta=b/r_0$  and generalizing the resulting expression to the 3-dimensional case, as was done in  \cite{Turyshev-Toth:2019-extend}, we have
{}
\begin{eqnarray}
\tilde r\theta&=&r\big(\theta+\delta\big)+{\cal O}(r^3/r_0^2)=|\vec{x}+{\vec x}_0+\beta{\vec x}'|+{\cal O}(r^3/r_0^2),
\label{eq:tildebeta}
\end{eqnarray}
where $\beta={\overline z}/{z_0}$ is from (\ref{eq:alpha-mu}).
This allows us to express the complex amplitudes $ {\cal A}_{\tt in/sc}(r,\theta,r_0)\rightarrow {\cal A}_{\tt in}({\vec x},{\vec x}_0,{\vec x}')$  as
{}
\begin{eqnarray}
{\cal A}_{\tt in/sc}({\vec x},{\vec x}_0,{\vec x}')&=&a_{\tt in/sc}({\vec x},{\vec x}_0,{\vec x}')
\exp\Big[-ik\Big\{\frac{1}{4\overline z}\big|\vec{x}+{\vec x}_0\big|\Big(\big|\vec{x}+{\vec x}_0+\beta{\vec x}'\big|\pm \sqrt{\big(\vec{x}+{\vec x}_0+\beta{\vec x}'\big)^2+8r_g \tilde r}\Big)-\nonumber\\
&&\hskip 80pt -\, r_g+2r_g\,
\ln \frac{k}{2}\Big(\big|\vec{x}+{\vec x}_0+\beta{\vec x}'\big|\pm\sqrt{\big(\vec{x}+{\vec x}_0+\beta{\vec x}'\big)^2+8r_g \tilde r}\Big)\Big\}\Big],
  \label{eq:Ain-3d}\\
a^2_{\tt in/sc}({\vec x},{\vec x}_0,{\vec x}')&=&\frac{\big({\textstyle\frac{1}{2}} (\sqrt{1+{8r_g \tilde r}/{(\vec{x}+{\vec x}_0+\beta{\vec x}')^2}}\pm1)\big)^2}{\sqrt{1+{8r_g\tilde r}/{(\vec{x}+{\vec x}_0+\beta{\vec x}')^2}}}.
\label{eq:a_insc2}
\end{eqnarray}

Clearly, these are rather complex expressions. However, in the case when displacements are large, $\rho_0 \gg \rho$ and $\beta\rho' \ll \rho_0$, we may use the approximation (\ref{eq:mod}), which allows us to expand (\ref{eq:Ain-3d}) and (\ref{eq:a_insc2}), to the first order in $\rho/\rho_0$ and $\beta\rho'/\rho_0$, yielding the following results:
{}
\begin{eqnarray}
{\cal A}_{\tt in}({\vec x},{\vec x}_0,{\vec x}')&=&
a_{\tt in}(\rho_0,\tilde r) \exp\Big[-i\Big(
\big(\xi_{\tt in}({\vec x}\cdot{\vec n}_0)+\eta_i({\vec x}\cdot{\vec n}_i)\big)+{\textstyle\frac{1}{2}}\xi_{\tt in}\beta\,({\vec x}'\cdot{\vec n}_0)\Big)\Big]e^{i\delta\varphi_{\tt in} (\rho_0,\tilde r)},
  \label{eq:amp-Ain}\\
  {\cal A}_{\tt sc}  ({\vec x},{\vec x}_0,{\vec x}')&=&
  a_{\tt sc}(\rho_0,\tilde r) \exp\Big[i\Big(  {\big(\xi_{\tt sc}({\vec x}\cdot{\vec n}_0)-\eta_i({\vec x}\cdot{\vec n}_i)\big)}+{{\textstyle\frac{1}{2}}\xi_{\tt sc}\beta\,({\vec x}'\cdot{\vec n}_0)}\Big)\Big]e^{i\delta\varphi_{\tt sc}  (\rho_0,\tilde r)},  \label{eq:amp-Asc}
\end{eqnarray}
where  the real-valued factors $a^2_{\tt in/sc}$ and phases $\delta\varphi_{\tt in/sc}(\rho_0,\tilde r)$ are given as
  \begin{eqnarray}
a^2_{\tt in/sc}(\rho_0,\tilde r) &=&
\frac{\big[{\textstyle\frac{1}{2}} (\sqrt{1+{8r_g\tilde r}/{\rho_0^2}}\pm1)\big]^2}{\sqrt{1+{8r_g\tilde r}/{\rho_0^2}}},
    \label{eq:a_insc}\\
    \delta\varphi_{\tt in/sc}
(\rho_0,\tilde r) &=& -k\Big(\frac{\rho^2_0}{4\tilde r}\Big(1\pm\sqrt{1+\frac{8r_g \tilde r}{\rho_0^2}}-\frac{4r_g \tilde r}{\rho_0^2}\Big)+2r_g\ln k\rho_0{\frac{1}{2}}\Big(\sqrt{1+\frac{8r_g \tilde r}{\rho^2_0}}\pm1\Big)\Big).
    \label{eq:Ain-d_ph}
\end{eqnarray}
Also, the spatial frequencies $\xi_{\tt in}$ and $\xi_{\tt sc}$ in (\ref{eq:amp-Ain}) and (\ref{eq:amp-Asc}), are defined as
{}
\begin{eqnarray}
\xi_{\tt in/sc}&=&k\Big(\sqrt{1+\frac{8r_g \tilde r}{\rho^2_0}}\pm1\Big)\frac{\rho_0}{2\tilde r},
\label{eq:betapm}
\end{eqnarray}
where, again, the upper sign is for $\xi_{\tt in}$ and the lower index is for $\xi_{\tt sc}$.

Note that in the case when angles $\theta$ are large, $\theta\gg\sqrt{2r_g/\tilde r}$ or  $\rho_0\gg \sqrt{2r_g\tilde r}$, the amplitude factors (\ref{eq:a_insc}) reduce to the known values (see, for instance, \cite{Turyshev-Toth:2019-extend}), namely
\begin{equation}
a_{\tt in}=1+{\cal O}(r_g^2), \qquad  a_{\tt sc}=\frac{2r_g\tilde r}{\rho_0^2}+{\cal O}(r_g^2), \qquad \rho_0 \gg \sqrt{2r_g\tilde r}.
\label{eq:a0_insc}
\end{equation}
However, the form of the expression (\ref{eq:a_insc}) allows us to study the case when $\rho_0\simeq \sqrt{2r_g\tilde r}$ and $\rho_0\lesssim \sqrt{2r_g\tilde r}$, which offers a description of the gravitational scattering of light in the transition region between the region of geometric optics and the weak interference region, and then toward the optical axis. This allows us to describe the entire process of gravitational scattering of light from the wave-optical standpoint.

To further emphasize the point above, we show the results that we obtained for the amplification factors $a^2_{\tt in/sc}$  and the spatial frequencies $\xi_{\tt in/sc}$, in relation to models that are used to describe gravitational microlensing. As we discussed in \cite{Turyshev-Toth:2019-image}, the spatial frequencies $\xi_{\tt in/sc}$ can be expressed as
{}
\begin{eqnarray}
\xi_{\tt in/sc}&=&k\Big(\sqrt{1+\frac{8r_g \tilde r}{\rho^2_0}}\pm1\Big)\frac{\rho_0}{2\tilde r}=k\theta_\pm, \qquad
\theta_\pm= {\textstyle\frac{1}{2}} \Big(\sqrt{\theta^2+4\theta_E^2}\pm\theta\Big),
\label{eq:xi}
\end{eqnarray}
where $\theta_E=\sqrt{{2r_g}/{\tilde r}}$ is the Einstein deflection angle and $\theta=\rho_0/{\tilde r}$.  The angles $\theta_\pm$ are the angles corresponding to the positions of the observed major and minor images \cite{Liebes:1964,Refsdal:1964,Schneider-Ehlers-Falco:1992}.
Furthermore, our results match the expressions used to describe light amplification observed in the microlensing experiments.
If  the  source  is  offset  from  the  optical axis by a  small amount, it is lensed into two images that appear in line with the source and the lens, and close to the Einstein ring.  Because the size of the Einstein ring is so small, the two images of the source are unresolved and the primary observable is their combined amplification. Using (\ref{eq:a_insc}) we obtain the combined light amplification, $A$, by adding the two amplification factors of the major and minor images, which yields the familiar expression
  \begin{eqnarray}
A=a^2_{\tt in}+a^2_{\tt sc}&=&
\frac{1+{4r_g\tilde r}/{\rho_0^2}}{\sqrt{1+{8r_g\tilde r}/{\rho_0^2}}}=\frac{u^2+2}{u\sqrt{u^2+4}},\qquad {\rm where}\qquad u=\frac{\theta}{\theta_E}.
    \label{eq:a12_amp}
\end{eqnarray}

Expressions (\ref{eq:xi})--(\ref{eq:a12_amp})  establish the correspondence between our analysis in this section and well-known models of microlensing \cite{Liebes:1964,Refsdal:1964,Schneider-Ehlers-Falco:1992}. Using our approach, we were able to present a previously unavailable description of microlensing phenomena using Maxwell's vector theory of the EM field. Our modeling approach can be further extended to incorporate other important features that allow for a better description of the source, the lens, and the backgrounds, including polarization of the incident EM wave, non-linear propagation effects, dispersion in the interstellar medium, contribution of the zodiacal background and others that are yet unavailable in the models of microlensing phenomena.

\subsection{Image EM field and intensity in the focal plane of the telescope}
\label{sec:image-EM-field}

With the expressions above, we may now develop the EM field that constitutes the image and evaluate its intensity in the focal plane of an imaging telescope.
To derive  the amplitudes of the EM field in the focal plane of the telescope that correspond to  (\ref{eq:amp-Ain}) and  (\ref{eq:amp-Asc}), we need to put these expressions in  (\ref{eq:amp-w-f}).  The corresponding integrals over $d^2{\vec x}$ are easy to evaluate. As a result, similarly to \cite{Turyshev-Toth:2019-image}, we derive the  amplitudes of the two EM waves on the optical telescope's focal plane in the following form:
{}
\begin{eqnarray}
{\cal A}_{\tt in}({\vec x}_i,{\vec x}_0,{\vec x}')&=&
\Big(\frac{kd^2}{8f}\Big)\, \Big\{a_{\tt in}\Big(\frac{
2J_1(v_+\frac{1}{2}d)}{v_+ \frac{1}{2}d}\Big)+{\cal O}(r_g^2)\Big\}e^{i\big(kf(1+{{\vec x}_i^2}/{2f^2})+\delta\varphi_{\tt in}(\rho_0,\tilde r) +\frac{\pi}{2}-{\textstyle\frac{1}{2}}\xi_{\tt in}\beta\rho'\cos(\phi'-\phi_0)\big)},
  \label{eq:amp-Aind}\\
  {\cal A}_{\tt sc}({\vec x}_i,{\vec x}_0,{\vec x}')&=&
\Big(\frac{kd^2}{8f}\Big)\Big\{
a_{\tt sc}\Big(
\frac{
2J_1(v_-\frac{1}{2}d)}{v_- \frac{1}{2}d}\Big)+{\cal O}(r^2_g)\Big\}e^{i\big(kf(1+{{\vec x}_i^2}/{2f^2})+\delta\varphi_{\tt sc}(\rho_0,\tilde r)
+\frac{\pi}{2}+{\textstyle\frac{1}{2}}\xi_{\tt sc}\beta\rho'\cos(\phi'-\phi_0)\big)},
  \label{eq:amp-Ascd}
\end{eqnarray}
where the spatial frequencies $v_\pm$ are defined as
{}
\begin{eqnarray}
v_+=\sqrt{\xi_{\tt in}^2+2\xi_{\tt in}\eta_i\cos\big(\phi_i-\phi_0\big)+\eta_i^2}\qquad {\rm and} \qquad
v_-=\sqrt{\xi_{\tt sc}^2-2\xi_{\tt sc}\eta_i\cos\big(\phi_i-\phi_0\big)+\eta_i^2}.\label{eq:vpms}
\end{eqnarray}

Remembering the time-dependent phase from (\ref{eq:DB-sol-in-cc}), we  substitute this expression in (\ref{eq:Pv}) and, after time averaging, we derive the Poynting vector of the EM wave in the focal plane of the imaging telescope.  As a result, in the region of the geometric optics, where only the incident EM wave is present, the intensity of the EM field in the focal plane sensor is derived using (\ref{eq:amp-Aind}), resulting in expression independent on $\rho'$ and $\phi'$:
 {}
\begin{eqnarray}
S_{\tt geom.opt.}({\vec x}_i,{\vec x}_0,{\vec x}')&=&
\frac{c}{8\pi}\frac{E_0^2}{z_0^2}
 \Big(\frac{kd^2}{8f}\Big)^2\Big\{a^2_{\tt in}\Big(\frac{
2J_1(v_+\frac{1}{2}d)}{v_+\frac{1}{2}d}\Big)^2+{\cal O}(r^2_g)\Big\}.
  \label{eq:FI-go}
\end{eqnarray}

As in the region of weak interference both incident and scattered waves are present, the field intensity in the focal plane of the imaging telescope is derived using the sum of the two solutions, (\ref{eq:amp-Aind}) and (\ref{eq:amp-Ascd}), yielding
{}
\begin{eqnarray}
S_{\tt weak.int.}({\vec x}_i,{\vec x}_0,{\vec x}')&=&
\frac{c}{8\pi}\frac{E_0^2}{z_0^2}
 \Big(\frac{kd^2}{8f}\Big)^2\Big\{a^2_{\tt in}\Big(\frac{
2J_1(v_+\frac{1}{2}d)}{v_+\frac{1}{2}d}\Big)^2+
a^2_{\tt sc}\Big(\frac{
2J_1(v_-\frac{1}{2}d)}{v_-\frac{1}{2}d}\Big)^2+\nonumber\\
&&\hskip -80pt +\,
2\cos\Big(\frac{k\rho_0}{2\tilde r}\sqrt{\rho_0^2+8r_g \tilde r}+2kr_g\ln \frac{\sqrt{\rho^2_0+8r_g \tilde r}+\rho_0}{\sqrt{\rho_0^2+8r_g \tilde r}-\rho_0}\Big)
a_{\tt in}a_{\tt sc}\Big(\frac{
2J_1(v_+\frac{1}{2}d)}{v_+\frac{1}{2}d}\Big)\bigg(\frac{
2J_1(v_-\frac{1}{2}d)}{v_-\frac{1}{2}d}\Big)+{\cal O}(r_g^2)
\Big\},~~~
  \label{eq:FI-ir}
\end{eqnarray}
also independent on $\rho'$ and $\phi'$.  Similar simplifying assumptions, based on the behavior of the ratios involving the Bessel function $2J_1(v_\pm\frac{1}{2}d)/{v_\pm\frac{1}{2}d}$ in these regions \cite{Turyshev-Toth:2019-image}, are applicable here. Therefore, the intensity distribution pattern in the weak interference region takes the following simplified form:
{}
\begin{eqnarray}
S_{\tt weak.int.}({\vec x}_i,{\vec x}_0,{\vec x}')&=&
\frac{c}{8\pi}\frac{E_0^2}{z_0^2}
 \Big(\frac{kd^2}{8f}\Big)^2\Big\{
 a^2_{\tt in}\Big(\frac{
2J_1(v_+\frac{1}{2}d)}{v_+\frac{1}{2}d}\Big)^2+
a^2_{\tt sc}\Big(\frac{
2J_1(v_-\frac{1}{2}d)}{v_-\frac{1}{2}d}\Big)^2+{\cal O}(r^2_g)
\Big\}.
  \label{eq:FI-ir+}
\end{eqnarray}

Substituting the resulting expressions (\ref{eq:FI-go}) and (\ref{eq:FI-ir+})  in (\ref{eq:psf}), we compute the convolved PSFs for the two regions:
 {}
\begin{eqnarray}
\mu_{\tt geom.opt.}({\vec x}_i,{\vec x}_0,{\vec x}')&=&
 \Big(\frac{kd^2}{8f}\Big)^2\Big\{a^2_{\tt in}\Big(\frac{
2J_1(v_+\frac{1}{2}d)}{v_+\frac{1}{2}d}\Big)^2+{\cal O}(r^2_g)\Big\},
  \label{eq:FI-go-mu}\\
\mu_{\tt weak.int.}({\vec x}_i,{\vec x}_0,{\vec x}')&=&
 \Big(\frac{kd^2}{8f}\Big)^2\Big\{a^2_{\tt in}\Big(\frac{
2J_1(v_+\frac{1}{2}d)}{v_+\frac{1}{2}d}\Big)^2+a^2_{\tt sc}\Big(\frac{
2J_1(v_-\frac{1}{2}d)}{v_-\frac{1}{2}d}\Big)^2+{\cal O}(r_g^2)\Big\}.
  \label{eq:FI-ir+-mu}
\end{eqnarray}

\begin{figure}
\includegraphics[width=0.3\linewidth]{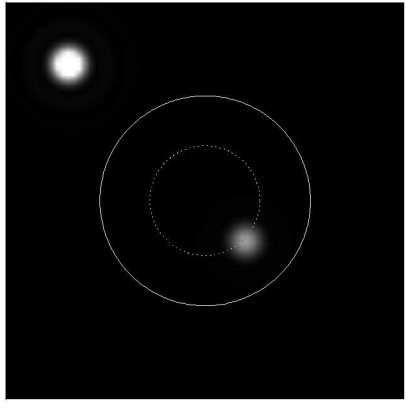}
\caption{\label{fig:images-go}Density plot simulating the image seen by the optical telescope when it is positioned in the region of weak interference, $\rho_0\gtrsim R_\odot$ from the optical axis, with the resulting minor and major images shown in accordance with Eq.~(\ref{eq:FI-ir+-Int}). The Sun is indicated with a dashed line, while the Einstein ring is shown as a solid line. Note that in the region of geometric optics only the major image remains, as described by Eq.~(\ref{eq:FI-go-Int}).
}
\end{figure}

Substituting this result (\ref{eq:psf-bl1}) into (\ref{eq:power}), we derive the expression that may be used to determine the intensity distribution for the signals received in these two regions. Again assuming uniform surface brightness, and noticing that (\ref{eq:FI-go-mu}) and (\ref{eq:FI-ir+-mu}) do not depend on $\rho'$ and $\phi'$, we can easily evaluate the integral. This results in the following intensities to be observed in the focal plane of the imaging telescope:
 {}
\begin{eqnarray}
I_{\tt geom.opt.}({\vec x}_i,{\vec x}_0)&=&\pi B_{\tt s}
 \Big(\frac{kd^2}{8f}\Big)^2 \frac{R^2_\oplus}{z_0^2}
 \Big\{a^2_{\tt in}\Big(\frac{
2J_1(v_+\frac{1}{2}d)}{v_+\frac{1}{2}d}\Big)^2+{\cal O}(r^2_g)\Big\},
  \label{eq:FI-go-Int}\\
I_{\tt weak.int.}({\vec x}_i,{\vec x}_0)&=&\pi
B_{\tt s}  \Big(\frac{kd^2}{8f}\Big)^2 \frac{R^2_\oplus}{z_0^2}\Big\{
a^2_{\tt in}\Big(\frac{
2J_1(v_+\frac{1}{2}d)}{v_+\frac{1}{2}d}\Big)^2+
a^2_{\tt sc}\Big(\frac{
2J_1(v_-\frac{1}{2}d)}{v_-\frac{1}{2}d}\Big)^2+{\cal O}(r^2_g)\Big\}.
  \label{eq:FI-ir+-Int}
\end{eqnarray}

Eqs.~(\ref{eq:FI-go-Int})--(\ref{eq:FI-ir+-Int}) describe the intensity distributions that correspond to the imaging in two different optical regions behind the Sun. They describe the spots of light corresponding to incident and scattered waves, that are given by the terms containing $v_+$ and $v_-$, correspondingly.  Examining (\ref{eq:FI-go-Int}) and  (\ref{eq:FI-ir+-Int}) in conjunction with (\ref{eq:vpms}), we see that these expressions nearly vanish for most values of $v_\pm$, except when  $v_\pm$  becomes zero which happens, when $\eta_i  \rightarrow \xi_{\tt in/sc}$. When this happens, we observe a spot that is outside the Einstein ring (for $\eta_i  \rightarrow \xi_{\tt in}$) describing the major image and the other one inside the ring (for $\eta_i  \rightarrow \xi_{\tt sc}$) describing the minor image. This approach provides a wave-optical treatment for the microlensing phenomena that is usually described by invoking the language of geometric optics \cite{Schneider-Ehlers-Falco:1992}.

Examining (\ref{eq:vpms}), we see that because the combinations $\xi_{\tt in}{\textstyle\frac{1}{2}}d$ and $\xi_{\tt sc}{\textstyle\frac{1}{2}}d$ are rather large,  expression (\ref{eq:FI-go-Int}) is almost zero everywhere except for one point where the argument of the Bessel function vanishes. Taking the limit  $\eta_i\rightarrow \xi_{\tt in}$ in (\ref{eq:FI-go-Int}), we obtain:
 {}
\begin{eqnarray}
I_{\tt geom.opt.}({\vec \xi}^{\tt in}_i,{\vec x}_0)&=&\pi
B_{\tt s}\Big(\frac{kd^2}{8f}\Big)^2 \frac{R^2_\oplus}{z_0^2}
 \Big\{\Big(\frac{
2J_1\big(\xi_{\tt in}d\cos{\textstyle\frac{1}{2}}(\phi_i-\phi_0)\big)}{\xi_{\tt in}d\cos{\textstyle\frac{1}{2}}(\phi_i-\phi_0)}\Big)^2+{\cal O}(r^2_g)\Big\},
  \label{eq:FI-go+}
\end{eqnarray}
where to show the dominant behavior of this expression in the geometric optics region, we used the value for $a_{\tt in}$ from (\ref{eq:a0_insc}).
This expression describes one peak corresponding to the incident wave whose intensity is not amplified by the SGL. It is for the major image corresponding $\xi_{\tt in}$, which appears always outside the Einstein ring. Similarly to (\ref{eq:FI-go+}), we take the limit in $\eta_i\rightarrow \xi_{\tt sc}$ in the expression (\ref{eq:FI-ir+-Int}) and obtain
{}
\begin{eqnarray}
I_{\tt weak.int.}({\vec \xi}^{\tt sc}_i,{\vec x}_0)&=&\pi
B_{\tt s} \Big(\frac{kd^2}{8f}\Big)^2\frac{R^2_\oplus}{z_0^2}\Big\{\Big(\frac{
2J_1\big(\xi_{\tt in}d\cos{\textstyle\frac{1}{2}}(\phi_i-\phi_0)\big)}{\xi_{\tt in}d\cos{\textstyle\frac{1}{2}}(\phi_i-\phi_0)}\Big)^2+ \Big(\frac{2r_g{\tilde r}}{\rho^2_0}  \Big)^2 \Big(\frac{
2J_1\big(\xi_{\tt sc}d\sin{\textstyle\frac{1}{2}}(\phi_i-\phi_0)\big)}{\xi_{\tt sc}d\sin{\textstyle\frac{1}{2}}(\phi_i-\phi_0)}\Big)^2\Big\},~~~
  \label{eq:FI-ir+*}
\end{eqnarray}
where to explicitly demonstrate the behavior of $I_{\tt weak.int}$, we used the values for $a_{\tt in/sc}$ from (\ref{eq:a0_insc}).

Eq.~(\ref{eq:FI-ir+*}) describes two images with uneven brightness, one depending on $v_+$ from (\ref{eq:vpms}), characteristic of the incident wave, that appears outside the Einstein ring and the other image given by the $v_-$-dependent term and scaled by the factor $(2r_g{\overline z}/\rho_0^2)^2$, corresponding to the scattered wave, that appears inside the Einstein ring.

\section{Power received at the image of the Einstein ring}
\label{sec:power}

Fig.~\ref{fig:images} shows the signals from the directly imaged region and from the rest of the source, as received at the Einstein ring at the focal plane of an optical telescope. The thickness of the Einstein ring is determined by the resolution of the diffraction-limited telescope, given as $\sim\lambda/d$ (from (\ref{eq:S_=0})). Eqs.~(\ref{eq:pow-dirD}) and  (\ref{eq:P-blur*off4*})  describe the intensities of light received from the directly imaged region, $I(\rho_i) $, and blur from the rest of the planet, $I_{\tt blur}({\vec x}_i,{\vec x}_0)$, correspondingly. These expressions describe the signal intensity.

In determining the useful area in the focal plane of an optical telescope, we observe that a meter-class telescope positioned in the strong interference region of the SGL will not be able to resolve the thickness of the Einstein ring given as $2r_\oplus=({\overline z}/z_0)2R_\oplus$; for that, a telescope aperture of $2r_\oplus$ would be required. However a meter-class telescope will be able to resolve the circumference of the ring, $\ell_{\tt ER}=2\pi \sqrt{2r_g/\overline z}$, at an angular resolution characterized by $\lambda/d$.

There are two natural ways to use the information present in the Einstein ring:
\begin{inparaenum}[1)]
\item to use the total power deposited within the Einstein ring, as seen by the diffraction-limited telescope, or
\item to measure brightness variations of the Einstein ring along its circumference.
\end{inparaenum}
Measuring the total power allows for a straightforward signal estimation. Measuring brightness variations along the Einstein ring represents another valuable observable that can help improve image quality and also reduce unwanted light contamination from nearby off-image sources. Here, we focus on the measuring the total power; we leave the topic of measuring brightness variations for a separate discussion.

As shown in Fig.~\ref{fig:images}, the Einstein ring is seen in the focal plane of an imaging telescope  as an annulus of unresolved width, with radius determined from (\ref{eq:alpha-mu}) as $\alpha=\eta_i$, yielding $\rho_{\tt ER}=f\sqrt{2r_g/{\overline z}}$. Therefore, the useful signal  received in the focal plane of a diffraction-limited telescope is received from the entire circumference of the Einstein ring that occupies the annulus within the two radii, $\rho_{\tt ER}^\pm$, given as
{}
\begin{eqnarray}
\rho_{\tt ER}^\pm=f\Big(\sqrt{\frac{2r_g}{\overline z}}\pm\frac{\lambda}{2d}\Big).
  \label{eq:ER-rho}
\end{eqnarray}

As a result, to estimate the power received in the focal plane of a diffraction-limited telescope from a distant, extended and resolved source, we need to integrate the intensities (\ref{eq:pow-dirD}) and (\ref{eq:P-blur*off4*}) over the  focal plane corresponding to the annulus between the radii (\ref{eq:ER-rho}).

\subsection{Power in the focal plane from the directly imaged region}
\label{sec:power-dir}

Before considering  the power deposited at the annulus around the Einstein ring corresponding to the signal received from the directly imaged region, we first compute the total power deposited by this signal in the entire focal plane. For this, we take (\ref{eq:pow-dirD}) and derive the following
{}
\begin{eqnarray}
P^0_{\tt fp.dir}&=&\int^{2\pi}_0 \hskip -4pt  d\phi_i \int_0^\infty
\hskip 0pt I_{\tt dir}(\rho_i) \rho_i d\rho_i = \nonumber\\
&=& \pi B_{\tt s} \Big(\frac{kd^2}{8f}\Big)^2\frac{\mu_0d^2}{4{\overline z}^2}
\int^{2\pi}_0 \hskip -4pt  d\phi_i \int_0^\infty
\hskip 0pt  \rho_i d\rho_i \Big(\frac{2}{(\alpha^2-\eta_i^2){\textstyle\frac{1}{2}}d}   \Big(\alpha J_0(\eta_i {\textstyle\frac{1}{2}}d) J_1(\alpha {\textstyle\frac{1}{2}}d)-\eta_i J_0(\alpha {\textstyle\frac{1}{2}}d) J_1(\eta_i {\textstyle\frac{1}{2}}d)\Big)\Big)^2.
  \label{eq:pow-fp0}
\end{eqnarray}
To evaluate this integral, we remember the identity
{}
\begin{eqnarray}
\int_0^{d/2} \hskip -5pt  \rho d \rho J_0(\alpha \rho)J_0(\eta_i \rho)= \Big(\frac{d}{2}\Big)^2
\frac{1}{(\alpha^2-\eta_i^2){\textstyle\frac{1}{2}}d}
 \Big(
 \alpha J_0(\eta_i {\textstyle\frac{1}{2}}d) J_1(\alpha {\textstyle\frac{1}{2}}d)-\eta_i J_0(\alpha {\textstyle\frac{1}{2}}d) J_1(\eta_i {\textstyle\frac{1}{2}}d)\Big).
  \label{eq:amp-w-2+}
\end{eqnarray}
With the help of (\ref{eq:amp-w-2+}) and (\ref{eq:alpha-mu}), we  present (\ref{eq:pow-fp0}) as
{}
\begin{eqnarray}
P^0_{\tt fp.dir}&=&\pi B_{\tt s} \frac{\mu_0d^2}{4{\overline z}^2}
\int_0^{d/2} \hskip -5pt  \rho d \rho J_0(\alpha \rho)
\int_0^{d/2}\hskip -5pt  \rho' d \rho' J_0(\alpha \rho')
\int^{2\pi}_0 \hskip -4pt  d\phi_i \int_0^\infty
\hskip 0pt  \eta_i d\eta_i J_0(\eta_i \rho)J_0(\eta_i \rho').
  \label{eq:pow-fp1}
\end{eqnarray}
The last integral in (\ref{eq:pow-fp1})
 is just the semi-infinite integral of a Fourier-Bessel transform (Hankel transform) that is bounded at $\rho\rightarrow 0$ and vanishes at $\rho\rightarrow\infty$, constituting the orthogonality relation on a semi-infinite interval \cite{deLeon:2014}:
{}
\begin{eqnarray}
\int_0^\infty \hskip -3pt  q dq J_n\big(q \rho\big)J_n\big(q\rho'\big)=\frac{\delta(\rho-\rho')}{\rho'}.
  \label{eq:fb}
\end{eqnarray}

Using (\ref{eq:fb}) in (\ref{eq:pow-fp1}), we have
{}
\begin{eqnarray}
P^0_{\tt fp.dir}&=&
B_{\tt s} \frac{\mu_0 \pi^2d^4}{16{\overline z}^2}\Big(J^2_0(\alpha{\textstyle\frac{1}{2}}d)+J^2_1(\alpha{\textstyle\frac{1}{2}}d)\Big)
\equiv P_{\tt dir}=\frac{B_s}{z_0^2}\pi({\textstyle\frac{1}{2}}d)^2\pi({\textstyle\frac{1}{2}}D)^2\frac{4 \sqrt{2r_g\overline z}}{d},
  \label{eq:pow2*}
\end{eqnarray}
where $P_{\tt dir}$ is the power of the EM field received from the directly imaged region of the resolved target and measured at the entrance of the telescope (just in front of the convex lens) as was derived in \cite{Turyshev-Toth:2019-blur} by integrating the energy density over the aperture. Eq.~(\ref{eq:pow2*}) confirms that in the case of imaging with the SGL, the total energy is conserved. This is despite the fact that the PSF (\ref{eq:S_z*6z-mu2}) diminishes as $\propto 1/\rho$ as the distance from its optical axis, $\rho$, increases \cite{Turyshev-Toth:2019-extend}.

Now we can estimate the power deposited at the annuals around the Einstein ring corresponding to the signal received from the directly imaged region, $P_{\tt fp.dir}$. For this, we take  (\ref{eq:pow-dirD}) and integrate it over the area seen by the diffraction-limited telescope
{}
\begin{eqnarray}
P_{\tt fp.dir}&=&\int^{2\pi}_0 \hskip -4pt  d\phi_i \int_{\rho^-_{\tt ER}}^{\rho_{\tt ER}^+}
\hskip 0pt I_{\tt dir}(\rho_i) \rho_i d\rho_i = \nonumber\\
&=& \pi B_{\tt s} \Big(\frac{kd^2}{8f}\Big)^2\frac{\mu_0d^2}{4{\overline z}^2}
\int^{2\pi}_0 \hskip -4pt  d\phi_i \int_{\rho^-_{\tt ER}}^{\rho_{\tt ER}^+}
\hskip 0pt  \rho_i d\rho_i \Big(\frac{2}{(\alpha^2-\eta_i^2){\textstyle\frac{1}{2}}d}   \Big(\alpha J_0(\eta_i {\textstyle\frac{1}{2}}d) J_1(\alpha {\textstyle\frac{1}{2}}d)-\eta_i J_0(\alpha {\textstyle\frac{1}{2}}d) J_1(\eta_i {\textstyle\frac{1}{2}}d)\Big)\Big)^2.
  \label{eq:pow-fp}
\end{eqnarray}

To consider practical applications of the SGL, it is convenient to represent $P_{\tt fp.dir}$ as a fraction of the total power incident at the telescope entrance, $P_{\tt dir}$, namely:
{}
\begin{eqnarray}
P_{\tt fp.dir}&=&\epsilon_{\tt dir}P_{\tt dir}.
  \label{eq:pow-frac}
\end{eqnarray}
The quantity $\epsilon_{\tt dir}$ is the encircled energy ratio that describes the ratio of the power deposited within the first few Airy rings of the diffraction pattern seen at the focal plane of a convex lens to the total energy incident on a telescope. Similarly, in our case,  $\epsilon_{\tt dir}$ describes the fraction of the total energy incident on the telescope from the directly imaged region that is deposited around the Einstein ring as seen  by a diffraction-limited telescope.

 To evaluate $\epsilon_{\tt dir}$, we introduce a new variable, $p_i$, and new integration limits corresponding to (\ref{eq:ER-rho}):
{}
\begin{eqnarray}
p_i=\eta_i {\textstyle\frac{1}{2}}d=\frac{\pi d}{\lambda f}\rho_i, \qquad {\rm and}
\qquad
p^\pm_{\tt ER}=\alpha {\textstyle\frac{1}{2}}d \pm {\textstyle\frac{\pi}{2}},
  \label{eq:p-eta}
\end{eqnarray}
where $\eta_i$ and $\alpha$ are from (\ref{eq:alpha-mu}). Then, from (\ref{eq:pow2*}) and (\ref{eq:pow-fp}), we have:
{}
\begin{eqnarray}
\epsilon_{\tt dir}&=&\frac{1}{2\Big(J^2_0(\alpha{\textstyle\frac{1}{2}}d)+J^2_1(\alpha{\textstyle\frac{1}{2}}d)\Big)}\int_{p^-_{\tt ER}}^{p_{\tt ER}^+}
\hskip 0pt  p_i dp_i \Big(\frac{2}{(\alpha{\textstyle\frac{1}{2}}d)^2-p_i^2}   \Big(\alpha{\textstyle\frac{1}{2}}d J_0(p_i) J_1(\alpha {\textstyle\frac{1}{2}}d)-p_i J_0(\alpha {\textstyle\frac{1}{2}}d) J_1(p_i)\Big)\Big)^2.
  \label{eq:een*}
\end{eqnarray}

\begin{figure}[t]
\includegraphics[scale=1.1]{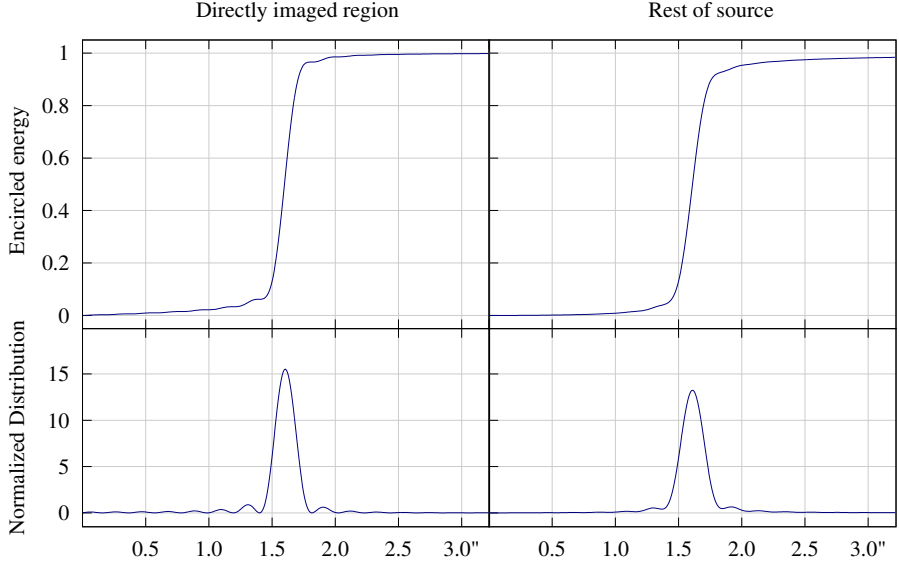}
\caption{\label{fig:s9099}Encircled energy and its normalized distribution for the directly imaged region (\ref{eq:een*}) and the rest of the source (\ref{eq:een*b}). Horizontal axis is in seconds of arc, as seen by a telescope positioned at ${\overline z}=650$~AU. The peak at $\sim 1.6''$ corresponds to the location of the Einstein ring.
}
\end{figure}

As the quantity $(\alpha{\textstyle\frac{1}{2}}d)$ is rather large, $\alpha{\textstyle\frac{1}{2}}d\simeq 24.49\,\big({1\,\mu{\rm m}}/{\lambda}\big)\big({d}/{1\,{\rm m}}\big)\big({650\,{\rm AU}}/{\overline z}\big)^\frac{1}{2}$, we may simplify (\ref{eq:een*}) by using the asymptotic approximation of the Bessel functions  (\ref{eq:BF}), which results in the following:
{}
\begin{eqnarray}
\epsilon_{\tt dir}&=&\frac{1}{\pi}\int_{\alpha {\textstyle\frac{1}{2}}d - {\textstyle\frac{\pi}{2}}}^{\alpha {\textstyle\frac{1}{2}}d + {\textstyle\frac{\pi}{2}}}
\hskip 0pt  dp_i \Big(\frac{\sin\big(\alpha{\textstyle\frac{1}{2}}d-p_i\big)}{\alpha{\textstyle\frac{1}{2}}d-p_i}  -\frac{\cos\big(\alpha{\textstyle\frac{1}{2}}d+p_i\big)}{\alpha{\textstyle\frac{1}{2}}d+p_i} \Big)^2\simeq 0.77,
  \label{eq:een*appr}
\end{eqnarray}
which indicates that only $\sim77\%$ of the energy incident on the telescope from the directly imaged region is deposited within the annulus with thickness of $\lambda/d$ centered at the Einstein ring.

As a result, the power received from the directly imaged region on a resolved exoplanet and measured at the Einstein ring in the focal plane of a diffraction-limited telescope, with $\epsilon_{\tt dir} $ from (\ref{eq:een*appr}),  may be given as
{}
\begin{eqnarray}
P_{\tt fp.dir}&=&\epsilon_{\tt dir} B_{\tt s} \frac{\mu_0\pi^2d^4}{16{\overline z}^2}\Big(J^2_0(\alpha{\textstyle\frac{1}{2}}d)+J^2_1(\alpha{\textstyle\frac{1}{2}}d)\Big)\simeq \epsilon_{\tt dir} B_{\tt s} \frac{\pi^2 d^3}{4{\overline z}}\sqrt{\frac{2r_g}{\overline z}},
  \label{eq:pow**}
\end{eqnarray}
where we used the approximations (\ref{eq:BF}) and the definitions (\ref{eq:alpha-mu}). We note that the power (\ref{eq:pow**}) is independent of the observing wavelength and the distance to the target; however it is a strong function of the telescope's aperture, as expected.

\subsection{Power in the focal plane due to blur from the rest of the planet}
\label{sec:foc-power-blur}

Similarly to the discussion on the signal from the directly imaged region, we first compute the total power deposited in the focal plane from the rest of the extended, resolved exoplanet. For this, we take (\ref{eq:P-blur*2*}) and form the quantity
{}
\begin{eqnarray}
P^0_{\tt fp.blur}({\vec x}_0)&=&\int^{2\pi}_0 \hskip -4pt  d\phi_i \int_0^\infty
\hskip 0pt I_{\tt blur}({\vec x}_i,{\vec x}_0) \rho_i d\rho_i = \frac{B_{\tt s}}{{\overline z}^2} \Big(\frac{kd^2}{8f}\Big)^2\frac{\mu_0 d}{2\alpha}\times\nonumber\\
&& \hskip -50pt \times\,\frac{1}{2\pi}
\int_0^{2\pi} \hskip -8pt d\phi'' \bigg(\frac{2r_\oplus}{d}
\sqrt{1-\big(\frac{\rho_0}{r_\oplus}\big)^2\sin^2\phi''}-1\bigg)
\int^{2\pi}_0 \hskip -4pt  d\phi_i \int_0^\infty
\hskip 0pt  \rho_i d\rho_i  \bigg(\Big( \frac{2
J_1\big(u_+\frac{1}{2}d\big)}{u_+\frac{1}{2}d}\Big)^2+\Big( \frac{2
J_1\big(u_-\frac{1}{2}d\big)}{u_-\frac{1}{2}d}\Big)^2\bigg).
  \label{eq:pow-fp0b}
\end{eqnarray}

Using the variable $p_i$ given by (\ref{eq:p-eta}), that yields
\begin{eqnarray}
u_\pm {\textstyle\frac{1}{2}}d=\sqrt{(\alpha {\textstyle\frac{1}{2}}d)^2\mp2 \alpha {\textstyle\frac{1}{2}}d \,p_i \cos(\phi_i-\phi'')+p_i^2},
  \label{eq:upm+}
\end{eqnarray}
the last integral in the expression (\ref{eq:pow-fp0b}) is evaluated as
{}
\begin{eqnarray}
\int^{2\pi}_0 \hskip -4pt  d\phi_i \int_0^\infty
\hskip 0pt  \rho_i d\rho_i  \Big( \frac{2
J_1\big(u_\pm\frac{1}{2}d\big)}{u_\pm\frac{1}{2}d}\Big)^2&=& \Big(\frac{\lambda f}{\pi d}\Big)^2\int^{2\pi}_0 \hskip -4pt  d\phi_i \int_0^\infty
\hskip 0pt  p_i dp_i  \Big( \frac{2
J_1\big(u_\pm\frac{1}{2}d\big)}{u_\pm\frac{1}{2}d}\Big)^2
=4\pi \Big(\frac{\lambda f}{\pi d}\Big)^2.
  \label{eq:int-fp0*}
\end{eqnarray}
The integrand in (\ref{eq:int-fp0*}) effectively behaves akin to a delta function as it predominantly selects points on the Einstein ring, as shown in (\ref{eq:Bess-rat}). This result allows us to express (\ref{eq:pow-fp0b}) as
{}
\begin{eqnarray}
P^0_{\tt fp.blur}({\vec x}_0)&=&B_{\tt s}   \frac{\pi^2 d^3}{4 {\overline z}^2}\frac{\mu_0}{\pi\alpha} \Big(\frac{2r_\oplus}{d}\epsilon(\rho_0)-1\Big)\equiv P_{\tt blur}({\vec x}_0),
  \label{eq:pow-fp0+}
\end{eqnarray}
where $\epsilon(\rho_0)$ is given by (\ref{eq:eps_r0}) and $P_{\tt blur}({\vec x}_0)$  is the total integrated flux (i.e., power) received from the area on the source which is outside the directly imaged region, as given by Eq.~(30) of \cite{Turyshev-Toth:2019-blur}. Therefore, our results describing the intensity distribution due to the blur at the focal plane of an imaging telescope (\ref{eq:P-blur*2*}) and those derived for photometric imaging in  \cite{Turyshev-Toth:2019-blur}, where we estimated the total power incident on the aperture of that telescope, are also equivalent.

Now, similarly to (\ref{eq:pow-fp0b}), we can estimate the power deposited at the annulus around the Einstein ring corresponding to the blur signal, $P_{\tt blur}$. For this, we take  (\ref{eq:P-blur*2*}) and integrate it over the area seen by the diffraction-limited telescope:
{}
\begin{eqnarray}
P_{\tt fp.blur}({\vec x}_0)&=&\int^{2\pi}_0 \hskip -4pt  d\phi_i \int_{\rho^-_{\tt ER}}^{\rho_{\tt ER}^+}
\hskip 0pt I_{\tt blur}({\vec x}_i,{\vec x}_0) \rho_i d\rho_i = \frac{B_{\tt s}}{{\overline z}^2} \Big(\frac{kd^2}{8f}\Big)^2\frac{\mu_0 d}{2\alpha}\times\nonumber\\
&& \hskip -50pt \times\,\frac{1}{2\pi}
\int_0^{2\pi} \hskip -8pt d\phi'' \bigg(\frac{2r_\oplus}{d}
\sqrt{1-\big(\frac{\rho_0}{r_\oplus}\big)^2\sin^2\phi''}-1\bigg)
\int^{2\pi}_0 \hskip -4pt  d\phi_i \int_{\rho^-_{\tt ER}}^{\rho_{\tt ER}^+}
\hskip 0pt  \rho_i d\rho_i  \bigg(\Big( \frac{2
J_1\big(u_+\frac{1}{2}d\big)}{u_+\frac{1}{2}d}\Big)^2+\Big( \frac{2
J_1\big(u_-\frac{1}{2}d\big)}{u_-\frac{1}{2}d}\Big)^2\bigg).
  \label{eq:pow-fp0*}
\end{eqnarray}

To simplify (\ref{eq:pow-fp0*}), similarly to (\ref{eq:pow-frac}), it is convenient to introduce the encircled energy factor, $\epsilon_{\tt dir}$, for the blur contribution
{}
\begin{eqnarray}
P_{\tt fp.blur}({\vec x}_0)&=&\epsilon_{\tt blur}P_{\tt blur}({\vec x}_0).
  \label{eq:pow-frac-bl}
\end{eqnarray}
As we integrate over $d\phi_i$ for the entire period of  $[0,2\pi]$, the factor $\epsilon_{\tt blur}$ may be given in a very concise form. Thus, with the help of (\ref{eq:pow-fp0+}), (\ref{eq:pow-fp0b}) and the variable $p_i$ from (\ref{eq:p-eta}) yielding $u_\pm {\textstyle\frac{1}{2}}d$ given by (\ref{eq:upm+}), after numerical integration, we have
{}
\begin{eqnarray}
\epsilon_{\tt blur}&=&\frac{1}{8\pi}\int^{2\pi}_0 \hskip -4pt  d\phi_i \int_{p^-_{\tt ER}}^{p_{\tt ER}^+}
\hskip 0pt  p_i dp_i \bigg(\Big( \frac{2
J_1\big(u_+\frac{1}{2}d\big)}{u_+\frac{1}{2}d}\Big)^2+\Big( \frac{2
J_1\big(u_-\frac{1}{2}d\big)}{u_-\frac{1}{2}d}\Big)^2\bigg)\simeq 0.69,
  \label{eq:een*b}
\end{eqnarray}
independent of the angle $\phi''$ present in (\ref{eq:upm+}). This result suggests that only $\sim69\%$ of the energy incident on the telescope from the the area outside the directly imaged region is deposited within the annulus with thickness of $\lambda/d$ centered at the Einstein ring. Because of the diffraction within the telescope, a  significant part of the remaining energy is deposited at the center of the focal plane and in the side lobes of the diffraction pattern, as seen in Fig.~\ref{fig:images}.

Therefore, the power received from outside the directly imaged region of a resolved source and measured at the Einstein ring in the focal plane of a diffraction-limited telescope, with $\epsilon_{\tt blur} $ from (\ref{eq:een*b}),  is given as
{}
\begin{eqnarray}
P_{\tt fp.blur}(\rho_0)&=&\epsilon_{\tt blur} B_{\tt s}   \frac{\pi^2 d^3}{4 {\overline z}^2}\frac{\mu_0}{\pi\alpha} \Big(\frac{2r_\oplus}{d}\epsilon(\rho_0)-1\Big)\simeq \epsilon_{\tt blur} B_{\tt s} \frac{\pi^2 d^3}{4{\overline z}}\sqrt{\frac{2r_g}{\overline z}}\Big(\frac{2R_\oplus}{d}\frac{\overline z}{z_0}\epsilon(\rho_0)-1\Big),
\label{eq:pow*b*}
\end{eqnarray}
where we used (\ref{eq:BF}) and (\ref{eq:alpha-mu}) to simplify the result. We note that the power (\ref{eq:pow*b*}) is also independent of the observing wavelength, but is inversely proportional to the distance to the source.

As a result, the total power received from the entire exoplanet,
\begin{eqnarray}
P_{\tt fp.exo}(\rho_0)=P_{\tt fp.dir}+P_{\tt fp.blur}(\rho_0),
  \label{eq:pow*tot}
\end{eqnarray}
at the location of the Einstein ring in the focal plane of a diffraction-limited telescope with the help of  (\ref{eq:pow**}) and (\ref{eq:pow*b*}) is given as
{}
\begin{eqnarray}
P_{\tt fp.exo}(\rho_0)&=&
B_{\tt s} \frac{\pi^2 d^3}{4{\overline z}}\sqrt{\frac{2r_g}{\overline z}}\Big(\epsilon_{\tt dir} +\epsilon_{\tt blur}  \Big(\frac{2r_\oplus}{d}\epsilon(\rho_0)-1\Big)\Big)\simeq
 \epsilon_{\tt blur} B_{\tt s}\pi^2 d^2  \frac{R_\oplus}{2z_0}\sqrt{\frac{2r_g}{\overline z}}\epsilon(\rho_0), \qquad 0\leq \rho_0 \leq r_\oplus,~~~~
  \label{eq:pow*exo*}
\end{eqnarray}
which is similar to the result obtained in \cite{Turyshev-Toth:2019-blur} for the case of photometric imaging of extended objects with the SGL.

\subsection{Power in the focal plane from an off-image source}
\label{sec:foc-power-off}

Similarly to (\ref{eq:pow-fp0*}), we may evaluate the energy received at the focal plane corresponding to intensity (\ref{eq:P-blur*off4*}).  We can do that by integrating (\ref{eq:P-blur*off4*}) over the focal plane of the imaging telescope, as we did for (\ref{eq:pow-dirD}) and (\ref{eq:P-blur*2*}), namely
{}
\begin{eqnarray}
P_{\tt fp.off}(\rho_0)&=&\int^{2\pi}_0 \hskip -4pt  d\phi_i \int_{\rho^-_{\tt ER}}^{\rho_{\tt ER}^+}
\hskip 0 pt I_{\tt blur}({\vec x}_i,{\vec x}_0) \rho_i d\rho_i=\frac{B_{\tt s}}{{z_0}^2} \Big(\frac{kd^2}{8f}\Big)^2\frac{2\mu_0 R_\oplus}{\alpha\beta}\times\nonumber\\
&&\hskip -30pt \times\,
\frac{1}{2\pi}
\int_{\phi_-}^{\phi_+} \hskip -8pt d\phi''
\Big(\sqrt{1-\big(\frac{\rho_0}{r_\oplus}\big)^2\sin^2\phi''}\Big)
\int^{2\pi}_0 \hskip -4pt  d\phi_i \int_{\rho^-_{\tt ER}}^{\rho_{\tt ER}^+}
\rho_i d\rho_i
 \bigg(\Big( \frac{2
J_1\big(u_+\frac{1}{2}d\big)}{u_+\frac{1}{2}d}\Big)^2+\Big( \frac{2
J_1\big(u_-\frac{1}{2}d\big)}{u_-\frac{1}{2}d}\Big)^2\bigg).
  \label{eq:pow*2+}
\end{eqnarray}
Similarly to the derivation of $P_{\tt fp.blur}(\rho_0)$ above, this expression  results in the following
{}
\begin{eqnarray}
P_{\tt fp.off}(\rho_0)&=&\epsilon_{\tt blur} B_{\tt s} \pi^2 d^2 \frac{R_\oplus}{2z_0}\sqrt{\frac{2r_g}{\overline z}}\beta(\rho_0),\qquad \rho_0 \geq r_\oplus,
  \label{eq:pow*2++}
\end{eqnarray}
which is equivalent to $\epsilon_{\tt blur} P_{\tt off}(\rho_0)$, where $P_{\tt off}$ is the power received for off-source pointing, as given by Eq.~(38) of \cite{Turyshev-Toth:2019-blur} and $\beta(\rho_0)$ is from (\ref{eq:beta_r0}). Therefore, our results describing the intensity distribution due to the blur at the focal plane of an imaging telescope for off-source pointing (\ref{eq:P-blur*off4*}) and those derived for photometric imaging in  \cite{Turyshev-Toth:2019-blur}, are complimentary.

Result (\ref{eq:pow*2++}) may be used, in particular, to model light contamination from the parent star, which, as shown in Fig.~\ref{fig:images} (right) contributes two spots at the Einstein ring that may be masked by an appropriate  management of the focal plane.

\subsection{Power in the focal plane at a large distance from the optical axis}
\label{sec:power-gowi}

Once we move far away from the optical axis, the power deposited in the focal plane of the optical telescope is computed with the intensity distributions (\ref{eq:FI-go-Int}) and (\ref{eq:FI-ir+-Int}) for the geometric optics and weak interference regions, correspondingly.   When we integrate over the focal plane, we see from Fig.~\ref{fig:images} that the two images corresponding to the incident and scattered waves are seen in the focal plane as unresolved circles, with radii determined from (\ref{eq:alpha-mu}) and (\ref{eq:betapm}) as $\eta_i=\xi_{\tt in/sc}$. Therefore, the useful signal received in the focal plane of a diffraction-limited telescope occupies the annulus between the two radii, $\rho_{\tt in/sc}^\pm$, that from (\ref{eq:betapm}) are given as
{}
\begin{eqnarray}
\rho_{\tt in}^\pm=f\Big(\Big(\sqrt{1+\frac{8r_g \tilde r}{\rho^2_0}}+1\Big)\frac{\rho_0}{2\tilde r}\pm\frac{\lambda}{2d}\Big)
\qquad{\rm and}\qquad
\rho_{\tt sc}^\pm=f\Big(\Big(\sqrt{1+\frac{8r_g \tilde r}{\rho^2_0}}-1\Big)\frac{\rho_0}{2\tilde r}\pm\frac{\lambda}{2d}\Big).
  \label{eq:insc-rho}
\end{eqnarray}
As a result, the variable $p_i$ from (\ref{eq:p-eta}) varies within different radii:
{}
\begin{eqnarray}
p^\pm_{\tt in}=\xi_{\tt in} {\textstyle\frac{1}{2}}d \pm {\textstyle\frac{\pi}{2}}
 \qquad {\rm and} \qquad
p^\pm_{\tt sc}=\xi_{\tt sc}  {\textstyle\frac{1}{2}}d \pm {\textstyle\frac{\pi}{2}}.
  \label{eq:p-eta-gi}
\end{eqnarray}

Following the approach that was developed in in Sec.~\ref{sec:foc-power-blur}, with the help of (\ref{eq:FI-go-Int}) and (\ref{eq:FI-ir+-Int}), we compute the power deposited in the focal plane in the geometric optics and weak interference regions, which take the form
{}
\begin{eqnarray}
P_{\tt fp.geom.opt}({\vec x}_0)&=&\int^{2\pi}_0 \hskip -4pt  d\phi_i \int_{\rho^-_{\tt in}}^{\rho_{\tt in}^+}
\hskip 0pt I_{\tt geom.opt}({\vec x}_i,{\vec x}_0) \rho_i d\rho_i = \epsilon_{\tt geom.opt} B_{\tt s} \pi^2d^2\frac{R^2_\oplus}{4z_0^2} a^2_{\tt in},
  \label{eq:fp.pow-go}\\
  P_{\tt fp.weak.int}({\vec x}_0)&=&\int^{2\pi}_0 \hskip -4pt  d\phi_i \int_{\rho^-_{\tt sc}}^{\rho_{\tt sc}^+}
\hskip 0pt I_{\tt weak.int}({\vec x}_i,{\vec x}_0) \rho_i d\rho_i = \epsilon_{\tt weak.int} B_{\tt s} \pi^2 d^2\frac{R^2_\oplus}{4z_0^2} \big(a^2_{\tt in}+a^2_{\tt sc}\big),
  \label{eq:fp.pow-wi}
\end{eqnarray}
where the encircled energies for these regions with the help of (\ref{eq:int-fp0*}) are given as
 {}
\begin{eqnarray}
\epsilon_{\tt geom.opt}&=&\frac{1}{4\pi}\int^{2\pi}_0 \hskip -4pt  d\phi_i \int_{p^-_{\tt in}}^{p_{\tt in}^+}
\hskip 0pt  p_i dp_i \Big( \frac{2
J_1\big(v_+\frac{1}{2}d\big)}{v_+\frac{1}{2}d}\Big)^2\simeq 0.69,
  \label{eq:ee-go}\\
\epsilon_{\tt weak.int}&=&
\frac{1}{4\pi (a^2_{\tt in}+a^2_{\tt sc})}\Big\{a^2_{\tt in}\int^{2\pi}_0 \hskip -4pt  d\phi_i \int_{p^-_{\tt in}}^{p_{\tt in}^+}
\hskip 0pt  p_i dp_i \Big( \frac{2
J_1\big(v_+\frac{1}{2}d\big)}{v_+\frac{1}{2}d}\Big)^2+a^2_{\tt sc}\int^{2\pi}_0 \hskip -4pt  d\phi_i \int_{p^-_{\tt sc}}^{p_{\tt sc}^+}
\hskip 0pt  p_i dp_i \Big( \frac{2
J_1\big(v_-\frac{1}{2}d\big)}{v_-\frac{1}{2}d}\Big)^2\Big\}\simeq\nonumber\\[4pt]
&&\simeq 0.69.
  \label{eq:ee-wi}
\end{eqnarray}
We see that the power deposited at the foal plane of the optical telescope is amplified by the factors $a^2_{\tt in}$ and $a^2_{\tt sc}$ which, according to (\ref{eq:a12_amp}), is getting larger as the deviation from the optical axis, $\rho_0$, decreases. Thus, as we move closer to the optical axis, amplification gets larger and once we enter the strong interference region it is given by  (\ref{eq:pow*exo*}).

Finally, we mention that sources at moderate distances from the parent star do not contribute to the signal measured at the Einstein ring. As their diffraction-limited images will be centered at the angles $\xi_{\tt in/sc}$ given by  (\ref{eq:betapm}), they will not bring light contamination to the Einstein ring and thus, they may be ignored in the relevant SNR analysis.

\subsection{Anticipated signals for imaging an exo-Earth}
\label{sec:signal-estimates}

We may now estimate the signals that could be expected from realistic targets when they are imaged with the SGL. We consider a planet identical to our Earth that orbits a star identical to our Sun.  The total flux received by such a target is the same as the solar irradiance at the top of Earth's atmosphere, given as  $I_0=1,366.83~{\rm W/m}^2$. Approximating the planet as a Lambertian sphere illuminated from the viewing direction yields a Bond spherical albedo \cite{Lester-etal:1979} of $2/3\pi$, and  the target's average surface brightness becomes $B_{\tt s}=({2}/{3\pi})\alpha I_0$, where we take Earth's broadband albedo to be $\alpha=0.3$ and assuming that we see a fully-illuminated planet at 0 phase angle.

With these parameters, the power, $P_{\tt fp.dir}$, and the photon flux, $Q_{\tt fp.dir}=P_{\tt fp.dir}(\lambda/hc)$, corresponding to the signal received from the directly imaged region of the planet is estimated from (\ref{eq:pow**}) to be
{}
\begin{eqnarray}
P_{\tt fp.dir}&=& \epsilon_{\tt dir} \alpha I_0 \frac{\pi d^3}{6{\overline z}}\sqrt{\frac{2r_g}{\overline z}}=1.33\times 10^{-17}\,\Big(\frac{d}{1\,{\rm m}}\Big)^3\Big(\frac{650\,{\rm AU}}{\overline z}\Big)^\frac{3}{2}~{\rm W},
  \label{eq:dir-sP}\\
Q_{\tt fp.dir}&=&
66.71\, \Big(\frac{d}{1\,{\rm m}}\Big)^3\Big(\frac{650\,{\rm AU}}{\overline z}\Big)^\frac{3}{2}\Big(\frac{\lambda}{1\,\mu{\rm m}}\Big)~{\rm photons/s},
  \label{eq:dir-s}
\end{eqnarray}
where we assumed that all light is transmitted at $\lambda=1~\mu$m and used $\epsilon_{\tt dir}=0.77$.

Similarly, assuming that the planet is positioned at $z_0=30$~pc away from us,  with the help of (\ref{eq:pow*b*}) (or, equivalently, from (\ref{eq:pow*exo*})) and using $\epsilon_{\tt blur}=0.69$,
we estimate the signal from the rest of the planet as
{}
\begin{eqnarray}
P_{\tt fp.blur}(\rho_0)&=&  \epsilon_{\tt blur} \alpha I_0 {\pi d^2}\frac{R_\oplus}{3z_0}\sqrt{\frac{2r_g}{\overline z}}\epsilon(\rho_0)= 1.59\times 10^{-14}\epsilon(\rho_0)
\Big(\frac{d}{1\,{\rm m}}\Big)^2\Big(\frac{650\,{\rm AU}}{\overline z}\Big)^\frac{1}{2}\Big(\frac{30\,{\rm pc}}{z_0}\Big)~{\rm W},
\label{eq:blur-sP}\\
Q_{\tt fp.blur}(\rho_0)&=&
8.01\times 10^4\epsilon(\rho_0) \Big(\frac{d}{1\,{\rm m}}\Big)^2\Big(\frac{650\,{\rm AU}}{\overline z}\Big)^\frac{1}{2}\Big(\frac{30\,{\rm pc}}{z_0}\Big)
\Big(\frac{\lambda}{1\,\mu{\rm m}}\Big)~{\rm photon/s}.~~~~~
\label{eq:blur-s}
\end{eqnarray}

For comparison, we can also compute the power observed by a regular telescope (unaided by the SGL). Using (\ref{eq:fp.pow-go}) and positioning the telescope at the distance $\rho_0=10R_\odot$ (so that $a^2_{\tt in}=1$) from the SGL optical axis, which corresponds to geometric optics regime, typically found in modern astronomical observations (with $\epsilon_{\tt geom.opt}=0.69$):
{}
\begin{eqnarray}
P_{\tt fp.geom.opt}(\rho_0)&=& \epsilon_{\tt geom.opt} \alpha I_0 {\pi d^2} \frac{R^2_\oplus}{6z_0^2} a^2_{\tt in}\simeq
7.03\times 10^{-21}~
\Big(\frac{d}{1\,{\rm m}}\Big)^2\Big(\frac{30\,{\rm pc}}{z_0}\Big)^2~{\rm W},  \label{eq:pow-go+phP}\\
Q_{\tt fp.geom.opt}(\rho_0)&=&
3.54\times 10^{-2}\Big(\frac{d}{1\,{\rm m}}\Big)^2\Big(\frac{30\,{\rm pc}}{z_0}\Big)^2\Big(\frac{\lambda}{1\,\mu{\rm m}}\Big)~{\rm photons/s}.
  \label{eq:pow-go+ph}
\end{eqnarray}

Using this estimate, we can compare the performance of a conventional telescope against one aided by the SGL. The angular resolution (\ref{eq:S_=0}) needed to resolve features of size $D$ given by (\ref{eq:Dd}) in the target plane requires a telescope with aperture $d_D\sim1.22\, (\lambda/D) z_0=1.22\, (\lambda/d) {\overline z}\simeq 1.19 \times 10^5~{\rm km}=18.60 R_\oplus$, which is not realistic. The photon flux of a $d=1$~m telescope can be calculated by scaling the result (\ref{eq:pow-go+ph}) by a factor of $(D/2R_\oplus)^2\simeq 5.57 \times 10^{-7}$, yielding the value of $1.97 \times 10^{-8}$
photons/s, which is extremely small. Comparing this flux with (\ref{eq:dir-s}), we see that the SGL, used in conjunction with a $d=1$~m telescope, amplifies the light from the directly imaged region (i.e., an unresolved source) by a factor of
$\sim 3.38\times10^9\,(d/1{\rm m})({650\,{\rm AU}}/{\overline z})^\frac{3}{2}(z_0/{30\,{\rm pc}})^2$.

\subsection{Noise from the solar corona and detection SNR}
\label{sec:sol-cor}

The Einstein ring corresponding to a distant target, as observed from a position in the SGL focal region, is seen through the bright solar corona, which represents an important noise contribution that must be considered. Noise from the solar corona can be mitigated by letting as little light from the corona to reach the instrument as possible. This is achieved by employing a suitably designed solar coronagraph, needed in any case to block direct light from the Sun, but which can also be used to reduce the noise from the solar corona.

Solar coronagraphy was invented by Lyot \cite{Lyot:1932} to study the solar corona by blocking out the Sun and reproducing solar eclipses artificially. Coronagraphs are also considered to block out light from point sources, such as the host star of an exoplanet imaged with conventional telescope \cite{Traub:2010}. The SGL coronagraph is different, as it needs to block the light from the Sun and the solar corona, leaving visible only those areas where the Einstein ring appears.

\begin{figure}
\includegraphics[scale=0.833]{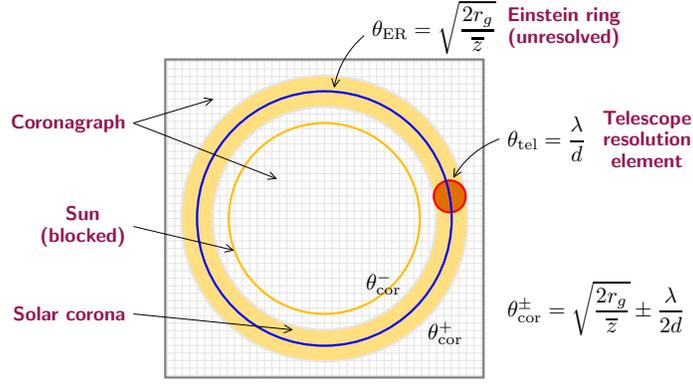}
\caption{\label{fig:foc-plane-det} The annular coronagraph concept. The coronagraph blocks light from both within and outside the Einstein ring. The thickness of the exposed area is determined by the diffraction limit of the optical telescope at its typical observational wavelength.}
\end{figure}

\begin{figure}
\includegraphics[width=0.40\linewidth]{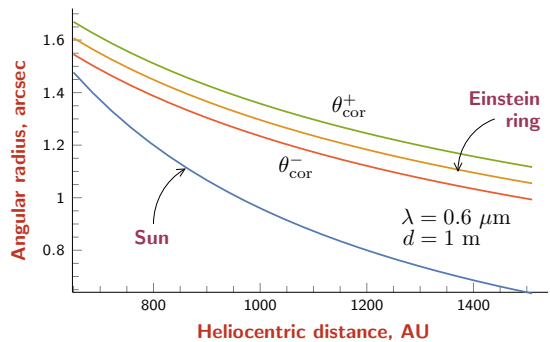}
\caption{\label{fig:sol-cor-bright}  Angular sizes of the Sun and the diffraction-limited view of the Einstein ring as functions of heliocentric distance (for $\lambda=0.6~\mu$m.) As the heliocentric distance increases, the Einstein ring (together with the entire imaged region) further separates from the Sun. A coronagraph may have to be able to compensate for decreasing angular sizes.}
\end{figure}

The already available design for the SGL coronagraph \cite{Zhou:2018} rejects sunlight with a contrast ratio of $\sim 10^7$. At this level of rejection, the light from the solar disk is completely blocked to the level comparable to the brightness of the solar corona.
Taking a further step, we consider two possible coronagraph concepts. A conventional coronagraph (which we call a ``disk coronagraph'') that blocks light only from the solar disk and the solar corona up to the inner boundary, $\theta_{\tt cor}^-$, of the $\lambda/d$ annulus centered on the Einstein ring, and a coronagraph that also blocks light outside the outer boundary, $\theta_{\tt cor}^+$, of the $\lambda/d$-annulus centered at the Einstein ring (the ``annular coronagraph'', shown in Fig.~\ref{fig:foc-plane-det}). Fig.~\ref{fig:sol-cor-bright} shows the relative angular sizes for the Sun and the Einstein ring, as heliocentric distance increases.

Compared to the disk coronagraph, the annular coronagraph reduces the noise contribution from the solar corona by an additional $\sim 10\%$. As the solar corona is quite bright compared to the Einstein ring, the use of an annular coronagraph is preferred for an SGL imaging instrument. Consequently, in the estimates that we develop for the corona contribution, we assume an annular coronagraph design.

In Appendix~\ref{sec:model}, we estimate the contribution from the solar corona. Integrating (\ref{eq:model-th}) over the observed width and circumference of the Einstein ring annulus, we obtain (\ref{eq:pow-fp=+*4}), which yields the following estimate (with $\epsilon_{\tt cor}\simeq0.60$):
{}
\begin{eqnarray}
P_{\tt fp.cor}
&=&19.48\,\epsilon_{\tt cor}\, \pi^2 \lambda d\,\frac{R_\odot}{\overline z} \Big(\frac{R_\odot}{\sqrt{2r_g\overline z}}\Big)^{6.8}\Big[1+1.89 \Big(\frac{R_\odot}{\sqrt{2r_g\overline z}}\Big)^{10.2}
+0.0284\Big(\frac{\sqrt{2r_g\overline z}}{R_\odot}\Big)^{5.3}
\Big]
=\nonumber\\
&=&4.56 \times 10^{-10}\,\Big[1+0.79 \Big(\frac{650\,{\rm AU}}{\overline z}\Big)^{5.1}
+0.05\Big(\frac{\overline z}{650\,{\rm AU}}\Big)^{2.65}
\Big]\Big(\frac{d}{1\,{\rm m}}\Big)\Big(\frac{650\,{\rm AU}}{\overline z}\Big)^{4.4}\Big(\frac{\lambda}{1\,\mu{\rm m}}\Big)
~~  {\rm W}.
  \label{eq:pow-fp=+*4+}
\end{eqnarray}
This corresponds to the corona photon flux, which is estimated to be
{}
\begin{align}
Q_{\tt fp.cor}
=2.29 \times 10^{9}\,\Big[1+0.79 \Big(\frac{650\,{\rm AU}}{\overline z}\Big)^{5.1}
+0.05\Big(\frac{\overline z}{650\,{\rm AU}}\Big)^{2.65}
\Big]\Big(\frac{d}{1\,{\rm m}}\Big)\Big(\frac{650\,{\rm AU}}{\overline z}\Big)^{4.4}\Big(\frac{\lambda}{1\,\mu{\rm m}}\Big)^2
~~  {\rm photons/s}.
  \label{eq:pow-fp=+*4+2}
\end{align}

Assuming that the contribution of the solar corona is removable (e.g., by observing the corona from a slightly different vantage point) and only stochastic (shot) noise remains, we estimate the resulting ${\rm SNR}_{\tt C}$  of detecting the signal (convolved with the SGL, thus, the subscript `{\tt C}')
in the solar corona dominated regime as
{}
\begin{equation}
{\rm SNR}_{\tt C}=\frac{Q_{\tt fp.blur}}{\sqrt{Q_{\tt fp.cor}}}=\frac{1.68\,\epsilon(\rho_0) }{\sqrt{1+0.79 \Big(\dfrac{650\,{\rm AU}}{\overline z}\Big)^{5.1}+0.05\Big(\dfrac{\overline z}{650\,{\rm AU}}\Big)^{2.65}
}}\Big(\frac{d}{1\,{\rm m}}\Big)^\frac{3}{2}\Big(\frac{30\,{\rm pc}}{z_0}\Big)\Big(\frac{\overline z}{650\,{\rm AU}}\Big)^{1.7}\,\sqrt{\frac{t}{1\,{\rm s}}}.
\label{eq:snr-cor}
\end{equation}

It is noteworthy to consider the behavior of this ${\rm SNR}_{\tt C}$ of (\ref{eq:snr-cor}) with respect to the several parameters involved:
\begin{inparaenum}[1)]
\item It does not depend on the wavelength. This is because for this estimate we assumed the presence of an annular coronagraph. The width of the annulus of such a coronagraph is $\propto \lambda/d$, thus canceling out the wavelength dependence. (A disk coronagraph would increase the noise contribution from the corona by $\sim 10$\% with a weak wavelength dependence.)
\item Within heliocentric ranges of interest, the ${\rm SNR}_{\tt C}$ improves almost linearly with the heliocentric distance. Although the angular size of the Einstein ring  decreases as $\propto 1/\sqrt{\overline z}$, the plasma contribution diminishes much faster, as $\propto 1/{\overline z}^{4.4}$. Combining these two factors results in the overall $\propto{\overline z}^{1.7}$ behavior of the ${\rm SNR}_{\tt C}$.
\item The ${\rm SNR}_{\tt C}$ has a rather strong dependence on the telescope aperture, behaving as $\propto d^\frac{3}{2}$. This is, again, due to our use of the annular coronagraph in deriving the  estimate of the solar corona signal.
\end{inparaenum}

\section{Image reconstruction with the SGL}
\label{sec:convolve}

In the preceding sections we developed analytical tools that are needed to estimate the signal levels from various distant targets. The next step is to understand how these signals can be measured and used to reconstruct the images of those targets. We also need to understand the actual circumstances of signal acquisition, the inevitable noise that accompanies these observations, and the implied constraints such as minimum integration times that are required to acquire signals of sufficient quality.

To address these questions, we need to study the role of the SGL PSF, ${ \mu}_{\tt SGL}$, from (\ref{eq:S_z*6z-mu2}) in image formation and how knowledge of the PSF makes image reconstruction possible.

\subsection{Image convolution by the SGL}
\label{sec:conv-integral}

We consider a photometric imaging process, in which a telescope is used to measure the power (yielding the signal amplitude) of the signal that enters a telescope with aperture diameter $d$. To compute the total power of the signal that is amplified by the SGL and is received by the telescope, we convolve the surface brightness of the source, $B_{\tt s}({\vec x}')$,  by the amplification factor of the SGL, ${ \mu}_{\tt SGL}$, given by (\ref{eq:S_z*6z-mu2}) and integrate over the aperture by way of the following quadruple integral (as was first given by Eq.~(8) in \cite{Turyshev-Toth:2019-blur}):
{}
\begin{eqnarray}
P({\vec x}_0)&=&
\frac{\mu_0}{z_0^2}\iint\displaylimits_{-\infty}^{+\infty}d^2{\vec x}'\, B_{\tt s}({\vec x}')
\hskip -5pt\iint\displaylimits_{|{\vec x}|^2\leq (\frac{1}{2}d)^2}\hskip -5pt d^2{\vec x}
\,J^2_0\big(\alpha|{\vec x}_0+{\vec x}+\beta{\vec x}'|\big),
\label{eq:power_rec2*}
\end{eqnarray}
where $\alpha$ and $\beta$ are given by (\ref{eq:alpha-mu}) and ${\vec x}_0$, as before, is the telescope's position in the image plane.
Equation (\ref{eq:power_rec2*}) describes the convolution of the extended source with the SGL and may be used to estimate the power of the anticipated photometric signals (see Sec.~\ref{sec:power} and \cite{Turyshev-Toth:2019-image}).  It describes a typical power transmission from an extended source through the medium with the gain of ${ \mu}_{\tt SGL}$, and with the $1/z_0^2$ distance dependence.

We observe that integration over $d^2\vec{x}$ in (\ref{eq:power_rec2*}) amounts to averaging of the SGL PSF (which is given after (\ref{eq:S_=0}) as ${ \mu}_{\tt SGL}/\mu_0=J^2_0\big(\alpha|{\vec x}_0+{\vec x}+\beta{\vec x}'|\big)$) over the telescope aperture, namely:
{}
\begin{eqnarray}
{\overline {\rm PSF}}(|{\vec x}_0+\beta{\vec x}'|)&=&\frac{1}{\pi({\textstyle\frac{1}{2}}d)^2 }
\hskip -5pt\iint\displaylimits_{|{\vec x}|^2\leq (\frac{1}{2}d)^2}\hskip -5pt d^2{\vec x}
\,J^2_0\big(\alpha|{\vec x}+{\vec x}_0+\beta{\vec x}'|\big).
\label{eq:power_rec2}
\end{eqnarray}
As the telescope aperture is expected to be significantly larger than the spatial wavelength of the PSF (i.e., $\alpha d\gg 1$, see relevant discussion in \cite{Turyshev-Toth:2019-image}), the integral in (\ref{eq:power_rec2*}) can be easily evaluated. For this, it is instructional to express the coordinates on the source plane, ${\vec x}'$, via those measured on the image plane, ${\vec x}''$, which can be done with the help of (\ref{eq:mapping}) and (\ref{eq:alpha-mu}), resulting in ${\vec x}'=-{\vec x}''/\beta$. Next, following  \cite{Turyshev-Toth:2019-blur}, we split the argument of the Bessel function into two intervals  $|{\vec x}_0-{\vec x}''|\ll |{\vec x}|< \frac{1}{2}d$ and $|{\vec x}_0-{\vec x}''|\geq \frac{1}{2}d$, which is equivalent to separating the integration over the directly-imaged region and the rest of the exoplanet done in preceding sections. Using the approach demonstrated in Appendix~\ref{sec:PSF-average}, we present the averaged SGL PSF in the form of  (\ref{eq:psf-mu}):
{}
\begin{eqnarray}
{\overline {\rm PSF}}(|{\vec x}_0+\beta{\vec x}'|)\equiv {\overline {\rm PSF}}\big(|{\vec x}_0-{\vec x}''|\big)&=&\frac{1}{\pi\alpha} \frac{4}{d}\, \mu(|{\vec x}_0-{\vec x}''|),
\label{eq:power_x02}
\end{eqnarray}
with the factor $\mu(|{\vec x}_0-{\vec x}''|)$ having the following form:
{}
\begin{eqnarray}
\mu(|{\vec x}_0-{\vec x}''|)&=&
 \bigg\{ \begin{aligned}
\epsilon(|{\vec x}_0-{\vec x}''|), \hskip 10pt 0\leq |{\vec x}_0-{\vec x}''|\leq {\textstyle\frac{1}{2}}d& \\
\beta(|{\vec x}_0-{\vec x}''|), \hskip 30pt |{\vec x}_0-{\vec x}''| > {\textstyle\frac{1}{2}}d& \\
  \end{aligned}\, ,
\label{eq:power_rec9*a}
\end{eqnarray}
and where $\epsilon(|{\vec x}_0-{\vec x}''|)$ and $\beta(|{\vec x}_0-{\vec x}''|)$ are from (\ref{eq:av3b}) and (\ref{eq:av3bb}), correspondingly:
{}
\begin{eqnarray}
\epsilon(|{\vec x}_0-{\vec x}''|)&=&\frac{2}{\pi}{\tt E}\Big[\Big(\frac{2|{\vec x}_0-{\vec x}''|}{d}\Big)^2\Big] \qquad{\rm and}\qquad
\beta(|{\vec x}_0-{\vec x}''|)=\frac{2}{\pi}{\tt E}\Big[\arcsin \Big(\frac{d}{2|{\vec x}_0-{\vec x}''|}\Big),\Big(\frac{2|{\vec x}_0-{\vec x}''|}{d}\Big)^2\Big],~~~~
\label{eq:av9b}
\end{eqnarray}
with ${\tt E}[x]$ and  ${\tt E}[a,x]$ being the elliptic and incomplete elliptic integrals \cite{Abramovitz-Stegun:1965}, respectively.

With this, (\ref{eq:power_rec2*}) transforms  equivalently:
\begin{eqnarray}
P({\vec x}_0)&=&
\frac{\mu_0}{z_0^2\beta^2} \pi({\textstyle\frac{1}{2}}d)^2 \frac{1}{\pi\alpha } \frac{4}{d} \iint\displaylimits_{-\infty}^{+\infty}d^2{\vec x}''\, B_{\tt s}\big(\hskip -2pt-{\vec x}''/\beta\big)\mu(|{\vec x}_0-{\vec x}''|).
\label{eq:power_rec3*}
\end{eqnarray}

Assuming uniform irradiance at the top of the exoplanet's atmosphere, $B_{\tt s}$, we may present the surface brightness of the source as $B_{\tt s}({\vec x}')=B_{\tt s}\alpha_{\tt s}({\vec x}')$, where $\alpha_{\tt s}({\vec x}')$ is the exoplanetary albedo. With this, (\ref{eq:power_rec3*}) takes the form
{}
\begin{eqnarray}
P({\vec x}_0)&=&P_{\tt dir}
\iint\displaylimits_{-\infty}^{+\infty} d^2{\vec x}''\, {\hat \alpha}_{\tt s}\big(\hskip -2pt-{\vec x}''/\beta\big)\mu(|{\vec x}_0-{\vec x}''|),
\label{eq:power_rec7a*}
\end{eqnarray}
where ${\hat \alpha}_{\tt s}\big(\hskip -2pt-{\vec x}''/\beta\big)={ \alpha}_{\tt s}\big(\hskip -2pt-{\vec x}''/\beta\big)/(\pi({\textstyle\frac{1}{2}}d)^2)$ is the albedo surface density within the source area selected by the telescope and $P_{\tt dir}$ is the power that would be received by the telescope at a particular position in the image plane from the source area with the diameter $D=b/\beta$ (as in (\ref{eq:pow**})):
{}
\begin{eqnarray}
P_{\tt dir}&=&
\frac{\mu_0}{z_0^2\beta^2} \pi({\textstyle\frac{1}{2}}d)^2 \frac{4}{d}\frac{1}{\pi\alpha } \pi({\textstyle\frac{1}{2}}d)^2  {B}_{\tt s}=B_{\tt s} \frac{\pi^2 d^3}{4{\overline z}}\sqrt{\frac{2r_g}{\overline z}}.
\label{eq:power_rec5a*}
\end{eqnarray}

Expression (\ref{eq:power_rec7a*}), together with (\ref{eq:power_rec9*a}) exhibits essentially the same structure as (\ref{eq:pow*tot}), where the total power received by the telescope is a sum two components: the power received from the directly-imaged region and that from the rest of the planet.  At any particular telescope position in the image plane, ${{\vec x}_0}_i$, the signal from the directly imaged region $P_{\tt dir}{\alpha_{\tt s}}_i$ is overwhelmed by the blur from the rest of the exoplanet and it is therefore not directly observable. However, as we shall discuss in the next subsection, it is recoverable after deconvolution.

For imaging purposes, we are interested in reconstructing the surface albedo, $\alpha({\vec x}')$, from a series of measurements of $P({\vec x}_0)$. This requires inverting the {\em convolution operator}, represented by the double integral in (\ref{eq:power_rec7a*}).

Computationally, this is best accomplished by way of the Fourier quotient method, taking advantage of the convolution theorem \cite{FILTERS07}, according to which the inverse can be carried out using simple division after a two-dimensional Fourier transform into the spatial frequency domain. This approach also makes it easy to make use of deblurring and spatial filtering algorithms that exist and are applicable for many deconvolution or image deblurring problems \cite{Hansen-etal:2006}.

Our present goal is more modest: We wish to estimate the ``deconvolution penalty'', the amount by which the deconvolution process amplifies noise.

\subsection{Deconvolution in matrix form and noise}
\label{sec:conv-matr}

To understand the effect of deconvolution on signal and noise, we first discretize the integral in (\ref{eq:power_rec7a*}) by replacing the infinite integration limits with a finite integration area that fully covers the source, $r^{\tt im}_\oplus \geq r_\oplus$ and then dividing this area into $N$ equal non-overlapping area elements of size $\sim d^2$, thus
$N=\pi r^{\tt im\, 2}_\oplus/(\pi({\textstyle\frac{1}{2}}d)^2)=(2r^{\tt im}_\oplus/d)^2$. We characterize the positions of each of these source elements projected in the image plane as $\vec{x}''_j$ ($1\le j\le N$). We define the mean surface albedo ${\overline \alpha}_{{\tt s}j}$ for the $j$-th surface element defined by the ${|{\vec x}''_j-{\vec x}''|< \frac{1}{2}d}$ distance from position ${\vec x}''_j$ as
{}
\begin{eqnarray}
{\overline \alpha}_{{\tt s}j}=\iint\displaylimits_{{|{\vec x}''_j-{\vec x}''|< \frac{1}{2}d}}d^2{\vec x}''\, {\hat \alpha}_{\tt s}\big(\hskip -2pt-{\vec x}''/\beta\big)\equiv \frac{1}{\pi({\textstyle\frac{1}{2}}d)^2 }\iint\displaylimits_{{|{\vec x}''_j-{\vec x}''|< \frac{1}{2}d}}d^2{\vec x}''\, { \alpha}_{\tt s}\big(\hskip -2pt-{\vec x}''/\beta\big).
\label{eq:alb}
\end{eqnarray}
Next, we choose $N$ measurement locations ${\vec{x}_0}_i$ in the image plane that satisfy ${\vec{x}_0}_i-\vec{x}_i''=0.$

With these notations, a discretized version of Eq.~(\ref{eq:power_rec7a*}) may be given as
{}
\begin{eqnarray}
P({{\vec x}_0}_i)&=&
P_{\tt dir}
\sum_{j=1}^N \Big(\delta_{ij}+\beta(|{{\vec x}_0}_i-{\vec x}_j''|)(1-\delta_{ij})\Big){\overline \alpha}_{{\tt s}j} =
P_{\tt dir}
\sum_{j=1}^N C_{ij}{\overline \alpha}_{{\tt s}j},~~~~
\label{eq:power0_r}
\end{eqnarray}
where we introduced the {\em convolution matrix}
\begin{align}
C_{ij}=\delta_{ij}+\beta(|{{\vec x}_0}_i-{\vec x}_j''|)(1-\delta_{ij}),
\label{eq:Cij+}
\end{align}
which, with the help of (\ref{eq:psf-mu*2}), may be given in the following approximate form:
\begin{align}
C_{ij}=\delta_{ij}+\frac{d}{4|{{\vec x}_0}_i-{\vec x}_j''|}(1-\delta_{ij})\qquad {\rm or} \qquad
C_{ij}=
\delta_{ij}\Big(1-\frac{d}{4|{{\vec x}_0}_i-{\vec x}_j''|}\Big)+\frac{d}{4|{{\vec x}_0}_i-{\vec x}_j''|}.
\label{eq:Cij}
\end{align}

The quantity  $|{{\vec x}_0}_i-{\vec x}_j''|$ here is distance between the $i$-th telescope location ${{\vec x}_0}_i$ and the projected directly imaged location ${\vec x}_j''$ (as introduced in Sec.~\ref{sec:conv-integral}) of the $j$-th source surface element, both located in the image plane.

As the relationship between the $P({{\vec x}_0}_i)$ and $\alpha_{{\tt s}j}$ is linear, recovering the latter from the former, that is, deconvolution, is accomplished easily in principle using matrix inversion:
\begin{align}
\alpha_{{\tt s}i}=\frac{1}{P_{\tt dir}}\sum_{j=1}^N\, C^{-1}_{ij}P({{\vec x}_0}_j).
\label{eq:CijDeconv}
\end{align}
In practice, this is not a viable approach given the extreme size of the convolution matrix (e.g., $10^{12}$ elements for a megapixel image) and the resulting computational burden and numerical instabilities. However, this representation of the deconvolution process permits us to study its properties and, in particular, its impact on noise.

We model measurement noise as uniform, uncorrelated Gaussian noise of magnitude $\sigma$. The contribution of noise is introduced in (\ref{eq:CijDeconv}) using root-mean-square addition, where the estimate for $\hat \alpha_{{\tt s}i}$ is obtained as
{}
\begin{align}
\hat \alpha_{{\tt s}i}
=\frac{1}{P_{\tt dir}}\bigg(\sum_{j=1}^N\, C^{-1}_{ij}P({{\vec x}_0}_j)+\Big({\sum_{j=1}^N\,(C^{-1}_{ij})^2}\Big)^\frac{1}{2}~\sigma\bigg)
=\alpha_{{\tt s}i}+\frac{1}{P_{\tt dir}}\Big({\sum_{j=1}^N\,(C^{-1}_{ij})^2}\Big)^\frac{1}{2}~\sigma,
\label{eq:CijDeconv1}
\end{align}
where $\hat \alpha_{{\tt s}i}$ now represents the estimate of the recovered signal in the presence of noise. We need to understand how this deconvolution process treats the signal $P({{\vec x}_0}_i)$ and the noise $\sigma$ differently. Specifically, given the observed SNR (again, as in (\ref{eq:snr-cor}), denoted with the subscript {\tt C} for convolved),
{}
\begin{align}
{\rm SNR}_{\tt C}=\frac{\langle P({{\vec x}_0}_i)\rangle}{\sigma},
\end{align}
we wish to estimate the {\rm SNR} of the recovered signal (denoted using the subscript {\tt R}) after deconvolution:
\begin{align}
{\rm SNR}_{\tt R}=\frac{\langle \alpha_{{\tt s}i}\rangle}{\displaystyle\frac{1}{P_{\tt dir}}\Big({\sum_{j=1}^N\,(C^{-1}_{ij})^2}\Big)^\frac{1}{2}~\sigma}.
\end{align}
To do so, we need to be able to estimate the behavior of the deconvolution matrix $C^{-1}_{ij}$.

\subsection{Approximating the deconvolution matrix to compute the SNR}

To approximate $C_{ij}$ (\ref{eq:Cij}), we first observe that its diagonal elements are identically 1. Its off-diagonal elements are all less than 1. The largest off-diagonal element is determined by the distance $d$ between adjacent area elements yielding the value $1/4$. The rest of the off-diagonal elements of $C_{ij}$ are smaller than this value. This leads us to approximate $C_{ij}$ by the form
\begin{align}
C_{ij}\rightarrow {\widetilde C}_{ij}= \mu\delta_{ij}+\nu U_{ij}, \qquad {\rm with} \qquad \mu=1-\nu,
\label{eq:fffs2s}
\end{align}
where $\nu\ll1$ is a constant, $\delta_{ij}$ is the unit matrix and $U_{ij}$ is the ``everywhere one'' matrix, every element of which is equal to 1. (Note that (\ref{eq:fffs2s}) resembles the structure of (\ref{eq:Cij})). We choose $\nu$ to be
\begin{align}
\nu=\langle C_{ij}\rangle_{i\ne j},
\end{align}
that is to say, $\nu$ is the average value of the off-diagonal elements of $C_{ij}$. We can easily compute $\nu$ for large $N$ by replacing the summation with an integral over the observable image area $A=N\pi({\textstyle\frac{1}{2}}d)^2$  (or $A=Nd^2$ if a square imaging area is used) corresponding to the source coordinates ${\vec x}''_i$ and the corresponding area $A$ for the image coordinates ${{\vec x}_0}_i$. Using the relevant components of the PSF from the matrix form (\ref{eq:Cij}) and that form (\ref{eq:psf-mu*2}), we compute
{}
\begin{eqnarray}
\nu=\frac{1}{N(N-1)}\Big(\sum_{i=1}^N\sum_{j=1}^NC_{ij}-\sum_{i=1}^NC_{ii}\Big)=
\frac{1}{A^2}\iint_{A} d^2{\vec{x}_0}\iint_{A} d^2\vec{x}''\frac{d}{4|{\vec x}_0-{\vec x}''|}\sim \frac{1}{a\sqrt{N}},
\end{eqnarray}
where the value of $a$ depends on the shape of the integration area $A$.  For a circular integration area, $a= 1.18$, while for a square integration area, it is $a= 1.35$.  

The inverse of $\widetilde {C}_{ij}$ from (\ref{eq:fffs2s}) is easily computed:
\begin{align}
\widetilde{C}_{ij}^{-1}=\frac{1}{\mu}\delta_{ij}-\frac{\nu}{\mu(\mu+\nu N)}U_{ij}.
\end{align}
This form allows us to estimate the effect of deconvolution on signal and noise. For this, we assume a uniform signal $P({\vec{x}_0}_i)={\langle P({{\vec x}_0}_i)\rangle}\equiv \langle P\rangle$ in (\ref{eq:CijDeconv1}):
{}
\begin{eqnarray}
\hat \alpha_{{\tt s}i}
=\frac{1}{P_{\tt dir}}\bigg(\sum_{j=1}^N\, C^{-1}_{ij}\langle P\rangle+\Big({\sum_{j=1}^N\,(C^{-1}_{ij})^2}\Big)^\frac{1}{2}~\sigma\bigg),
\label{eq:CijDeconv1*}
\end{eqnarray}
and thus the post-deconvolution ${\rm SNR}_{\tt R}$ is calculated as
\begin{eqnarray}
{\rm SNR}_{\tt R}=\frac{\displaystyle\frac{1}{N}\sum_{i=1}^N\sum_{j=1}^N\, C^{-1}_{ij}}{\Big({\displaystyle\frac{1}{N}\sum_{i=1}^N\sum_{j=1}^N\,(C^{-1}_{ij})^2}\Big)^\frac{1}{2}~}\frac{\langle P\rangle}{\sigma}.
\end{eqnarray}
Replacing $C_{ij}^{-1}$ with $\widetilde{C}_{ij}^{-1}$, we estimate the deconvolution penalty in the limit of large $N$:
\begin{eqnarray}
\frac{{\rm SNR}_{\tt R}}{{\rm SNR}_{\tt C}}=
\frac{\displaystyle\frac{1}{N}\sum_{i=1}^N\sum_{j=1}^N\, \widetilde{C}^{-1}_{ij}}{\Big({\displaystyle\frac{1}{N}\sum_{i=1}^N\sum_{j=1}^N\,(\widetilde{C}^{-1}_{ij})^2}\Big)^\frac{1}{2}~}
= \frac{\mu}{\nu N}\sim\frac{a}{\sqrt N}.
\label{eq:penalty}
\end{eqnarray}

This deconvolution penalty arises unavoidably, as a consequence of how the deconvolution process affects signal versus noise. However, the estimate (\ref{eq:penalty}) with either $a=1.18$ or $a=1.35$ is rather conservative. Our numerical simulations confirm that even a simple filter in the frequency domain, introduced as part of the deconvolution algorithm, especially when applied to realistic planetary images, can improve the result such that $a={\cal O}(10)$ or better. Further improvements are expected with the use of advanced spatial filtering and deblurring techniques. These are currently being investigated and results, when available, will be reported. For now, we treat $a=10$ as a conservative estimate and use it in the next section to evaluate realistic SNRs and corresponding integration times.

\subsection{Towards realistic imaging of exoplanets}
\label{sec:SNR-estim}

To assess the value of the estimates obtained in the processing section, we need to consider them in the context of realistic imaging scenarios.

We take (\ref{eq:blur-s}) to represent the estimate of the total convolved signal received from a uniformly illuminated source and measured at a particular location in the image plane, namely $\left<Q_i\right>=Q_{\tt fp.blur}(\rho_0).$ Accounting for the fact that photons obey Poisson statistics, we estimate the variance of the signal as being $\sigma(Q_i)=\sqrt{Q_{\tt fp.blur}(\rho_0)}$, resulting in the {\rm SNR} of the convolved image as ${\rm SNR}^0_{\tt C}=\left<Q_i\right>/\sigma(Q_i)=\sqrt{Q_{\tt fp.blur}(\rho_0)}.$ Using this result in (\ref{eq:penalty}) with $a=10$, we obtain the SNR of the deconvolved signal:
{}
\begin{eqnarray}
{\rm SNR}_{\tt R}\geq
\frac{10}
{\sqrt{N}}\sqrt{Q_{\tt fp.blur}(\rho_0)}\sqrt{\frac{t}{1\,{\rm s}}}.  \label{eq:snr*o*}
\end{eqnarray}
Given the desired ${\rm SNR}_{\tt R}$, equation (\ref{eq:snr*o*}) allows us to estimate the per-pixel integration time, $t_{\tt pix}$:
 {}
\begin{eqnarray}
t_{\tt pix}\leq 10^{-2}
N \frac{{\rm SNR}^2_{\tt R}}{Q_{\tt fp.blur}}&=&1.25\times 10^{-7} \,
N \, {\rm SNR}^2_{\tt R}\, \Big(\frac{1\,{\rm m}}{d}\Big)^2\Big(\frac{\overline z}{650\,{\rm AU}}\Big)^\frac{1}{2}\Big(\frac{30\,{\rm pc}}{z_0}\Big)\Big(\frac{1\,\mu{\rm m}}{\lambda}\Big)~{\rm s}.
  \label{eq:tin0_pix}
\end{eqnarray}
Therefore, from (\ref{eq:tin0_pix}) we determine that in the signal dominated regime it takes $\sim11\,{\rm s}$ of integration time to reach ${\rm SNR}_{\tt R} = 7$. With $t_{\tt tot}= t_{\tt pix} N$ to be the total integration time needed collect data for the entire $N$-pixel image, using (\ref{eq:tin0_pix}) we see that to recover a high-resolution image with $N=1024\times 1024$ pixels, we need $\sim 4.5$ months of integration time. A 2-m telescope would compete this task in less than 50 days.

The short integration times resulting from (\ref{eq:snr*o*}) are possible for bright exoplanets or other luminous objects, where the solar corona contribution in not a significant part of the overall noise budget. However, as we discussed in Sec.~\ref{sec:sol-cor}, the brightness of the solar corona affects the performance of the SGL in a significant way.  Thus, in the presence of the solar corona, an estimate similar to (\ref{eq:snr*o*}) may be obtained directly from the SNR for the signal in the presence of the solar corona ${\rm SNR}_{\tt C}$ given by (\ref{eq:snr-cor}). Using this result in (\ref{eq:penalty}) we obtain an estimate for the SNR of the deconvolved image in the presence of the solar corona as
 {}
\begin{eqnarray}
{\rm SNR}_{\tt R}\geq
\frac{10}{\sqrt{N}}\frac{Q_{\tt fp.blur}}{\sqrt{Q_{\tt fp.cor}}}\sqrt{\frac{t}{1\,{\rm s}}}.  \label{eq:snr*rec_cor*}
\end{eqnarray}

This expression yields the following per-pixel integration time, $t_{\tt pix}$, in the presence of the solar corona noise:
 {}
\begin{eqnarray}
t_{\tt pix}&\leq& 10^{-2} N
\frac{{Q_{\tt fp.cor}}{\rm SNR}^2_{\tt R}}{Q^2_{\tt fp.blur}}=\nonumber\\
&=& 3.54\times 10^{-3} \,N\, {\rm SNR}^2_{\tt R}\Big(1+0.79 \Big(\dfrac{650\,{\rm AU}}{\overline z}\Big)^{5.1}
+0.05\Big(\dfrac{\overline z}{650\,{\rm AU}}\Big)^{2.65}\Big)
\Big(\frac{1\,{\rm m}}{d}\Big)^3\Big(\frac{z_0}{30\,{\rm pc}}\Big)^2\Big(\frac{650\,{\rm AU}}{\overline z}\Big)^{3.4}~{\rm s}.~~~~~
  \label{eq:tin_cor_pix}
\end{eqnarray}

Result (\ref{eq:tin_cor_pix}) suggests that for $d=1$~m it could take up to $\sim 3\times 10^3$ sec of  integration time per pixel to reach the ${\rm SNR}_{\tt R}=7$ for an image of $N=100\times 100=10^4$ pixels. For ${\overline z}=650\,{\rm AU}$, this translates in a $t_{\tt tot}= t_{\tt pix}N\sim 1$ year of total integration time needed to recover the entire $100\times 100$ pixel image of an exoplanet at 30 pc. Using for this purpose a larger telescope, say $d=2\,{\rm m}$,  the per-pixel integration time drops to  $390$ sec, reducing the integration time required to recover an image with the same number of pixels to $\lesssim 1.5$ months  of integration time. Use of a 5~m telescope implies a per-pixel integration time of $\sim 150$~s on the a $250\times 250$ pixel image, for a total integration time of $\sim 110$ days. Collecting more, redundant data will allow us to account for the diurnal rotation of the exoplanet and its variable cloud cover. To compensate for the diurnal rotation, we may also benefit from a multitelescope architecture that can reduce the total integration time \cite{Turyshev-etal:2018}, while matching the temporal behavior of the target. However, if the direct spectroscopy of an exoplanet atmosphere is the main mission objective, this can be achieved with a single spacecraft.
We emphasize that direct imaging and spectroscopy of an exoplanet at such resolutions are impossible using any of the conventional astronomical instruments, either telescopes or interferometers; the SGL is the only means to obtain such results.

\subsection{Image reconstruction in the presence of noise}

Our estimate for the SNR deconvolution penalty (\ref{eq:penalty})
can be directly compared against simulated exoplanet image reconstruction at various levels of noise. Since the PSF of the SGL is known, convolution and deconvolution of a simulated image is a relatively straightforward process \cite{Toth-Turyshev:2020-deconv}.

\begin{figure}[t]
\includegraphics[width=0.3\linewidth]{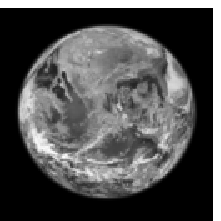}~\includegraphics[width=0.3\linewidth]{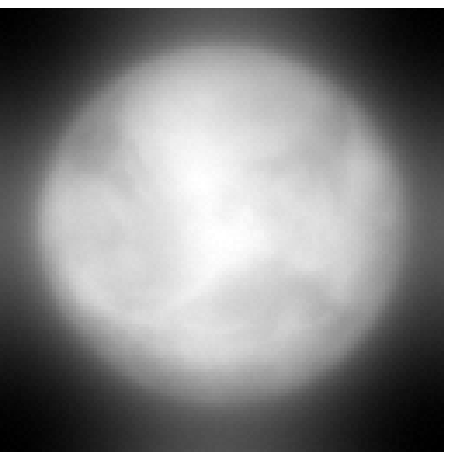}~\includegraphics[width=0.3\linewidth]{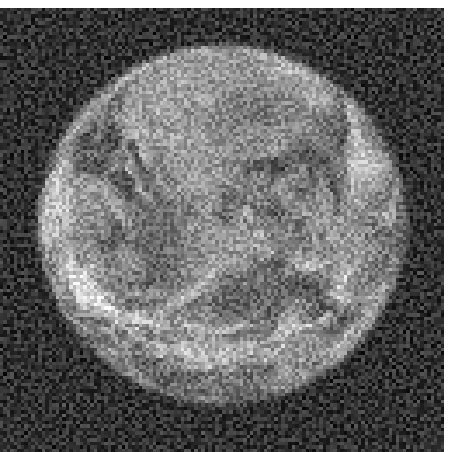}
\vskip 1pt
\includegraphics[width=0.3\linewidth]{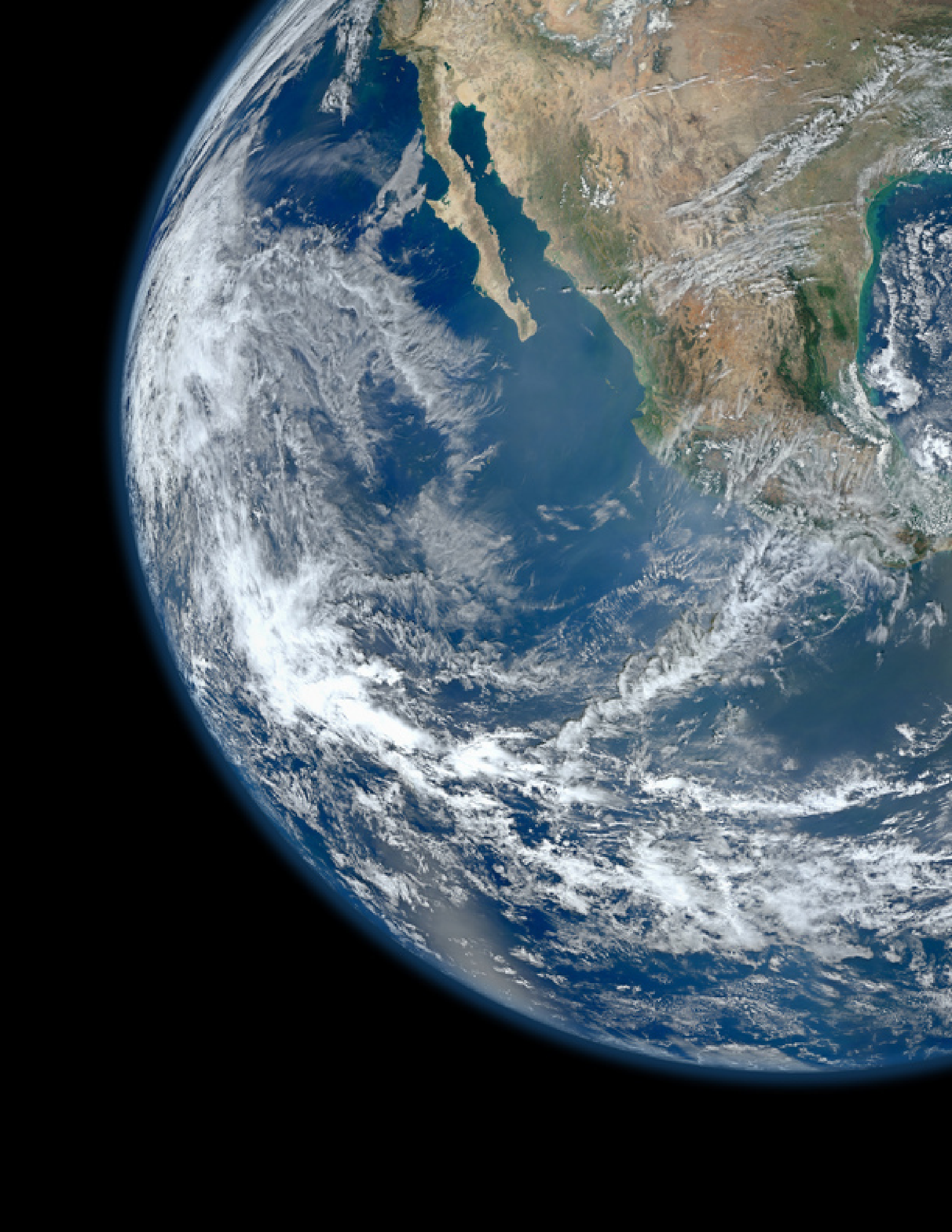}~\includegraphics[width=0.3\linewidth]{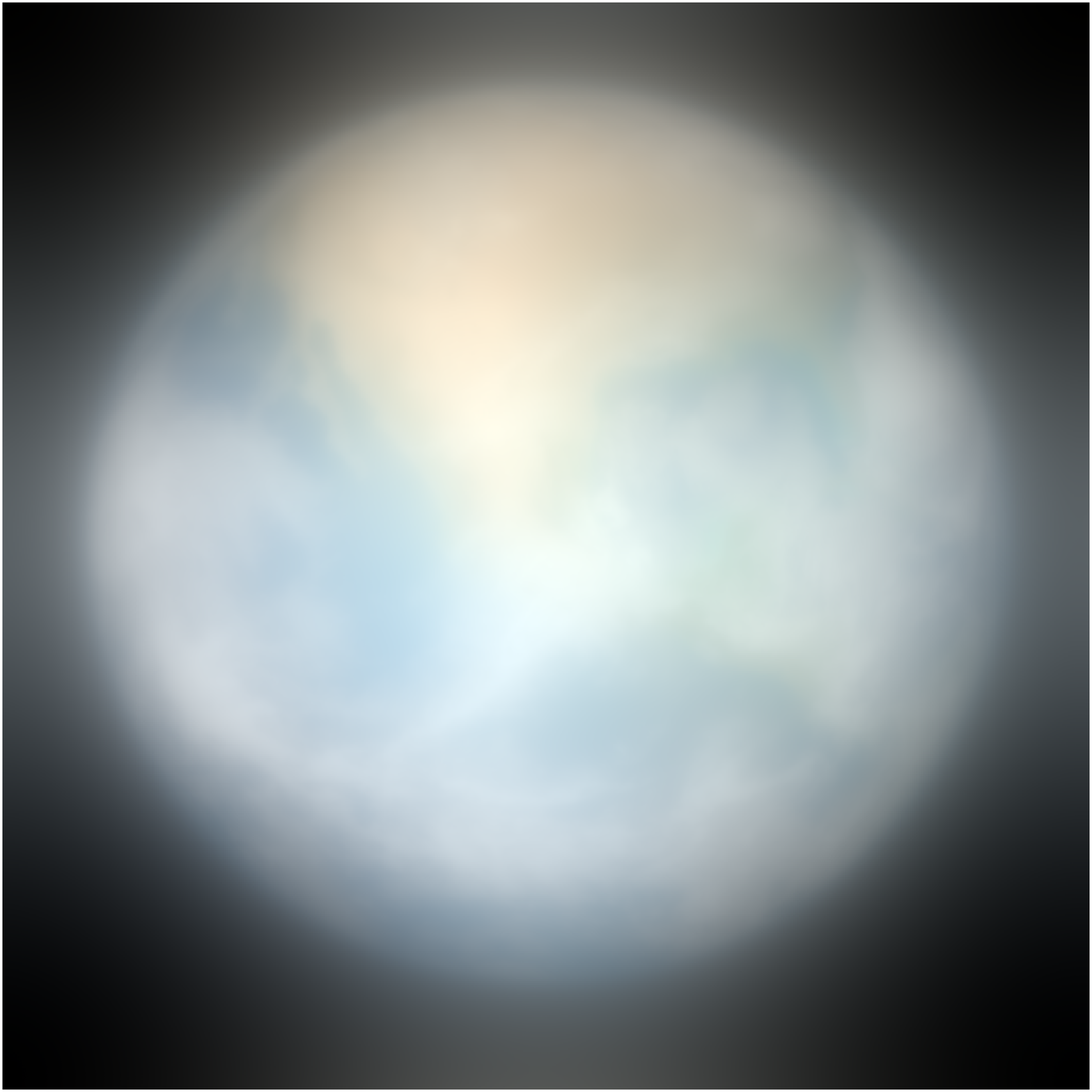}~\includegraphics[width=0.3\linewidth]{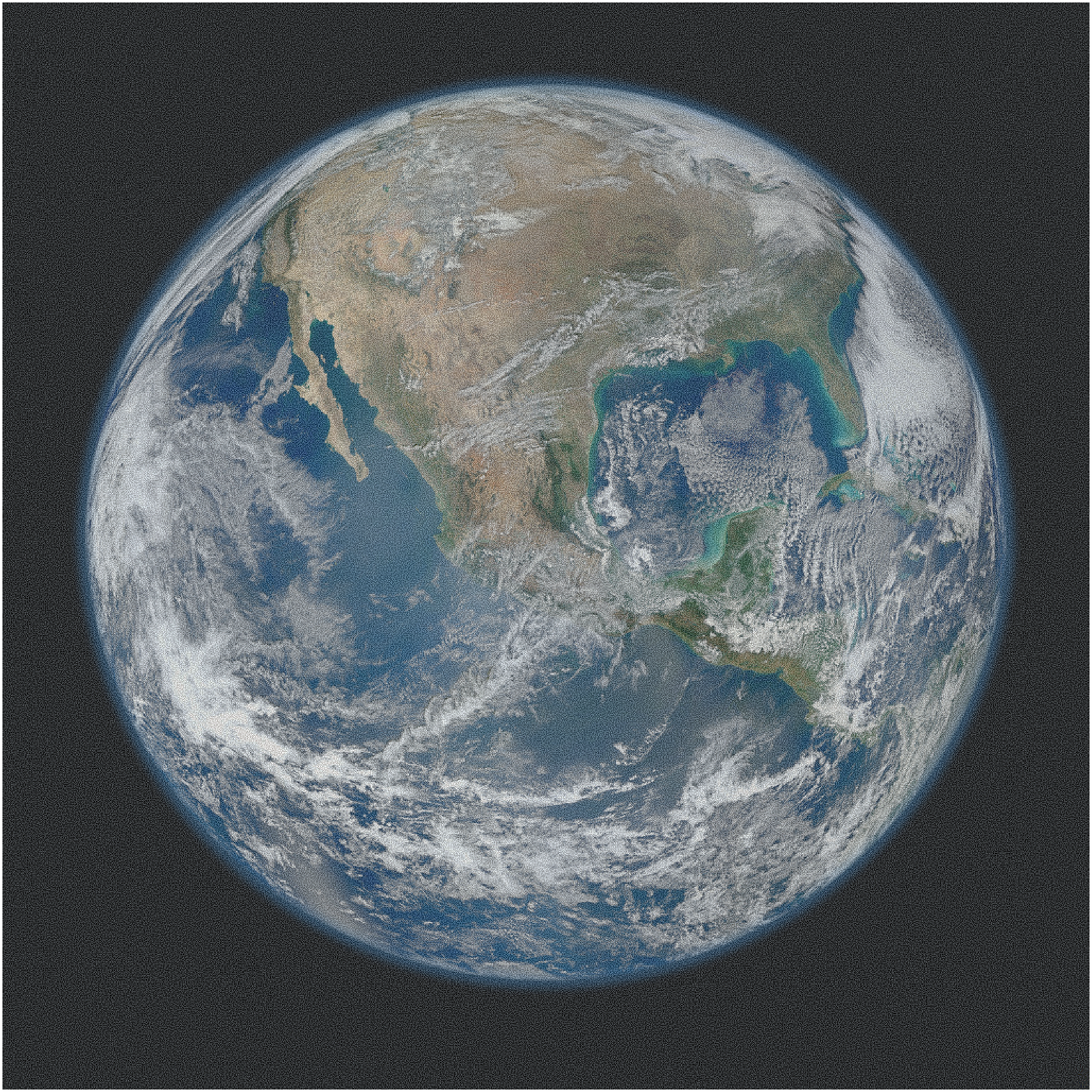}\\
\caption{\label{fig:earth}A simulation of the effects of the monopole solar gravitational lens on an Earth-like exoplanet image.
Top row, left: a monochrome image, sampled at 128$\times$128 pixels; center: blurred image; right: deconvolution at ${\rm SNR}\sim 4.5$. From \cite{Toth-Turyshev:2020-deconv}.
Bottom row, left: original RGB color image with a 1024$\times$1024 pixel resolution; center: image blurred by the SGL; right: the result of image deconvolution at an SNR of $\sim$5.2 per color channel, or combined SNR of $\sim$9.
}
\end{figure}

In Fig.~\ref{fig:earth}, we show the results of a simulated convolution of an Earth-like exoplanet image with the SGL PSF and subsequent deconvolution. The top row depicts the result of deconvolution of a monochrome image of an exo-Earth, using modest image resolution ($128\times 128$ image pixels), reconstructed with an ${\rm SNR}\sim 4.5$ after deconvolution. According to Eq.~(\ref{eq:tin_cor_pix}), an image of this quality may be achievable in $\sim 1.1$ years of cumulative integration time even for a source at a distance of 30~pc, using only a single $d=1$~m telescope, situated at 650~AU from the Sun.

Clearly, the SNR and the resulting image quality can be much improved by using a larger telescope, conducting an observational campaign at a greater distance from the Sun, and of course, using multiple instruments. A much more ambitious image reconstruction is depicted in the bottom row of Fig.~\ref{fig:earth}: a high-resolution (megapixel) RGB-color image of an exo-Earth, reconstructed at ${\rm SNR}\sim 5.2$ per color channel, for a combined ${\rm SNR}\sim 9$ for the color image. Even this image quality is within the realm of the feasible if we consider a target at $z_0=3$~pc, observed through the SGL using $d=2.5$~m telescopes at 1000~AU from the Sun. The cumulative integration time needed to obtain this image is less than 8 years with a single instrument.

These estimates demonstrate that utilizing the SGL to obtain a good quality resolved image of an exoplanet of interest within 30~pc from the Earth is firmly within the realm of the possible.

\section{Discussion and Conclusions}
\label{sec:disc}

We investigated the image formation process with the SGL. For that, we analyzed the EM field originating from an extended, resolved source and received in the focal plane of an imaging telescope, represented by a thin convex lens.

The complex amplitude of the EM signal in the telescope's focal plane can be modeled by splitting the signal into two parts: light from the directly imaged region (the spot on the distant source that geometrically corresponds to the imaging telescope's aperture) and the blur signal that is received by the telescope from the rest of the source. Assuming uniform surface brightness within the directly imaged spot, (\ref{eq:pow-dirD+}) describes the image of an Einstein ring in the imaging telescope's focal plane, as expected. The expression for blur (\ref{eq:pow-blur}) is given in integral form and cannot be evaluated analytically in the general case, when the surface brightness of the imaged source is nonuniform and an arbitrary function of the source plane coordinates. We have, however, endeavored to evaluate this integral in the special case when the source is a disk of uniform surface brightness. Being able to estimate the magnitude of the blur in this case in the form of expression (\ref{eq:P-blur*2}) provides useful limits when evaluating the magnitude of the signal and the anticipated SNR of measurements to be performed with the SGL.

Far away from the SGL's optical axis, in the region of weak interference, we recovered an expression that, as expected, corresponds to two spots of light of uneven brightness that are seen by the imaging telescope: one outside, one inside the nominal radius of the Einstein ring (which are know as the major and minor images, correspondingly; see \cite{Schneider-Ehlers-Falco:1992}). These correspond to the incident and scattered wavefronts, respectively, that are produced by the SGL. In the geometric optics region, the spot corresponding to the scattered wavefront (i.e., the minor image) vanishes, as this light is blocked by the opaque spherical Sun.

The results in this paper extend those obtained in \cite{Turyshev-Toth:2019-image} where a similar analysis was performed for the case of imaging of point sources.   The new results extend our understanding of the image formation process to the case of extended, resolved sources positioned at large, but finite distances from the Sun. In addition, these results are also in good agreement with those reported in \cite{Turyshev-Toth:2019-blur} for the case of photometric imaging where the goal is to measure the total power received by a telescope as it is positioned at various locations in the SGL image plane (i.e., the ``light bucket'' approach). Here we extended those results all the way to the focal plane of an optical telescope.

An azimuthally resolved picture of the Einstein ring due to an extended source opens new possibilities. If the surface brightness of the source is not uniform, this can produce variations in brightness along the Einstein ring (as described by (\ref{eq:pow-dirD}) and (\ref{eq:P-blur*2*})). This information on the azimuthally varying Einstein ring's brightness may help improve the effectiveness of image deconvolution. Similarly, light contamination due to nearby off-image sources (e.g., the parent star of an exoplanet being imaged) can contribute to the Einstein ring at specific spots (the case, that is captured by (\ref{eq:P-blur*off4*})). In these cases, it makes sense to collect light not from the entirety of the Einstein ring but only from specific sections that are less affected by contamination (Fig.~\ref{fig:images}). Similarly, light not coming from the immediate vicinity of the Einstein ring can be largely ignored by appropriate sampling the Einstein ring in the telescope focal plane.

We were also able to investigate the most significant source of noise, the solar corona. We have shown that it is possible to obtain a detailed image of a distant exoplanet with integration times consistent with a realistic SGL mission even in the presence of this noise. We developed a semianalytical model of the deconvolution process in order to understand the impact of deconvolution on noise. We showed that deconvolution amplifies measurement noise, thus reducing sensitivity. Nevertheless, even for very distant exoplanets located up to 30~pc from us, a telescope located in the strong interference region of the SGL can obtain multipixel images with the realistic mission lifetimes. We also note that with the use of multiple spacecraft, integration times can be significantly reduced, allowing investigations even in the presence of temporal variability of the target due to diurnal rotation or changing surface features (e.g., varying cloud cover).  At the same time, even a single spacecraft may be sufficient to obtain spectroscopic data that can be used to confirm the presence of active organic processes on that exoplanet.

The analytical tools developed here may be used to evaluate the anticipated signal levels from various targets of interest and sources of local light contamination, as well as compare these signals against background noise. These results are important for the design of future imaging missions to the focal region of the SGL, as they provide important insight into the various factors that may affect the performance of these projects.

The properties of the exoplanet (size, distance, albedo, parent star brightness, etc.), telescope parameters (aperture size, optical throughput, etc.), coronagraph parameters (annular vs. disk, contrast ratio, etc.), increasing heliocentric distance (as the spacecraft travels along the optical axis),  use of multiple telescopes, spectral filtering and other factors may improve the SNR estimates. However, already at this level, the analysis that we presented demonstrates that utilizing the SGL for the purposes of resolved imaging of distant exoplanets is feasible, providing unique capabilities not available through other means. As such, the SGL should be further investigated to determine its most optical practical applications. This work is ongoing and results, when available, will be reported elsewhere.

\begin{acknowledgments}
This work in part was performed at the Jet Propulsion Laboratory, California Institute of Technology, under a contract with the National Aeronautics and Space Administration.
VTT acknowledges the generous support of Plamen Vasilev and other Patreon patrons.

\end{acknowledgments}

\appendix

\section{Modeling the solar corona  signal in the focal plane}
\label{sec:model}

To develop reliable sensitivity estimates for imaging with the SGL, we need to consider the solar corona, which is the largest source of photometric noise \cite{Turyshev-Toth:2019}. For that, we model the solar corona as a 2-dimensional surface containing a collection of point emitters. Each point ${\vec x}'$ emits a spherical wave, the behavior of which is determined by $\propto e^{i(kr-\omega t)}/r$, where $r$ is the distance from a point with heliocentric coordinates $({\vec x'}, 0)$ in the corona plane to a point $({\vec x}_0+{\vec x},{\overline z})$ in the image plane: $r=\sqrt{{\overline z}^2+({\vec x}+{\vec x}_0-{\vec x}')^2}$. In the case of imaging with the SGL, the characteristic behavior of ${\vec x}_0$ is given as $|{\vec x}_0|=\rho_0\simeq r_\oplus=R_\oplus {\overline z}/z_0=1.3 \, ({\overline z}/650\,{\rm AU})(30\,{\rm pc}/z_0)$\,km. Also, accounting for the solar coronagraph \cite{Turyshev-etal:2018}, the distance $|{\vec x}'|$ is rather large, being  $|{\vec x}'|=\rho'\geq R_\odot$. With these assumptions and keeping only the linear terms, the distance $r$ may be expanded as
$r\simeq{\overline z}-({\vec x}\cdot{\vec x}')/{\overline z}+{\cal O}\big(({\rho^2},{\rho\rho_0},{\rho'^2})/{{\overline z}^2}\big)$, yielding the factor $\propto e^{i(k{\overline z}-k({\vec x}\cdot{\vec x}')/{\overline z}-\omega t)}/{\overline z}$.

Using these assumptions, we consider a spherical EM wave propagating from a point source in the corona plane towards the image plane. In the paraxial approximation, in a cylindrical coordinate system $(\rho,\phi,z)$, this wave may be given as
{}
\begin{eqnarray}
    \left( \begin{aligned}
{E}_\rho& \\
{H}_\rho& \\
  \end{aligned} \right) =    \left( \begin{aligned}
{H}_\phi& \\
-{E}_\phi& \\
  \end{aligned} \right)&=&
  \frac{E_0}{{\overline z}}  e^{i(k{\overline z}-\omega t)}
    \exp\big[-ik\frac{({\vec x}\cdot{\vec x}')}{\overline z}\big]
 \left( \begin{aligned}
 \cos\phi& \\
 \sin\phi& \\
  \end{aligned} \right),
  \qquad     \left( \begin{aligned}
{E}_z& \\
{H}_z& \\
  \end{aligned} \right) =    0.
  \label{eq:DB-sol-rho-s}
\end{eqnarray}
From this expression, similarly to (\ref{eq:amp-w}), we identify the complex amplitude of the EM wave just in front of the telescope aperture, which now is given only by the phase factor that is essentially independent on ${\vec x}_0$. This amplitude allows us to present (\ref{eq:amp-w-f0}) as the amplitude of the EM wave in the focal plane of the optical telescope:
{}
\begin{eqnarray}
{\cal A}_{\tt cor}({\vec x}_i,{\vec x}')&=&
-  \frac{e^{ikf(1+{{\vec x}_i^2}/{2f^2})}}{i\lambda f}\iint\displaylimits_{|{\vec x}|^2\leq (\frac{1}{2}d)^2} d^2{\vec x}\,
 e^{-i\frac{k}{\overline z}({\vec x}\cdot{\vec x}')} e^{-i\frac{k}{f}({\vec x}\cdot{\vec x}_i)}.
  \label{eq:amp-w-f+}
\end{eqnarray}
With this amplitude, similarly to (\ref{eq:amp-w-f})--(\ref{eq:DB-sol-rho2}), the EM field in the focal plane of the telescope is given as
{}
\begin{eqnarray}
    \left( \begin{aligned}
{E}_\rho& \\
{H}_\rho& \\
  \end{aligned} \right)_{\tt \hskip -3pt {\vec x}_i} =    \left( \begin{aligned}
{H}_\phi& \\
-{E}_\phi& \\
  \end{aligned} \right)_{\tt \hskip -3pt \vec x_i} &=&\frac{{E}_0({\vec x}')}{\overline z}
  {\cal A}_{\tt cor}({\vec x}_i,  {\vec x}')
    e^{i(k{\overline z}-\omega t)}
 \left( \begin{aligned}
 \cos\phi& \\
 \sin\phi& \\
  \end{aligned} \right).
  \label{eq:DB-sol-rho2*}
\end{eqnarray}
The phase of the integral in (\ref{eq:amp-w-f+}) may be expressed as
$\varphi({\vec x})=-{k}\big(({\vec x}\cdot{\vec x}')/{\overline z} +({\vec x}\cdot{\vec x}_i)/f\big)=-u\rho\cos\big(\phi-\epsilon\big)+{\cal O}(\rho^2),$  where use used (\ref{eq:x'})--(\ref{eq:alpha-mu}),  introduced the corona spatial frequency,
$\alpha_c=k{\rho'}/{\overline z},$
and defined $u$ as
{}
\begin{eqnarray}
u=\sqrt{\alpha_c^2+2\alpha_c\eta_i\cos\big(\phi'-\phi_i\big)+\eta_i^2},
\qquad
\cos\epsilon=u^{-1}\big({\alpha_c  \cos\phi'+\eta_i\cos\phi_i}\big),
\qquad
\sin\epsilon=u^{-1}\big({\alpha_c  \sin\phi'+\eta_i\sin\phi_i}\big).~~~
  \label{eq:upmA}
\end{eqnarray}

With these definitions, the integral in (\ref{eq:amp-w-f+}) can be easily evaluated, yielding
{ }
\begin{eqnarray}
{\cal A}_{\tt cor}({\vec x}_i,{\vec x}')&=&
ie^{ikf(1+{{\vec x}_i^2}/{2f^2})}
 \Big(\frac{kd^2}{8f}\Big) \Big( \frac{2
J_1\big(u\frac{1}{2}d\big)}{u\frac{1}{2}d}\Big).
  \label{eq:amp-blur3+}
\end{eqnarray}

\begin{figure}
\includegraphics[width=0.40\linewidth]{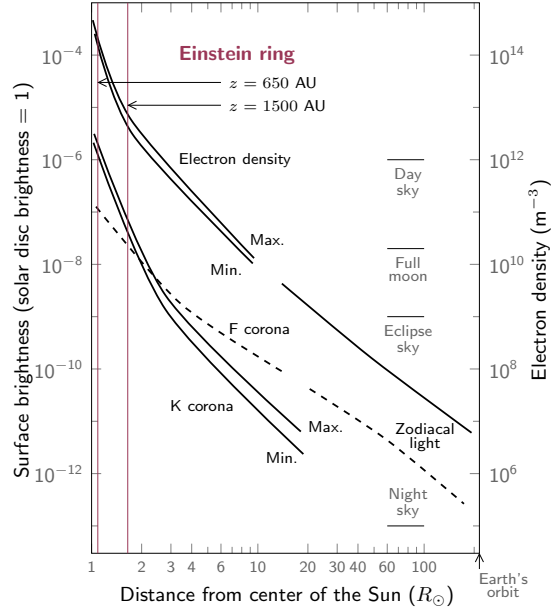}
\caption{\label{fig:sol-cor-bright2}  Solar corona brightness from \cite{Lang-ebook:2010}. As heliocentric distances increase, the Einstein ring further separates from the Sun. Positions of the Einstein ring for $z=600$~AU and $z=1,500$ AU are shown, both corresponding to distances from the center of the Sun of $\leq 2R_\odot$. For such solar separations, the K corona dominates.
}
\end{figure}

We may now compute the Poynting vector for this EM wave.  For this, we substitute (\ref{eq:amp-blur3+}) into (\ref{eq:Pv}) and (\ref{eq:psf}) to recover the conventional PSF of a regular optical telescope \cite{Born-Wolf:1999,Goodman:2017}, which we use to determine the intensity distribution of the corona signal received in the focal plane of the optical telescope:
  {}
\begin{eqnarray}
I_{\tt cor}({\vec x}_i) =\frac{1}{ {\overline z}^2}
\iint d^2{\vec x}'  B_{\tt cor}({\vec x}')  \mu_{\tt cor}({\vec x}_i,
{\vec x}')=\frac{1}{ {\overline z}^2}\Big(\frac{kd^2}{8f}\Big)^2
\iint d^2{\vec x}'  B_{\tt cor}({\vec x}')  \Big( \frac{2
J_1\big(u\frac{1}{2}d\big)}{u\frac{1}{2}d}\Big)^2,
  \label{eq:pow-cor}
\end{eqnarray}
where $B_{\tt cor}\simeq {E}_{\tt cor}^2$ is the surface brightness of the solar corona. We use a recent model for the solar corona \cite{November:1996}, which is slightly more conservative (predicting a slightly higher photon flux) in the region of the corona that is of interest to us, in comparison to the widely used Baumbach model \cite{Baumbach:1937,vandeHulst:1947,vandeHulst:1950,Golub-Pasachoff-book:2017}:
{}
\begin{equation}
B_{\tt cor}(\rho)= 20.09\Big[3.670 \Big(\frac{R_\odot}{\rho}\Big)^{18}+1.989\Big(\frac{R_\odot}{\rho}\Big)^{7.8}+ 5.51\times 10^{-2} \Big(\frac{R_\odot}{\rho}\Big)^{2.5}\Big]  ~~   \frac{\rm W}{{\rm m}^2\,{\rm sr}}.
\label{eq:model-cor}
\end{equation}
This surface brightness distribution strictly applies only to the K-corona, which dominates the brightness within the heliocentric ranges $\rho\in[R_\odot,2R_\odot]$ (see \cite{Lang-ebook:2010} and Fig.~\ref{fig:sol-cor-bright2}).

A coronagraph can be used to block sunlight everywhere, except for the annulus surrounding the Einstein ring with thickness of $\lambda/d$. Therefore, the useful signal will be received from the annulus within the two radii $\rho_{\tt cor}^\pm$, which correspond to the angles $\theta_{\tt cor}^\pm$, given as
{}
\begin{eqnarray}
\rho_{\tt cor}^\pm={\overline z} \Big(\sqrt{\frac{2r_g}{\overline z}}\pm\frac{\lambda}{2d}\Big), \qquad
\theta_{\tt cor}^\pm=\frac{\rho_{\tt cor}^\pm}{{\overline z}}=
\sqrt{\frac{2r_g}{\overline z}}\pm\frac{\lambda}{2d}.
  \label{eq:cor-rho}
\end{eqnarray}

As a result, the intensity distribution in the focal plane of the imaging telescope (\ref{eq:pow-blur}) takes the form
  {}
\begin{eqnarray}
I_{\tt cor}({\vec x}_i) =\frac{1}{{\overline z}^2}\Big(\frac{kd^2}{8f}\Big)^2
\int_0^{2\pi}\hskip -4pt d\phi'\int_{\rho_{\tt cor}^-}^{\rho_{\tt cor}^+}\hskip -4pt \rho' d\rho'  \, B_{\tt cor}(\rho')  \Big( \frac{2
J_1\big(u\frac{1}{2}d\big)}{u\frac{1}{2}d}\Big)^2.
  \label{eq:pow-cor*}
\end{eqnarray}

To compute the corresponding power deposited by the corona in the focal plane, $P_{\tt cor}$, we recognize that the Einstein ring in the focal plane is an unresolved circle with radius determined from (\ref{eq:alpha-mu}) as $\alpha=\eta_i$, yielding $\rho_{\tt ER}=f\sqrt{2r_g/{\overline z}}$. Therefore, the useful signal  received in the focal plane of a diffraction-limited telescope occupies the annulus between the radii $\rho_{\tt ER}^\pm$ (\ref{eq:ER-rho}).  Therefore, we take  (\ref{eq:pow-cor*}) and integrate it over the area seen by the diffraction-limited telescope:
{}
\begin{eqnarray}
P_{\tt fp.cor}&=&\int^{2\pi}_0 \hskip -4pt  d\phi_i \int_{\rho^-_{\tt ER}}^{\rho_{\tt ER}^+}
\hskip 0pt I_{\tt cor}({\vec x}_i) \rho_i d\rho_i = \frac{1}{ {\overline z}^2}\Big(\frac{kd^2}{8f}\Big)^2
\int_0^{2\pi}\hskip -4pt d\phi'\int_{\rho_{\tt cor}^-}^{\rho_{\tt cor}^+}\hskip -4pt \rho' d\rho'  \, B_{\tt cor}(\rho')
\int^{2\pi}_0 \hskip -4pt  d\phi_i \int_{\rho^-_{\tt ER}}^{\rho_{\tt ER}^+} \hskip -4pt
\rho_i d\rho_i
\Big( \frac{2
J_1\big(u\frac{1}{2}d\big)}{u\frac{1}{2}d}\Big)^2.~~~~~~~~
  \label{eq:pow-fp-cor*7}
\end{eqnarray}
Thus, to determine $P_{\tt fp.cor}$ we need to evaluate the two double integrals, which can be done numerically.  However, for estimation purposes, we may  simplify this expression. Considering the parameters involved in the imaging with the SGL, we may  present this expression (\ref{eq:pow-fp-cor*7}) as
{}
\begin{eqnarray}
P_{\tt fp.cor}&=&\epsilon_{\tt cor}P_{\tt cor},
  \label{eq:pow-frac-blA}
\end{eqnarray}
where $P_{\tt cor}$ is the  total energy deposited in the focal plane of the optical telescope given as
{}
\begin{eqnarray}
P_{\tt cor}&=&\int^{2\pi}_0 \hskip -4pt  d\phi_i \int_0^\infty
\hskip 0pt I_{\tt cor}({\vec x}_i)\rho_i d\rho_i =
\Big(\frac{d}{4{\overline z}^2}\Big)^2
\int_0^{2\pi}\hskip -4pt d\phi'\int_{\rho_{\tt cor}^-}^{\rho_{\tt cor}^+}\hskip -4pt \rho' d\rho'  \, B_{\tt cor}(\rho'),
  \label{eq:pow-fp0*A}
\end{eqnarray}
where we used the variable $p_i$ from (\ref{eq:p-eta}) and also the relationship (\ref{eq:int-fp0*}).

The quantity  $\epsilon_{\tt cor}$ introduced in (\ref{eq:pow-frac-blA}) is the encircled energy factor defined as $\epsilon_{\tt cor}={P_{\tt fp.cor}}/{P_{\tt cor}}$, yielding
{}
\begin{eqnarray}
\epsilon_{\tt cor}&=&
\frac{1}{4\pi}
\int_0^{2\pi}\hskip -6pt d\phi'\int_{\rho_{\tt cor}^-}^{\rho_{\tt cor}^+}\hskip -4pt \rho' d\rho'  \, B_{\tt cor}(\rho')
\int^{2\pi}_0 \hskip -6pt  d\phi_i \int_{p^-_{\tt ER}}^{p_{\tt ER}^+} \hskip -4pt
p_i dp_i
\Big( \frac{2
J_1\big(u\frac{1}{2}d\big)}{u\frac{1}{2}d}\Big)^2\Big/
\int_0^{2\pi}\hskip -6pt d\phi'\int_{\rho_{\tt cor}^-}^{\rho_{\tt cor}^+}\hskip -4pt \rho' d\rho'  \, B_{\tt cor}(\rho').~~~~~~
  \label{eq:een0*}
\end{eqnarray}

As the argument of $J_1$ here is a function of ${\vec x}'$, $u {\textstyle\frac{1}{2}}d=\big[(k\frac{\rho'}{\overline z} {\textstyle\frac{1}{2}}d)^2+ 2k\frac{\rho'}{\overline z} {\textstyle\frac{1}{2}}d \,p_i \cos(\phi_i-\phi')+p_i^2\big]^\frac{1}{2}$, in general, $\epsilon_{\tt cor}$, requires evaluation of  two double integrals in (\ref{eq:een0*}). In our case $\rho'$  is rather large, $\rho'\gtrsim \sqrt{2r_g\overline z}$, varying within narrow integration limits (\ref{eq:cor-rho}), corresponding to a coronagraph that blocks out not just the solar disk but also parts of the solar corona. Therefore, $\alpha_c {\textstyle\frac{1}{2}}d$ is also constrained to behave as $(\alpha_c {\textstyle\frac{1}{2}}d)=(k\frac{\rho'}{\overline z} {\textstyle\frac{1}{2}}d)\simeq
\alpha {\textstyle\frac{1}{2}}d \pm {\textstyle\frac{\pi}{2}},$
where $\alpha$ is from (\ref{eq:alpha-mu}).  Taking the mean value yields $u{\textstyle\frac{1}{2}}d\simeq \big((\alpha {\textstyle\frac{1}{2}}d)^2+2 \alpha {\textstyle\frac{1}{2}}d \,p_i \cos(\phi_i-\phi')+p_i^2\big)^\frac{1}{2}=u_+{\textstyle\frac{1}{2}}d. $ Consequently, the expression for $u$ is now independent of $\rho'$ and the two double integrals may be evaluated separately, allowing us to integrate the numerator of (\ref{eq:een0*}) over $d\phi_i$.  Numerical evaluation of the remaining terms (similarly to (\ref{eq:een*b}) and (\ref{eq:ee-go}), yields the value $\epsilon_{\tt cor}\simeq 0.69$. This is comparable to the value of $\epsilon_{\tt cor}\simeq 0.60$ obtained by direct numerical integration of (\ref{eq:een0*}). In addition, we can also evaluate (\ref{eq:een0*}) numerically by letting $\rho^+_{\tt cor}\to\infty$, representing a coronagraph that blocks only the solar disk; the result is $\epsilon_{\tt cor}=0.36$. These two coronagraph designs will differ in engineering complexity, but it is clear that the annular coronagraph will block more corona light, thus it is preferred for imaging with the SGL.

\begin{figure}
\includegraphics[width=0.40\linewidth]{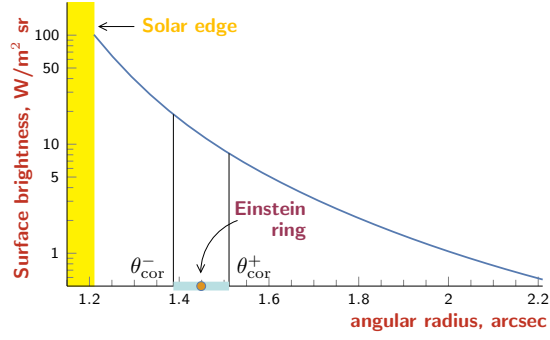}
\caption{\label{fig:sol-cor-bright3}  Typical surface brightness of the solar corona (from \cite{November:1996}) as seen at 800 AU, as given by (\ref{eq:model-th}).
}
\end{figure}

As a result, the power received from the solar corona within the annulus surrounding the Einstein ring around the Sun formed by the light from an exoplanet  and measured at the region occupied by the image of that Einstein ring in the focal plane of a diffraction-limited telescope is given as
{}
\begin{eqnarray}
P_{\tt fp.cor}&=&\epsilon_{\tt cor}\, \Big(\frac{\pi d^2}{4{\overline z}^2}\Big)
\int_0^{2\pi}\hskip -4pt d\phi'\int_{\rho_{\tt cor}^-}^{\rho_{\tt cor}^+}\hskip -4pt \rho' d\rho'  \, B_{\tt cor}(\rho').
  \label{eq:pow-fp7c*}
\end{eqnarray}
By changing the integration variable from $\rho'$ to $\theta=\rho'/{\overline z}$ and using (\ref{eq:cor-rho}), we present (\ref{eq:pow-fp7c*}) in the equivalent form
{}
\begin{eqnarray}
P_{\tt fp.cor}
&=&\epsilon_{\tt cor}\,\pi({\textstyle\frac{1}{2}}d)^2
\int_0^{2\pi}\hskip -4pt d\phi'
\int_{\theta_{\tt cor}^-}^{\theta_{\tt cor}^+}\hskip -4pt \theta' d\theta'  \, B_{\tt cor}(\theta'),
  \label{eq:pow-fp=+*}
\end{eqnarray}
where the surface brightness $B_{\tt cor}(\theta)$ is developed from the expression (\ref{eq:model-cor}) by expressing $R_\odot/\rho'$  via  a new variable $\theta=\rho'/{\overline z}$ and $\theta_0=R_\odot/{\overline z}$, which yields the following expression for $B_{\tt cor}(\theta)$:
{}
\begin{equation}
B_{\tt cor}(\theta)= 20.09\Big[3.670 \Big(\frac{\theta_0}{\theta}\Big)^{18}+1.939\Big(\frac{\theta_0}{\theta}\Big)^{7.8}+ 5.51\times 10^{-2} \Big(\frac{\theta_0}{\theta}\Big)^{2.5}\Big]  ~~   \frac{\rm W}{{\rm m}^2\,{\rm sr}}.
\label{eq:model-th}
\end{equation}
Fig.~\ref{fig:sol-cor-bright3} shows the typical surface brightness of the solar corona from (\ref{eq:model-th}) as seen at 800 AU. It also gives the position of the Einstein disk as used in the relevant estimates of the noise from the corona surface brightness.

We can now take the advantage of the corona model discussed above. We recognize that the two terms in the expressions for $\theta^\pm_{\tt cor}$ given by (\ref{eq:cor-rho}) have very different magnitudes, namely $\theta_{\tt ER}=\sqrt{2r_g/{\overline z}}\simeq 7.795 \times 10^{-6}\,(650\,{\rm AU}/{\overline z})^\frac{1}{2}$ and $\lambda/2d\simeq 5\times 10^{-7} (\lambda/1\,\mu{\rm m})(1\,{\rm m}/d)$. This allows us to integrate (\ref{eq:pow-fp=+*})  together with (\ref{eq:model-th}) and expand the results in series of the small parameter $\lambda/(2 d)/ \theta_{\tt ER}$. For heliocentric ranges $\overline z\in[547.8, 2500]\,{\rm AU}$, we may keep only the leading term with respect to this parameter, yielding
{}
\begin{eqnarray}
P_{\tt fp.cor}
&=&10.04\, \epsilon_{\tt cor}\,\pi^2 \lambda d \, \frac{R_\odot}{\overline z} \Big[3.67 \Big(\frac{R_\odot}{\sqrt{2r_g\overline z}}\Big)^{17}+1.94\Big(\frac{R_\odot}{\sqrt{2r_g\overline z}}\Big)^{6.8}+ 5.51\times 10^{-2}\Big(\frac{R_\odot}{\sqrt{2r_g\overline z}}\Big)^{1.5}\Big] ~~    \frac{\rm W}{{\rm m}^2}.~~~~
  \label{eq:pow-fp=+*3}
\end{eqnarray}

We can rewrite this expression emphasizing that it is the middle term within the square brackets that dominates the region of our interest; the first term becomes significant for impact parameters less than 650~AU, whereas the third term only becomes relevant at 1000~AU and beyond:
\begin{eqnarray}
P_{\tt fp.cor}
&=&19.48\,\epsilon_{\tt cor}\, \pi^2 \lambda d\,\frac{R_\odot}{\overline z} \Big(\frac{R_\odot}{\sqrt{2r_g\overline z}}\Big)^{6.8}\Big[1+1.89 \Big(\frac{R_\odot}{\sqrt{2r_g\overline z}}\Big)^{10.2}+2.84\times 10^{-2}\Big(\frac{\sqrt{2r_g\overline z}}{R_\odot}\Big)^{5.3}\Big]
~~  \frac{\rm W}{{\rm m}^2}.
  \label{eq:pow-fp=+*4}
\end{eqnarray}

These results may now be used to estimate the power deposited by the solar corona in the focal plane of an imaging telescope. As such, they allow one to develop SNR estimates for various imaging scenarios involving the SGL.

\section{Averaging the PSF of the SGL}
\label{sec:PSF-average}

As derived in \cite{Turyshev-Toth:2017}, the point spread function (PSF)  of the SGL, ${\rm PSF}={ \mu}_{\tt SGL}({\vec x},{\vec y})/\mu_0$ (as given by  (\ref{eq:S_z*6z-mu2})), has the form:
{}
\begin{eqnarray}
{\rm PSF}({\vec x},{\vec y})&=&J^2_0\big(\alpha|{\vec y}-{\vec x}|\big),
\label{eq:psf*0}
\end{eqnarray}
where $\alpha$ from (\ref{eq:alpha-mu}) is given as
\begin{equation}
\alpha=k\sqrt{\frac{2r_g}{\overline z}}= 48.976 \Big(\frac{1\,\mu{\rm m}}{\lambda}\Big)\Big(\frac{650\,{\rm AU}}{\overline z}\Big)^\frac{1}{2}~{\rm m}^{-1}.
\end{equation}

As $\alpha$ is rather large, there are at least 16 oscillations of $J^2_0(\alpha |{\vec x}|)$ contained within 1 meter. Thus, unless we use a telescope whose aperture  $d$ is very small satisfying the condition $\alpha d\lesssim10$ or $d\lesssim 10/\alpha=0.2$~m (see \cite{Turyshev-Toth:2019-image} for discussion), a moderate-size telescope will not see those oscillations, but will average them. Therefore, instead of using the PSF given by (\ref{eq:psf*0}) we introduce the PSF averaged over the telescope aperture:
{}
\begin{eqnarray}
\overline{\rm PSF}({\vec x})&=&\frac{4}{\pi d^2}\iint\displaylimits_{|{\vec y}|^2\leq (\frac{1}{2}d)^2}\hskip -5pt d^2{\vec y}\, J^2_0\big(\alpha|{\vec y}-{\vec x}|\big).
\label{eq:psf_average}
\end{eqnarray}

To integrate (\ref{eq:psf_average}), we split the integral in two parts, namely i) for $|{\vec y}-{\vec x}|\leq \frac{1}{2}d$, or when the integration is conducted within the aperture  $d$, and ii) for $|{\vec y}-{\vec x}|> \frac{1}{2}d$, or when the integration is outside $d$.  We introduce a new variable ${\vec y}-{\vec x}={\vec u}$, which in the polar coordinate system has the from ${\vec u}=(u,\phi)$.

For the first integration interval (i.e., with ${\vec x}$ is within the aperture or $|{\vec x}|\equiv r\leq \frac{1}{2}d$),  $u$ and $\phi$ vary within the following limits: $\phi\in[0,2\pi]$ and  $u\in[0,\rho(\phi)]$, where, similarly to the discussion in Sec.~\ref{sec:blur-in} (see (\ref{eq:rho+})),  $\rho(\phi)$ is given as
{}
\begin{eqnarray}
\rho(\phi)&=&\sqrt{({\textstyle\frac{1}{2}}d)^2-r^2\sin^2\phi}-r\cos\phi.
\label{eq:rho}
\end{eqnarray}
With these notations, (\ref{eq:psf_average}) takes the form
{}
\begin{eqnarray}
\overline{\rm PSF}_{\tt in}({\vec x})&=&\frac{4}{\pi d^2}\iint\displaylimits_{|{\vec y}|^2\leq (\frac{1}{2}d)^2}\hskip -5pt d^2{\vec y}\, J^2_0\big(\alpha|{\vec y}-{\vec x}|\big)=\frac{4}{\pi d^2}\int_0^{2\pi}d\phi \int_0^{\rho(\phi)}\, udu J^2_0\big(\alpha u\big)=
\nonumber\\
&=&\frac{1}{2\pi}\int_0^{2\pi}d\phi \, \Big[\sqrt{1-\Big(\frac{2r}{d}\Big)^2\sin^2\phi}-\frac{2r}{d}\cos\phi \Big]^2\times \nonumber\\\
&&\hskip 10pt \times\,
\Big\{J^2_0\Big(\alpha {\textstyle\frac{1}{2}}d\Big[\sqrt{1-\Big(\frac{2r}{d}\Big)^2\sin^2\phi}-\frac{2r}{d}\cos\phi \Big]\Big)+J^2_1\Big(\alpha {\textstyle\frac{1}{2}}d\Big[\sqrt{1-\Big(\frac{2r}{d}\Big)^2\sin^2\phi}-\frac{2r}{d}\cos\phi \Big]\Big)\Big\}.~~~
\label{eq:av3}
\end{eqnarray}

Now we consider the second integration interval where ${\vec x}$ is outside the aperture or $r> \frac{1}{2}d$. In this case, similarly to the discussion in Sec.~\ref{sec:extend-photo-vic}, $u$ and $\phi$ vary within different limits, given as $\phi\in[\phi_-, \phi_+]$, where $\phi_\pm=\pm \arcsin(d/2r)$  and  $u\in[\rho_-(\phi),\rho_+(\phi)]$, where  the quantity $\rho_\pm(\phi)$ (analogous to (\ref{eq:rho++})) is given as
{}
\begin{eqnarray}
\rho_\pm(\phi)&=&\pm\sqrt{({\textstyle\frac{1}{2}}d)^2-r^2\sin^2\phi}+r\cos\phi.
\label{eq:rho-pm}
\end{eqnarray}
With these notations, (\ref{eq:psf_average}) may be integrated:
{}
\begin{eqnarray}
\overline{\rm PSF}_{\tt out}({\vec x})&=&\frac{4}{\pi d^2}\iint\displaylimits_{|{\vec y}|^2> (\frac{1}{2}d)^2}\hskip -5pt d^2{\vec y}\, J^2_0\big(\alpha|{\vec y}-{\vec x}|\big)=\frac{4}{\pi d^2}\int_{\phi_-}^{\phi_+}d\phi \int_{\rho_-(\phi)}^{\rho_+(\phi)}\, udu J^2_0\big(\alpha u\big)=
\nonumber\\
&=&\frac{1}{2\pi}
\frac{4}{d^2}\int_{\phi_-}^{\phi_+}d\phi \, \Big\{\rho_+^2(\phi)\Big(J^2_0\big(\alpha \rho_+(\phi)\big)+J^2_1\big(\alpha \rho_+(\phi)\big)\Big)-\rho_-^2(\phi)\Big(J^2_0\big(\alpha \rho_-(\phi)\big)+J^2_1\big(\alpha \rho_-(\phi)\big)\Big)\Big\},
\label{eq:av4}
\end{eqnarray}
where $\phi_\pm=\pm \arcsin(d/2r)$ and $\rho_\pm(\phi)$ is given by (\ref{eq:rho-pm}).

\begin{figure}[h]
\includegraphics[scale=0.8]{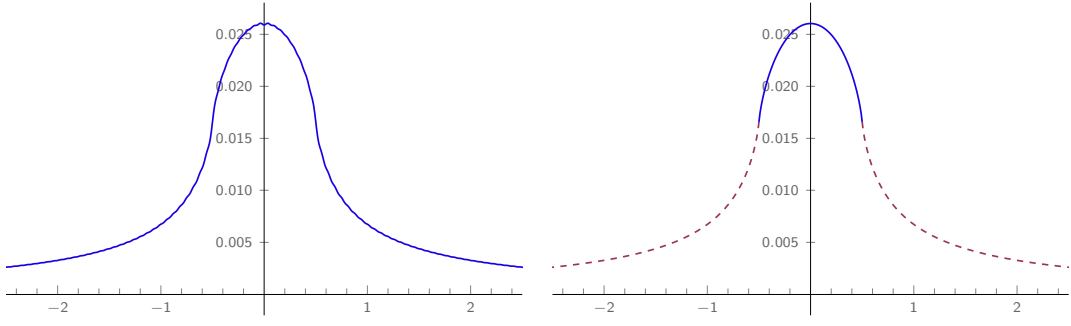}\\
\caption{\label{fig:sgl-num-appr3} Behavior of the averaged PSF of the SGL. Left: numerical integration of Eq.~(\ref{eq:psf_average}). Evaluating the analytical expression for the averaged PSF given by the combination of  Eqs.~(\ref{eq:av3})  and (\ref{eq:av4}) yields an identical plot. Right: the PSF from Eq.~(\ref{eq:psf-mu}) given by $\epsilon(r)$ (blue) and $\beta(r)$ (red, dashed). The plots are nearly identical. Note that a minor oscillatory behavior evident on the left is absent on the right. Horizontal axis is distance from the center of the aperture in meters.
}
\end{figure}

Given the fact that the arguments of the Bessel functions in (\ref{eq:av3})  and (\ref{eq:av4}) are large (this is especially true for  (\ref{eq:av4})), we may use the approximations for the Bessel functions for large arguments (\ref{eq:BF}) and simplify these two expressions. Thus, for (\ref{eq:av4}) we have
{}
\begin{eqnarray}
\overline{\rm PSF}_{\tt in}({\vec x})&=&
\frac{1}{\pi \alpha}\frac{4}{d}\frac{1}{2\pi}\int_0^{2\pi}d\phi \,\sqrt{1-\Big(\frac{2r}{d}\Big)^2\sin^2\phi}=\frac{1}{\pi \alpha}\frac{4}{d}\epsilon(r),
\label{eq:av3a}
\end{eqnarray}
where $\epsilon(r)$ is equivalent to (\ref{eq:eps_r0})
{}
\begin{eqnarray}
\epsilon(r)&=&\frac{1}{2\pi}\int_0^{2\pi}d\phi \,\sqrt{1-\Big(\frac{2r}{d}\Big)^2\sin^2\phi}=\frac{2}{\pi}{\tt E}\Big[\Big(\frac{2r}{d}\Big)^2\Big],
\label{eq:av3b}
\end{eqnarray}
with ${\tt E}[x]$ being the elliptic integral \cite{Abramovitz-Stegun:1965}, which is similar to (\ref{eq:eps_r0}) obtained for a uniform surface brightness.

Similarly, we have for (\ref{eq:av4}):
{}
\begin{eqnarray}
\overline{\rm PSF}_{\tt out}({\vec x})&=&
\frac{1}{2\pi}\frac{4}{d^2}\frac{2}{\pi \alpha}
\int_{\phi_-}^{\phi_+}d\phi \, \Big(\rho_+(\phi)-\rho_-(\phi)\Big)=
\frac{1}{\pi \alpha}\frac{4}{d}\frac{1}{\pi}\int_{\phi_-}^{\phi_+}d\phi \, \sqrt{1-\Big(\frac{2r}{d}\Big)^2\sin^2\phi}=
\frac{1}{\pi \alpha}\frac{4}{d}\beta(r),
\label{eq:av4b}
\end{eqnarray}
where $\beta(r)$ is equivalent to (\ref{eq:beta_r0})
{}
\begin{eqnarray}
\beta(r)&=&\frac{1}{\pi}\int_{\phi_-}^{\phi_+}d\phi \, \sqrt{1-\Big(\frac{2r}{d}\Big)^2\sin^2\phi}=\frac{2}{\pi}{\tt E}\Big[\arcsin \Big(\frac{d}{2r}\Big),\Big(\frac{2r}{d}\Big)^2\Big],
\label{eq:av3bb}
\end{eqnarray}
with ${\tt E}[a,x]$ being the incomplete elliptic integral \cite{Abramovitz-Stegun:1965}. This result is similar to (\ref{eq:beta_r0}), which was obtained for a uniform surface brightness and an off-image telescope pointing.

The similarities between Eqs.~(\ref{eq:av3b}) and (\ref{eq:av3bb}), on the one hand, and Eqs.~(\ref{eq:eps_r0}) and (\ref{eq:beta_r0}) on the other, though striking, should not be surprising. The fundamental geometry of the problem of mapping light from a uniformly illuminated disk to a location in the image plane vs. the geometry of mapping light from a point source to the uniformly sampled, finite, circular area of a telescope aperture in the image plane are identical.

Thus, the averaged PSF takes the form:
{}
\begin{eqnarray}
\overline{\rm PSF}({\vec x})=\overline{\rm PSF}_{\tt in}({\vec x})+\overline{\rm PSF}_{\tt out}({\vec x})=\frac{1}{\pi \alpha}\frac{4}{d}\,\mu(r),
\qquad{\rm with}\qquad
\mu(r)&=&
 \bigg\{ \begin{aligned}
\epsilon(r), \hskip 10pt 0\leq r \leq  {\textstyle\frac{1}{2}}d& \\
\beta(r), \hskip 30pt r > {\textstyle\frac{1}{2}}d& \\
  \end{aligned}\,.
  \label{eq:psf-mu}
\end{eqnarray}

Figure~\ref{fig:sgl-num-appr3} shows that this expression (\ref{eq:psf-mu}) is a very good approximation of the averaged PSF (\ref{eq:psf_average}). Apart from the mild oscillatory behavior in (\ref{eq:psf_average}) (which arises due to random phases of the Bessel function at the integration boundary), which is absent from (\ref{eq:psf-mu}), the two representations are identical. Eq.~(\ref{eq:psf-mu}), therefore, is a suitable representation of the SGL PSF in high-fidelity numerical approximations.

 \begin{figure}[h]
\includegraphics[scale=0.8]{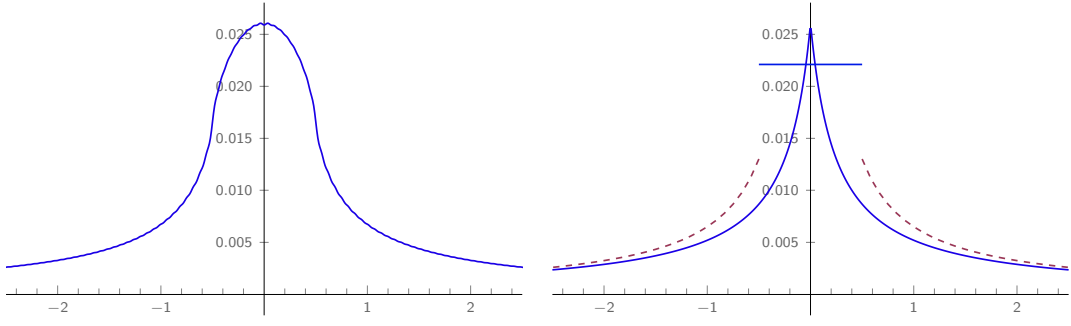}\\
\caption{\label{fig:sgl-num-appr5} Behavior of the averaged PSF of the SGL. Left: numerical integration of Eq.~(\ref{eq:psf_average}). Right:
the piecewise defined PSF from Eqs.~(\ref{eq:psf-mu*}) and the smoothed PSF from Eq.~(\ref{eq:psf-av3}).
Horizontal axis is distance from the center of the aperture in meters.}
\end{figure}

Although the expression (\ref{eq:psf-mu}) is much simpler than Eq.~(\ref{eq:psf_average}), it is still not very convenient for estimating changes in the SNR during deconvolution. For that, instead of $\epsilon(r)$ from (\ref{eq:av3b}), we take its mean value within the aperture:
{}
\begin{eqnarray}
\overline \epsilon=\frac{4}{\pi d^2}\int_0^{2\pi}d\phi' \int_0^{\frac{1}{2}d}r dr \epsilon(r)&=&\frac{1}{\pi} \int_0^{2\pi}d\phi \,\int_0^1q dq \sqrt{1-q^2\sin^2\phi}=\frac{8}{3\pi}.
\label{eq:av5a}
\end{eqnarray}
In addition, (\ref{eq:av3bb}) may be approximated as
{}
\begin{eqnarray}
\beta(r)&=&\frac{1}{\pi}\int_{\phi_-}^{\phi_+}d\phi \, \sqrt{1-\Big(\frac{2r}{d}\Big)^2\sin^2\phi}\simeq \frac{d}{4r}.
\label{eq:av5b}
\end{eqnarray}

With these approximations, the averaged PSF (\ref{eq:psf-mu}) may be given as
{}
\begin{eqnarray}
\overline{\rm PSF}({\vec x})=\frac{1}{\pi \alpha}\frac{4}{d}\,\mu(r),
\qquad{\rm with}\qquad
\mu(r)&=&
 \Bigg\{ \begin{aligned}
\frac{8}{3\pi}, \hskip 10pt 0\leq r \leq {\textstyle\frac{1}{2}}d& \\
\frac{d}{4r}, \hskip 30pt r > {\textstyle\frac{1}{2}}d& \\
  \end{aligned}\, .
  \label{eq:psf-mu*}
\end{eqnarray}
Fig.~\ref{fig:sgl-num-appr5} shows the result (\ref{eq:psf-mu*}) comparing it to the numerically integrated (\ref{eq:psf_average}).

Alternatively, $\epsilon(r)$ may be approximated by its value at the center of the aperture, $\epsilon(0)=1$,
yielding
{}
\begin{eqnarray}
\overline{\rm PSF}({\vec x})=\frac{1}{\pi \alpha}\frac{4}{d}\,\mu(r),
\qquad{\rm with}\qquad
\mu(r)&=&
 \Bigg\{ \begin{aligned}
1, \hskip 10pt 0\leq r \leq {\textstyle\frac{1}{2}}d& \\
\frac{d}{4r}, \hskip 30pt r > {\textstyle\frac{1}{2}}d& \\
  \end{aligned}\,,
  \label{eq:psf-mu*2}
\end{eqnarray}
which slightly overestimates the contribution from the directly-imaged region.

Note that expression (\ref{eq:psf-mu*2}) is the form of the averaged PSF that we implicitly used in  \cite{Turyshev-Toth:2019-blur,Turyshev-Toth:2019-image} to derive the power from the directly-imaged region and that from the rest of the exoplanet.

The piecewise-defined result given by Eq.~(\ref{eq:psf-mu*2}) consists of two discontinuous parts, representing the two regions where the corresponding solutions were obtained, namely $r\leq \frac{1}{2}d$ and $r> \frac{1}{2}d$. To derive continuous version of the $\overline{\rm PSF}({\vec x})$, we combine these expressions to form
{}
\begin{eqnarray}
\overline{\rm PSF}({\vec x})&\simeq&\frac{4}{\pi\alpha }\frac{1}{4r+d}.
\label{eq:psf-av3}
\end{eqnarray}

Result (\ref{eq:psf-av3}) is not perfect, but still a good approximation of (\ref{eq:psf_average}).  This can be seen from Fig.~\ref{fig:sgl-num-appr5} that shows the result of a numerical integration of (\ref{eq:psf_average}) and the behavior of the smoothed PSF from (\ref{eq:psf-av3}). The two solutions are quite different within the aperture, but match each other quite well for $r/d\gg 1$.

\end{document}